\documentclass[traditabstract]{aa}
\usepackage{graphicx}
\usepackage{txfonts}
\usepackage{longtable}
\usepackage{natbib}
\usepackage{hyperref}
\usepackage{url}
\usepackage{float,color}
\usepackage[figuresright]{rotating}
\bibpunct{(}{)}{;}{a}{}{,}
\newcommand{\teff}{T_{\rm eff}}
\newcommand{\logg}{\log{g}}
\newcommand{\feh}{[\rm{Fe}/\rm{H}]}
\newcommand{\Vmic}{\xi_{\rm t}}

\newcommand{\hi}{H\,{\textsc i}\rm}

\newcommand{\mni}{Mn\,{\textsc i}\rm}

\newcommand{\mnii}{Mn\,{\textsc {ii}}\rm}
\newcommand{\mniii}{Mn\,{\textsc {iii}}\rm}
\newcommand{\Mn}[5]{\mbox{$#1\,^#2{\rm #3}^{{\rm #4}}_{\rm #5}$}}

\newcommand{\Elow}{E_{\rm low}}

\newcommand{\opd}{\log \tau_{\rm 5000}}

\newcommand{\tda}{$\langle{\rm 3D}\rangle$}
\newcommand{\multitd}{MULTI3D}

\hypersetup{
  colorlinks, 
  citecolor=blue, 
  filecolor=blue, 
  linkcolor=blue,
  breaklinks=true, 
  plainpages=false,
  urlcolor=blue 
}
\begin{document}

\title{Observational constraints on the origin of the elements}
\subtitle{I. 3D NLTE formation of Mn lines in late-type stars}
\author{Maria Bergemann\inst{1} \and 
       Andrew J. Gallagher\inst{1} \and 
       Philipp Eitner\inst{1,2} \and 
       Manuel Bautista\inst{3} \and 
       Remo Collet\inst{4} \and 
       Svetlana A. Yakovleva\inst{5} \and 
       Anja Mayriedl \inst{6} \and 
       Bertrand Plez\inst{7} \and 
       Mats Carlsson\inst{8,9} \and
       Jorrit Leenaarts\inst{10} \and 
       Andrey K. Belyaev\inst{5} \and
       Camilla Hansen\inst{1}}
\institute{Max Planck Institute for Astronomy, 69117, Heidelberg, Germany \email{bergemann@mpia-hd.mpg.de}
\label{inst1}
\and
Ruprecht-Karls-Universit\"at, Grabengasse 1, 69117 Heidelberg, Germany 
\label{inst2}
\and
Department of Physics, Western Michigan University, Kalamazoo, Michigan 49008, USA
\label{inst3}
\and
Stellar Astrophysics Centre, Department of Physics and Astronomy, Aarhus University, DK-8000 Aarhus C, Denmark
\label{inst4}
\and
Department of Theoretical Physics and Astronomy, Herzen University, St. Petersburg 191186, Russia
\label{inst5}
\and
Montessori-Schule Dachau, Geschwister-Scholl-Str. 2, 85221 Dachau, Germany
\label{inst6}
\and
LUPM, UMR 5299, Universit\'{e} de Montpellier, CNRS, 34095 Montpellier, France 
\label{inst7}
\and
Rosseland Centre for Solar Physics, University of Oslo, P.O. Box 1029 Blindern, NO-0315 Oslo, Norway
\label{inst8}
\and
Institute of Theoretical Astrophysics, University of Oslo, P.O. Box 1029 Blindern, NO-0315 Oslo, Norway
\label{inst9}
\and
Institute for Solar Physics, Department of Astronomy, Stockholm University, AlbaNova University Centre, SE-106 91 Stockholm, Sweden
\label{inst10}
}

\date{Received date / Accepted date}
\abstract{Manganese (Mn) is a key Fe-group elements, commonly employed in stellar population and nucleosynthesis studies to explore the role of SN Ia. We have developed a new non-local thermodynamic equilibrium (NLTE) model of Mn, including new photo-ionisation cross-sections and new transition rates caused by collisions with H and H$^-$ atoms. We applied the model in combination with 1-dimensional (1D) LTE model atmospheres and 3D hydrodynamical simulations of stellar convection to quantify the impact of NLTE and convection on the line formation. We show that the effects of NLTE are present in Mn I and, to a lesser degree, in Mn II lines, and these increase with metallicity and with effective temperature of a model. Employing 3D NLTE radiative transfer, we derive new abundance of Mn in the Sun, A(Mn)$=5.52 \pm 0.03$ dex, consistent with the element abundance in C I meteorites. We also apply our methods to the analysis of three metal-poor benchmark stars. We find that 3D NLTE abundances are significantly higher than 1D LTE. For dwarfs, the differences between 1D NLTE and 3D NLTE abundances are typically within $0.15$ dex, however, the effects are much larger in the atmospheres of giants owing to their more vigorous convection. We show that 3D NLTE successfully solves the ionisation and excitation balance for the RGB star HD 122563 that cannot be achieved by 1D LTE or 1D NLTE modelling. For HD 84937 and HD 140283, the ionisation balance is satisfied, however, the resonance Mn I triplet lines still show somewhat lower abundances compared to the high-excitation lines. Our results for the benchmark stars confirm that 1D LTE modelling leads to significant systematic biases in Mn abundances across the full wavelength range from the blue to the IR. We also produce a list of Mn lines that are not significantly biased by 3D and can be reliably, within the $0.1$ dex uncertainty, modelled in 1D NLTE.}

\titlerunning{Mn}

\authorrunning{Bergemann et al.}

\maketitle
\section{Introduction}
Manganese (Mn) is a prominent member of the iron-group family that has interesting connections to several topics in astrophysics. In particular, from the point of view of stellar nucleosynthesis, this element is very sensitive to the physical conditions in supernovae Type Ia (SNIa) \citep{Seitenzahl2013}. Hence, the abundances of Mn in metal-poor stars provide powerful constraints on the progenitors and explosion mechanism of this important class of SNe.

Mn displays a large number of \mni~lines spanning a range of excitation potentials in the optical spectra of late-type stars \citep{bergemann2007}. Also a few lines of \mnii~can be detected in the blue at $\sim 350$ nm, and some strong lines of \mni~are available in the infra-red at 1.52 $\mu$m. Owing to the large number of observable lines, Mn is a useful element to test the excitation and ionisation equilibria in stellar atmospheres. The lines of both ionisation stages are  affected by hyperfine splitting (HFS), and some are also very sensitive to stellar activity. For example, the resonance \mni~line at 5394 \AA\ is known to vary across the solar cycle \citep{Vitas2009,Danilovic2016}.

A large number of studies over the past years have been devoted to the analysis of Mn abundances in the context of stellar population studies and nucleosynthesis. Most of these works have assumed local thermodynamic equilibrium (LTE). There is, however, evidence for the breakdown of the LTE assumption. \citet{Johnson2002} reported a systematic ionisation imbalance of \mni~and\mnii~in metal-poor stars. \citet{Bonifacio2009} found a $0.2$ dex offset between the abundances of Mn in metal-poor dwarfs and giants. They also observe a significant excitation imbalance, with strong \mni~resonance lines resulting in significantly lower abundances compared to the high-excitation features. \citet{Sneden2016} confirm the excitation imbalance in LTE, but they also find that the ionisation balance is satisfied, if one relies on the high-excitation \mni~lines only. However, that study employed one star only, HD 84937, which can make it difficult to generalise these conclusions to a large sample. \citet{Mishenina2015} also employed LTE models to analyse a large sample of main-sequence stars in the metallicity range from $-1$ to $+0.3$. Their abundances suggests a modest systematic correlation with $\teff$, signifying potential departures from LTE and 1D hydrostatic equilibrium. 

In earlier studies \citep{bergemann2007, Bergemann2008}, we showed that Mn is very sensitive to departures from LTE, also known as Non-LTE (hereafter, NLTE) effects. This is an element of the Fe-group, and is expected to be similar to Fe in terms of line formation properties. However, given its two orders of magnitude lower abundance (in the cosmic abundance scale) compared to Fe, but also significantly higher photo-ionisation cross-sections, and a peculiar atomic structure, with a very large number of strong radiative transitions between energy levels with excitation potentials of 2 and 4 eV, Mn is prone to stronger NLTE effects than Fe. In particular, it was shown, on the basis of detailed statistical equilibrium (SE) calculations, that NLTE Mn abundances are significantly higher compared to LTE, and this effect increases with decreasing metallicity of a star and increasing its $\teff$, but also with decreasing $\log g$.

Line formation across the solar granulation has been extensively discussed in the literature, in particular in the series of seminal papers by \citet{Dravins1981, Dravins1987, Dravins1990a, Dravins1990b, Nordlund1990}, but see also more recent studies of the solar center-to-limb variation \citep[e.g.][]{Lind2017} and solar abundances \citep[e.g.][]{Caffau2008, Caffau2009, Asplund2009, Caffau2010, Caffau2011,Amarsi2017, Amarsi2018a,Amarsi2019b}. Recently, this work has been extended towards 3D NLTE modelling of spectral line formation in other stars and applied to the lines of H, O, Si, and Al \citep{Amarsi2016, Nordlander2017,Amarsi2018b,Amarsi2019a}.

Given the interest in the impact of NLTE and 3D diagnostic on the abundance measurements, we present a re-analysis of Mn abundances in a small sample of well-studied FGK stars using an updated NLTE model atom, and 1D hydrostatic and 3D hydrodynamical model atmospheres. We use new atomic data, including transition probabilities, photo-ionisation cross-sections, and rate coefficients for the transitions caused by the inelastic collisions of \mni~and \mnii~ions with H atoms. We compare the results of two 1D statistical equilibrium codes, DETAIL and MULTI2.3 that are both widely used in the community for NLTE analyses of chemical abundances. We also perform full 3D NLTE radiative transfer calculations with the \multitd\ code \citep{Leenaarts2009} to derive Mn abundance from the high-resolution spectra of the Sun and several metal-poor stars.

The paper is organised as follows: In Sect. \ref{sec:obs}, we describe the observed spectra. The LTE and NLTE calculations in 1D and 3D, spectrum synthesis, and abundance analysis are documented in Sect. \ref{sec:analysis}. We present a considerable amount of details about the methods of calculations, because this is very important for a judgement of the resulting abundances. The  results are presented in Sect. \ref{sec:results}. They include the discussion of 1D NLTE and 3D NLTE abundance corrections, the analysis of the solar Mn abundance, a comparison between 3D LTE and 3D NLTE line profiles, and the excitation-ionisation balance of \mni$/$\mnii~in benchmark metal-poor stars. We close with a summary of the results and conclusions in Sect. \ref{sec:conclusions}.
\section{Observations}\label{sec:obs}
For the Sun, we use the high-resolution flux atlas taken with the KPNO facility \citep{kurucz1984}. The atlas has a resolving power $R \sim 400\,000$. Recently, solar spectra taken with the PEPSI instrument at Large Binocular Telescope (LBT) \citep{strassmeier2018} and with the Fourier transform spectrograph operated by the Institut fuer Astrophysik in Goettingen \citep{Reiners2016} were released. However, the profiles of Mn lines are very similar in all these atlases. Hence, we employ the KPNO spectrum in this work.

We also include several metal-poor benchmark stars (HD 84937, HD 140283, HD 122563) from \citet{Bergemann2012}. Their spectra are taken from the UVES-POP database \citep{Bagnulo2003}. These are the Gaia benchmark stars with $\teff$ and $\log g$  determined using interferometry and astrometry. The estimates of $\feh$ and micro-turbulence were derived using NLTE radiative transfer for Fe lines \citep{Bergemann2012}. The effective temperatures of two of these stars were recently revised \citep{Karovicova2018}. The new estimates, based on the CHARA angular diameters, are $\teff = 5787 \pm 48$ K for HD 140283 and $\teff = 4636 \pm 37$ K for HD 122563. These estimates are fully consistent with the values we adopted in \citet{Bergemann2012}. \citet{Creevey2019} propose a new asteroseismic surface gravity for HD 122563, $\log g = 1.39 \pm 0.01$ dex. This is a substantial downward revision of this parameter. However, we tested the effect of $\log g$ on the abundance estimates from Mn lines, but found that the abundances change by only $0.05$ dex. Hence we do not recompute the model and use the standard models employed in \citet{Bergemann2012}.
\section{Analysis}{\label{sec:analysis}}
\subsection{Model atom and diagnostic lines of Mn}

\begin{figure}
\includegraphics[width=0.45\textwidth, angle=0]{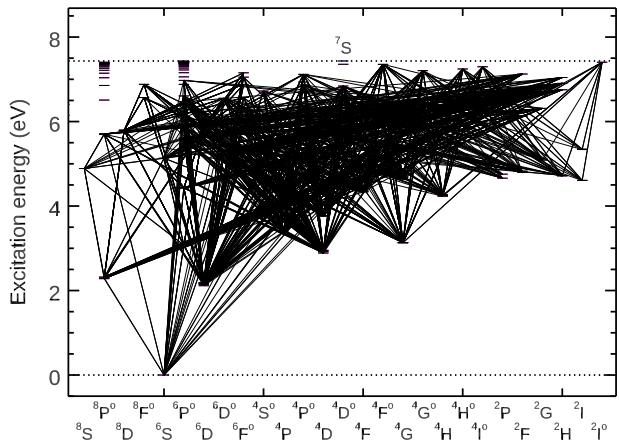} 
\caption{The Grotrian diagram of Mn I atomic system. The energy levels of \mni~ are shown with black dashes. The levels are connected by radiative transitions (solid black lines).}
\label{fig:grotrian}
\end{figure}

The model comprises three ionisation stages and $281$ energy levels, with 198 levels of \mni, 81 levels of \mnii, and the model is closed by the \mniii~ground state. The radiative transitions are taken from the Kurucz compilation \footnote{\url{http://kurucz.harvard.edu/atoms/2500/}}, which includes theoretically predicted and experimental estimates of the oscillator strengths, with the latter given a preference over theoretical estimates. The \mni~part of the model atom is shown in Fig. \ref{fig:grotrian}. We do not show the \mnii~system in this plot. The ionised specie has a very high ionisation potential and the bulk of \mnii~lines, which connect the levels do not play any role in the SE of the element \citep[see also the discussion of model atom completeness in][]{Bergemann2008}. In contrast to the latter study, we do not include fine structure for most of the \mnii\ levels, except those, which are relevant for the \mnii\ near-UV lines used in detailed abundance measurements. The full atom is provided in the MULTI2.3 format as a supplementary material in this paper (see online version).

We include fine structure levels for all energies up to $47\,300$ cm$^{-1}$, however, we also test the results using a compact model atom, which is devoid of fine structure for the majority of levels. This is important for our test calculations with full 3D simulations of stellar convection. 

The atomic data for the \mni~lines, which we employ in the abundance calculations are given in Table \ref{lines}. For all of them, the hyperfine structure (HFS) is included in the model atom directly. That is, the HFS structure of spectral lines was computed for all diagnostic lines of \mni~and \mnii, and included in the radiative transition part of the MULTI model atom. We employed the magnetic dipole constants $A$ and electric quadrupole constants $B$ assembled by \citet{Bergemann2008}, complementing these with the data from \citet{Holt1999} for the relevant \mnii~levels, \Mn{a}{5}{D}{}{} and \Mn{z}{5}{P}{\circ}{}. The full HFS patterns for each Mn line is provided in the Appendix. In the statistical equilibrium (SE) calculations, we treat the diagnostic \mni~ and \mnii~ lines with Voigt profiles, while all other Mn lines were computed with a Gaussian profile with $13$ frequency points. Our tests show that increasing the number of frequency points does not change the statistical equilibrium of the ion, on the other hand with this choice we still have a reasonable frequency quadrature to accurately represent each line profile.

We use the new experimental transition probabilities, where these are available. Most data are taken from \citet{Blackwell-Whitehead2007} and \citet{Denhartog2011}. For some of the lines, the new $\log gf$ values are typically $0.05$ to $0.1$ dex lower than the old values, that ultimately leads to slightly higher abundances compared to our earlier work. The broadening due to elastic collisions with H atoms is adopted from \citet{Barklem2000}, where available. These data are derived using the 2$^{\rm nd}$ order Rayleigh-Schroedinger perturbation theory as formulated by \citet{Brueckner1971} and later generalised to transitions between different azimuthal quantum number states by \citet{Omara1976}, \citet{Anstee1991}, and \citet{Anstee1995}. This theory offers a more accurate representation of H $+$ atom collision broadening than the theory by \citet{Unsold1927,Unsold1955}. The latter theory assumes that only collisions at large separations between atoms strongly influence the line broadening, thus severely under-estimating the line strengths.
\begin{table}
\begin{minipage}{\linewidth}
\renewcommand{\footnoterule}{} 
\caption{Parameters of lines used for abundance calculation.}
 \label{lines}     
\begin{center}
\begin{tabular}{lrrlccc}
\noalign{\smallskip}\hline\noalign{\smallskip}  ~~~~~~$\lambda$ & M &
$N_{\rm HFS}$ & $\Elow$ & Lower & Upper & $\log gf$ \\
     ~~~~[\AA] &  &  & [eV] & level & level &   \\
\noalign{\smallskip}\hline\noalign{\smallskip}
  3488.68  &  -- &  3  &  1.85  &  \Mn{a}{5}{D}{ }{1}   & \Mn{z}{5}{P}{o}{1}    &  --0.937 \\
  3496.81  &  -- &  2  &  1.83  &  \Mn{a}{5}{D}{ }{2}   & \Mn{z}{5}{P}{o}{3}    &  --1.779 \\
  3497.53  &  -- &  3  &  1.85  &  \Mn{a}{5}{D}{ }{1}   & \Mn{z}{5}{P}{o}{2}    &  --1.418 \\
  4018.10   &  9  & 5  &  2.12  &  \Mn{a}{6}{D}{ }{9/2} & \Mn{z}{6}{D}{o}{7/2}  &  --0.311 \\
  4030.76   &  4  & 5  &  0.00  &  \Mn{a}{6}{S}{ }{5/2} & \Mn{z}{6}{P}{o}{7/2}  &  --0.497 \\
  4033.07   &  4  & 4  &  0.00  &  \Mn{a}{6}{S}{ }{5/2} & \Mn{z}{6}{P}{o}{5/2}  &  --0.647 \\
  4034.49   &  4  & 3  &  0.00  &  \Mn{a}{6}{S}{ }{5/2} & \Mn{z}{6}{P}{o}{3/2}  &  --0.843 \\
  4055.54   &  9  & 4  &  2.14  &  \Mn{a}{6}{D}{ }{7/2} & \Mn{z}{6}{D}{o}{7/2}   & --0.077  \\
  4070.28   &  9  & 3  &  2.19  &  \Mn{a}{6}{D}{ }{1/2} & \Mn{z}{6}{D}{o}{1/2}   & --1.039  \\
  4451.58   &  24 & 3  &  2.89  &  \Mn{a}{4}{D}{ }{7/2} & \Mn{z}{4}{D}{o}{7/2}   &   0.278  \\
  4498.90   &  24 & 1  &  2.94  &  \Mn{a}{4}{D}{ }{2/2} & \Mn{z}{4}{D}{o}{5/2}   &   0.343  \\
  4502.22   &  24 & 2  &  2.92  &  \Mn{a}{4}{D}{ }{5/2} & \Mn{z}{4}{D}{o}{7/2}   & --0.345  \\
  4754.03  &  18 & 5  &  2.27  &  \Mn{z}{8}{P}{o}{5/2} & \Mn{e}{8}{S}{ }{7/2} &  --0.086 \\  
  4761.52  &  23 & 4  &  2.94  &  \Mn{a}{4}{D}{ }{1/2} & \Mn{z}{4}{F}{o}{3/2} &  --0.138 \\  
  4762.37  &  23 & 5  &  2.88  &  \Mn{a}{4}{D}{ }{7/2} & \Mn{z}{4}{F}{o}{9/2} &    0.425 \\  
  4765.86  &  23 & 3  &  2.93  &  \Mn{a}{4}{D}{ }{3/2} & \Mn{z}{4}{F}{o}{5/2} &  --0.080 \\  
  4766.42  &  23 & 4  &  2.91  &  \Mn{a}{4}{D}{ }{5/2} & \Mn{z}{4}{F}{o}{7/2} &    0.100 \\  
  4783.42  &  18 & 5  &  2.29  &  \Mn{z}{8}{P}{o}{7/2} & \Mn{e}{8}{S}{ }{7/2} &    0.042 \\  
  4823.51  &  18 & 6  &  2.31  &  \Mn{z}{8}{P}{o}{9/2} & \Mn{e}{8}{S}{ }{7/2} &    0.144 \\  
  5004.89  &  22 & 4  &  2.91  &  \Mn{a}{4}{D}{ }{5/2} & \Mn{z}{6}{F}{o}{7/2} &  --1.630 \\  
  5117.93  &  39 & 3  &  3.12  &  \Mn{a}{4}{G}{ }{5/2} & \Mn{z}{4}{F}{o}{3/2} &  --1.140 \\  
  5255.31  &  39 & 6  &  3.12  &  \Mn{a}{4}{G}{ }{11/2}& \Mn{z}{4}{F}{o}{9/2} &  --0.763 \\  
  5394.67  &  2  & 6  &  0.00  &  \Mn{a}{6}{S}{ }{5/2} & \Mn{z}{8}{P}{o}{7/2} &  --3.503 \\  
  5407.42  &  8  & 10 &  2.13  &  \Mn{a}{6}{D}{ }{7/2} & \Mn{y}{6}{P}{o}{7/2} &  --1.743 \\  
  5420.35  &  8  & 9  &  2.13  &  \Mn{a}{6}{D}{ }{7/2} & \Mn{y}{6}{P}{o}{5/2} &  --1.462 \\  
  5432.54  &  2  & 5  &  0.00  &  \Mn{a}{6}{S}{ }{5/2} & \Mn{z}{8}{P}{o}{5/2} &  --3.795 \\  
  5470.63  &  4  & 8  &  2.15  &  \Mn{a}{6}{D}{ }{5/2} & \Mn{y}{6}{P}{o}{5/2} &  --1.702 \\  
  5516.77  &  8  & 8  &  2.17  &  \Mn{a}{6}{D}{ }{3/2} & \Mn{y}{6}{P}{o}{3/2} &  --1.847 \\  
  5537.75  &  4  & 5  &  2.18  &  \Mn{a}{6}{D}{ }{1/2} & \Mn{y}{6}{P}{o}{3/2} &  --2.017 \\  
  6013.49  &  32 & 6  &  3.06  &  \Mn{z}{6}{P}{o}{3/2} & \Mn{e}{6}{S}{ }{5/2} &  --0.251 \\  
  6016.64  &  32 & 6  &  3.06  &  \Mn{z}{6}{P}{o}{5/2} & \Mn{e}{6}{S}{ }{5/2} &  --0.216 \\  
  6021.79  &  32 & 6  &  3.06  &  \Mn{z}{6}{P}{o}{7/2} & \Mn{e}{6}{S}{ }{5/2} &    0.034 \\  
\noalign{\smallskip}\hline\noalign{\smallskip}
\end{tabular}
\end{center}
\end{minipage}
\end{table}

The main difference in this work with respect to our earlier studies \citep{Bergemann2008} is the implementation of the new photo-ionisation cross-sections for \mni~ and the new rates of inelastic collisions due to the interactions of \mni~ with \hi~atoms. 
\subsubsection{Photoionisation}

We adopt new quantum-mechanical photoionisation cross-sections for $84$ LS  terms of \mni, which belong to the configurations $3d^6$, $3d^54s$, $3d^54p$, $3d^44s^2$. The photo-ionisation cross-sections for dipole allowed transitions in \mni~were computed using the R-matrix method for atomic scattering calculations \citep{Berrington1987}. The solution of the Schr\"odinger equation for the N$+1$ electron system is found on the basis of the close-coupling expansion of the wavefunction as
\begin{equation}
\Psi(E; SL\pi) = A \sum_i \chi_i\theta_i + \sum_j c_j\phi_j,
\end{equation}
where $A$ is the anti-symmetrisation operator, $\chi_i$ is the target ion wavefunctions in the target state $(SL\pi)_i$, $\theta_j$ is the wavefunction for the free electron, and $\phi_j$ are short range correlation functions for the bound $(e^- +$ion) system.

The calculations are done in LS-coupling and include all states with valence electron excitations up to the principal quantum number $n=10$. The single electron orthogonal orbitals that represent the atomic structure of the Mn$^+$ target, were derived using the {\sc AUTOSTRUCTURE} code \citep{Badnell1997}. The code employs scaled Thomas-Fermi-Dirac-Amaldi central-field potential. We adopt  configuration interactions expansion with spectroscopic orbitals 1s, 2s, 2p, 3s, 3p 3d, 4s, 4p, 4d, 5s, 5p. The configurations and scaling parameters for the orbitals are presented in Table \ref{table:AUTOconfigs} in the Appendix. LS terms of the target \mnii~ion included in the close coupling expansion are provided in Table \ref{fe2target} in the Appendix.

The cross-sections are sampled at $5\,000$ evenly-spaced energy points between zero and $0.8$ Ryd above the first ionisation threshold, followed by $250$ points from $0.8$ Ryd to $2.0$ Ryd. This mesh is also preserved in our NLTE calculations with DETAIL and with MULTI2.3, such that all resonances are fully accounted for in the statistical equilibrium  calculations. These resonances are mostly caused by the photo-excitation of the core and dominate the cross-sections of the majority of \mni~states. For the other levels, we employ hydrogenic cross-sections sampled on a regular mesh. The hydrogenic cross-sections were computed using the effective principal quantum number. 

\begin{figure}
\includegraphics[width=0.35\textwidth, angle=-90]{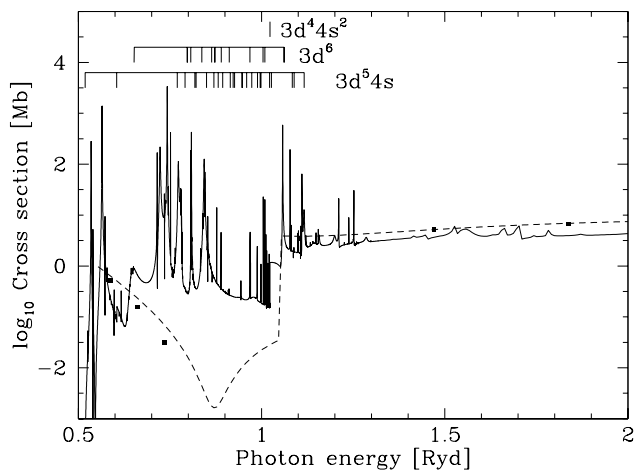}
\caption{Photo-ionisation cross-section of the 3d$^5$4s$^2$~~a~$^6$S ground term of \mni. The present R-matrix cross section is depicted by the solid line. The central-field cross sections of Verner \& Yakovlev (1995) and Reilman \& Manson (1979) are indicated by the dashed line and square dots, respectively. Above the figure we show the energies of all 3d$^5$4s, 3d$^6$, and 3d$^4$4s$^2$ thresholds.}
\label{fig:photo}
\end{figure}

Figure \ref{fig:photo} shows the total photo-ionisation cross-section for the ground state of \mni, compared with the central field approximation results of \citet{Verner1995} and \citet{Reilman1979}. The close coupling expansion accounts for the photoionisation of the outer 4s sub-shell, as well as, the open inner 3d sub-shell of the ground state 3d$^5$4s$^2$~$^6$S of Mn. The coupling of all relevant photo-ionisation channels results in extensive auto-ionisation structures and ensures that no sharp discontinuity in the cross-section at the 3d inner-shell ionisation edge is present. By contrast, the central field approximation, which misses channel couplings, yields a cross-section, which is severely underestimated up to the opening of the 3d sub-shell, where a discontinuity appears. The importance of channel couplings in representing low energy photoionisation cross-sections of iron-peak elements is well known \citep{Bautista1995}. For energies beyond the 3d sub-shell, the central field cross-sections agree very well with our data, which gives additional confidence on the accuracy of our results. Fig. \ref{fig:photo_app} in Appendix also show the comparison of the cross-sections with the hydrogenic data. Clearly the differences are substantial and amount to several orders of magnitude in the background, but also all resonances that contribute to the over-ionisation at longer wavelengths are missing in the hydrogenic data.

DETAIL does not have a provision for including the partial ionisation channels to specific states of the target ion. Hence, we adopt the total photo-ionisation cross-sections, computed by adding the partial cross-sections for each \mni~state.
\subsubsection{Inelastic collisions with H atoms}
The rate coefficients for the bound-bound transitions in \mni~ caused by collisions with H atoms, as well as for \mnii~ collisions with $\rm{H^-}$, were taken from \citet{Belyaev2017}, but we also compute new rates for these processes in this work.

The data from \citet{Belyaev2017} are available for the transitions between 19 levels\footnote{Molecular states $\rm{Mn(3d^64p\,\,z^6D^{\circ}) + H}$ and $\rm{Mn(3d^54s4p\,\,y^6D^{\circ}) + H}$ are excluded from the calculations because they have $^7\Sigma^-$ symmetry, while other considered states have $^7\Sigma^+$ symmetry.} of \mni~interacting with H and the ground state of \mnii~interacting with $\rm{H}^-$. They represent  collisional excitation, de-excitation, mutual neutralisation, and ion-pair formation\footnote{We emphasise that all these processes are bound-bound transitions, in which electrons remain bound to Mn or to H atoms, and are referred this way in the text.} due to the transitions between $^7\Sigma^+$ molecular states.

Here we present new calculations of the \hi~collision rates for $71$ additional levels of \mni~ interacting with H and the first excited state of \mnii~ interacting with $\rm{H}^-$. The first excited ionic state of MnH molecule has the $^5\Sigma^+$ symmetry and only covalent molecular states of the same symmetry are considered in the non-adiabatic nuclear dynamical calculations. These states are listed in Table \ref{tab:states_MnH} in the Appendix.

All calculations are performed within the simplified quantum model \citep{BelyaevYakovleva:2017aa-I, BelyaevYakovleva:2017aa-II}, which allows to find a rate coefficient for a particular process using general dependences of the reduced rate coefficients on the electron binding energies. The binding energies are calculated from different ionic limits for the cases of non-adiabatic transitions between $^7\Sigma^+$ molecular states and between $^5\Sigma^+$ states. The rate coefficients for the excitation and de-excitation processes are summed over molecular symmetry, when the initial and the final state of the process have both $^7\Sigma^+$ and $^5\Sigma^+$ symmetries.

\begin{figure}[h]
\includegraphics[width=\linewidth]{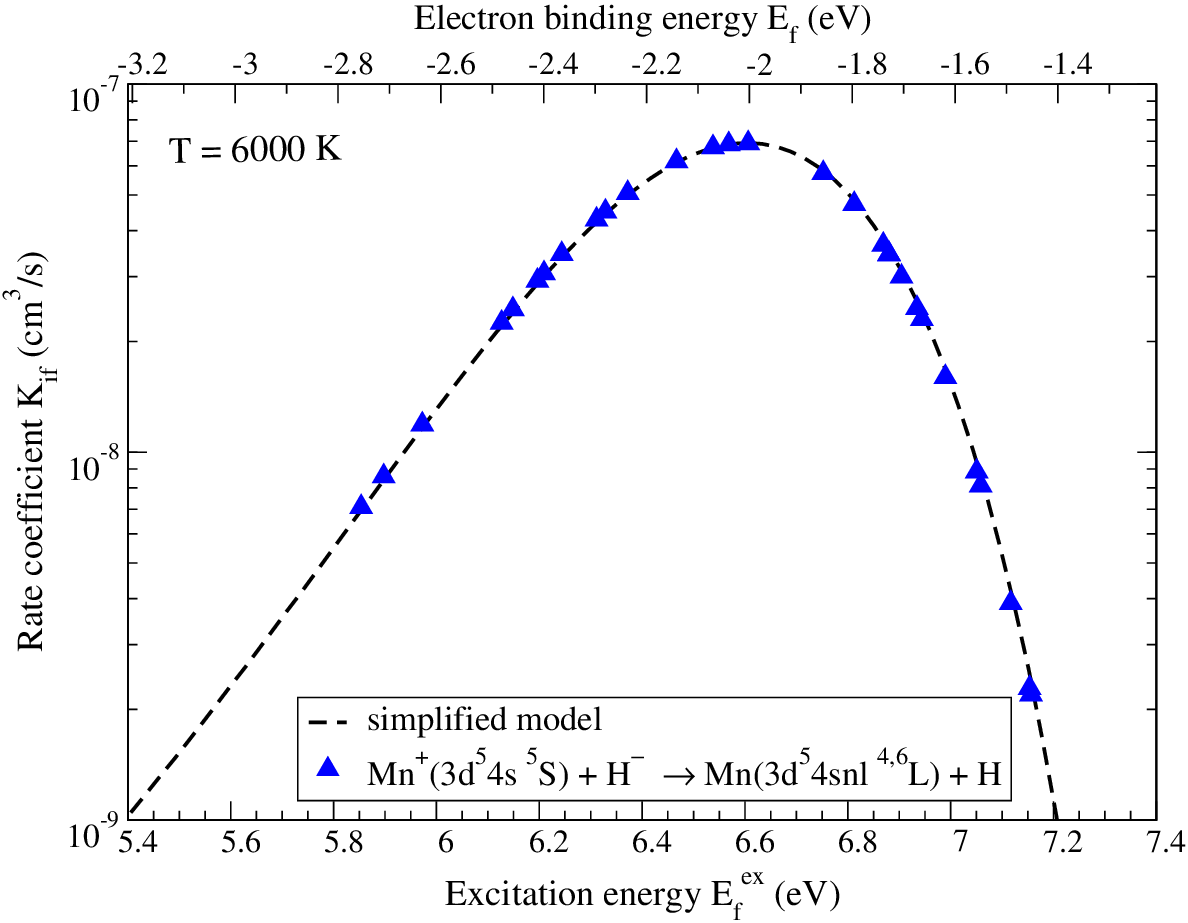}
\caption{Neutralization rate coefficients for $\rm{Mn^+(3d^54s\,^5S) + H^-}$ collisions as a function of the electronic energy in different excited states of \mni~. Dashed line represents the reduced rate coefficient given by the simplified model.}
\label{fig:rates_MnH_2}
\end{figure}

Neutralisation rate coefficients for collisions of $\rm{Mn^+(3d^54s\,^5S) + H^-}$ are presented in Fig.~\ref{fig:rates_MnH_2} as a function of the electron binding energy. For the case of MnH collisional system involving quintet molecular states, the largest rate coefficients at $6000$ K correspond to the mutual neutralisation processes to $\rm{Mn(3d^54s4p \,\,u^6P^{\circ}) + H}$, $\rm{Mn(3d^54s5d \,\,f^6D ) + H}$, $\rm{Mn(3d^54s4f \,\,w^6F^{\circ}) + H}$, $\rm{Mn(3d^54s6p \,\,t^6P^{\circ}) + H}$ states and they have values $\sim 6\times10^{-8}$ cm$^3$/s. The rate coefficients for the (de-) excitation processes are, at least, one order of magnitude lower than the rates for the neutralisation/ion-pair formation processes, as found in previous calculations for other chemical elements \citep{Belyaev_2014, Belyaev_2017, Yakovleva_2017}.

\begin{figure}[h]
\includegraphics[width=\linewidth]{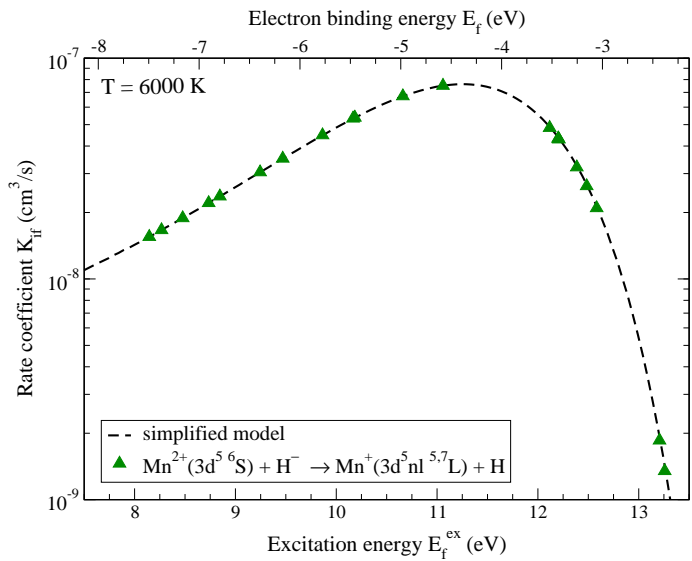}
\caption{Neutralisation rate coefficients for $\rm{Mn^{2+}(3d^5\,^6S) + H^-}$ collisions as a function of the electronic energy in different excited states of \mnii~. Dashed line represents the reduced rate coefficient given by the simplified model.}
\label{fig:rates_MnH+}
\end{figure}

We also derive new rate coefficients for the $42$ levels of \mnii~interacting with H and for the ground state of \mniii~interacting with $\rm{H}^-$. These calculations are performed for the transitions between $^6\Sigma^+$ molecular states as the ionic state of MnH$^+$ has $^6\Sigma^+$ symmetry. The states are presented in Table \ref{tab:states_MnH+} in the Appendix. Neutralisation rate coefficients for collisions $\rm{Mn^{2+}(3d^5\,^6S) + H^-}$ as a function of the electron bound energy are shown in Fig. \ref{fig:rates_MnH+}. The largest rate coefficient for the case of MnH$^+$ collisions at 6000 K with the value of $7.5\times10^{-8}$ cm$^3$/s corresponds to the mutual neutralisation process $\rm{Mn^{2+}(3d^5\,^6S) + H^-} \to \rm{Mn^+(3d^55p\,\,v^5P^{\circ}) + H}$.

Our data apply to $J$-averaged energy states, but the NLTE model atom includes fine structure. We have tested different recipes that are used in the literature to deal with this case. In particular, \citet{Barklem2007} propose to divide the fine-structure summed rate coefficient by the number of the targets states. However, we found that the effect of distributing the collision rate coefficients across the target states is virtually null. In particular, for the high-excitation \mni~lines in the model of a metal-poor dwarf, this leads to an error in the line equivalent width of less than 0.1 $\%$, which is negligibly small for abundance determinations. We, hence, assign the same rate coefficient for each fine structure level of a given term\footnote{Both Mn model atoms, with and without distributed rates, can be obtained from the corresponding author by request}. This is analogous to our handling of the photo-ionisation data, which are also provided for a given term. The new rate coefficients are available in the supplementary material.
 
 In the model atom, we tabulate the rates of exothermic processes for the bound-bound reactions, i.e. the transitions accompanied by the release of energy ($E_j > E_i$, where the transition occurs from $j$ to $i$), because these are almost independent of temperature, which minimises interpolation errors. For the charge transfer, the reverse is true, hence we tabulate the rates of endothermic processes. The reverse rates are computed from the detailed balance internally within the code, i.e. see eq. 5 in \citealt{Belyaev2017} (note that neither $n_{\rm H}$ nor $n_{\rm H^{-}}$ explicitly enter these equations):
\begin{equation}
\label{eq:1}
  r_{ij} = r_{ji} \dfrac{g_j}{g_i} \mathrm{exp} \dfrac{-\Delta E_{ji}}{k_{\mathrm B} T}
\end{equation}
where r$_{ij}$ and r$_{ji}$ are rate coefficients for the transition from the lower energy level $i$ to the higher energy level $j$, $g_i$ and $g_j$ the statistical weights of these levels, and $E_{ji}$ the energy difference ("energy defect") between the energies of the two states.

Apart from the radial avoided crossing mechanism, one can estimate additional rate coefficients using the free electron model, which is expected to include other inelastic mechanisms except the long-range ionic covalent mechanism \citep{Barklem2016, Barklem2018, Amarsi2019b}. We hence also supplement the model atom with collision rates for all \mni~states computed using the scattering-length approximation\footnote{These data were calculated with assistance from Anish Amarsi and Paul Barklem.} according to eq. 18 of \citet{Kaulakys1991} using the code of  \citet{Barklem2017}. As described \citet{Osorio2015} and \citet{Barklem2016}, the rate coefficients computed in this way need to be redistributed over all possible final spine states. Here, we assume that all transitions have two possible final spin states, and that each final spin state is equally likely, so that the rate coefficients were reduced by a factor of two. The error associated with this assumption is less than a factor of two. The Kaulakys model is developed for application to Rydberg molecular states, hence, our implementation shall be viewed as a limiting case with strong collisional binding, and, hence, might under-estimate NLTE effects.
\begin{figure}[!ht]
\includegraphics[width=0.5\textwidth, angle=0]{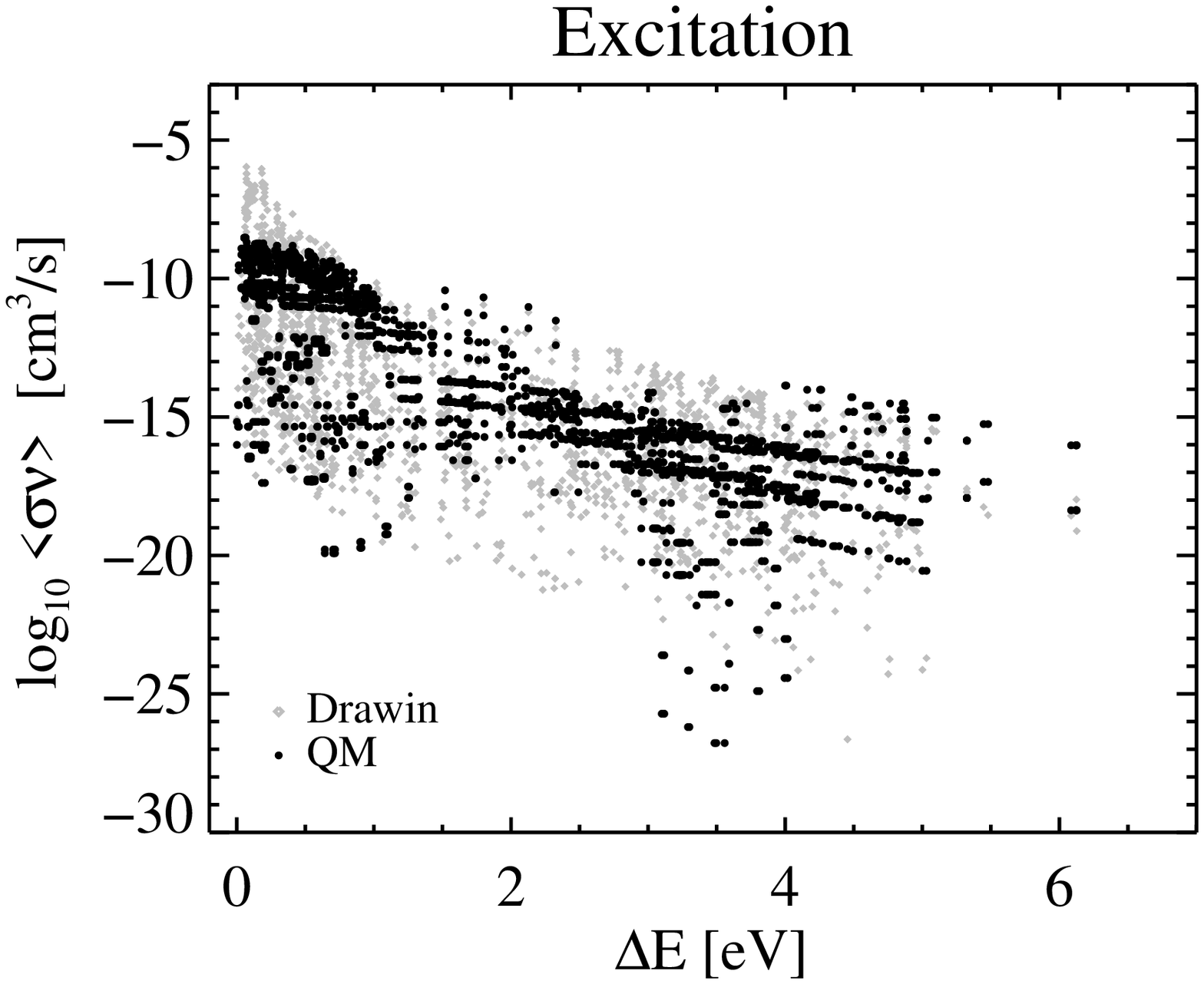} 
\includegraphics[width=0.5\textwidth, angle=0]{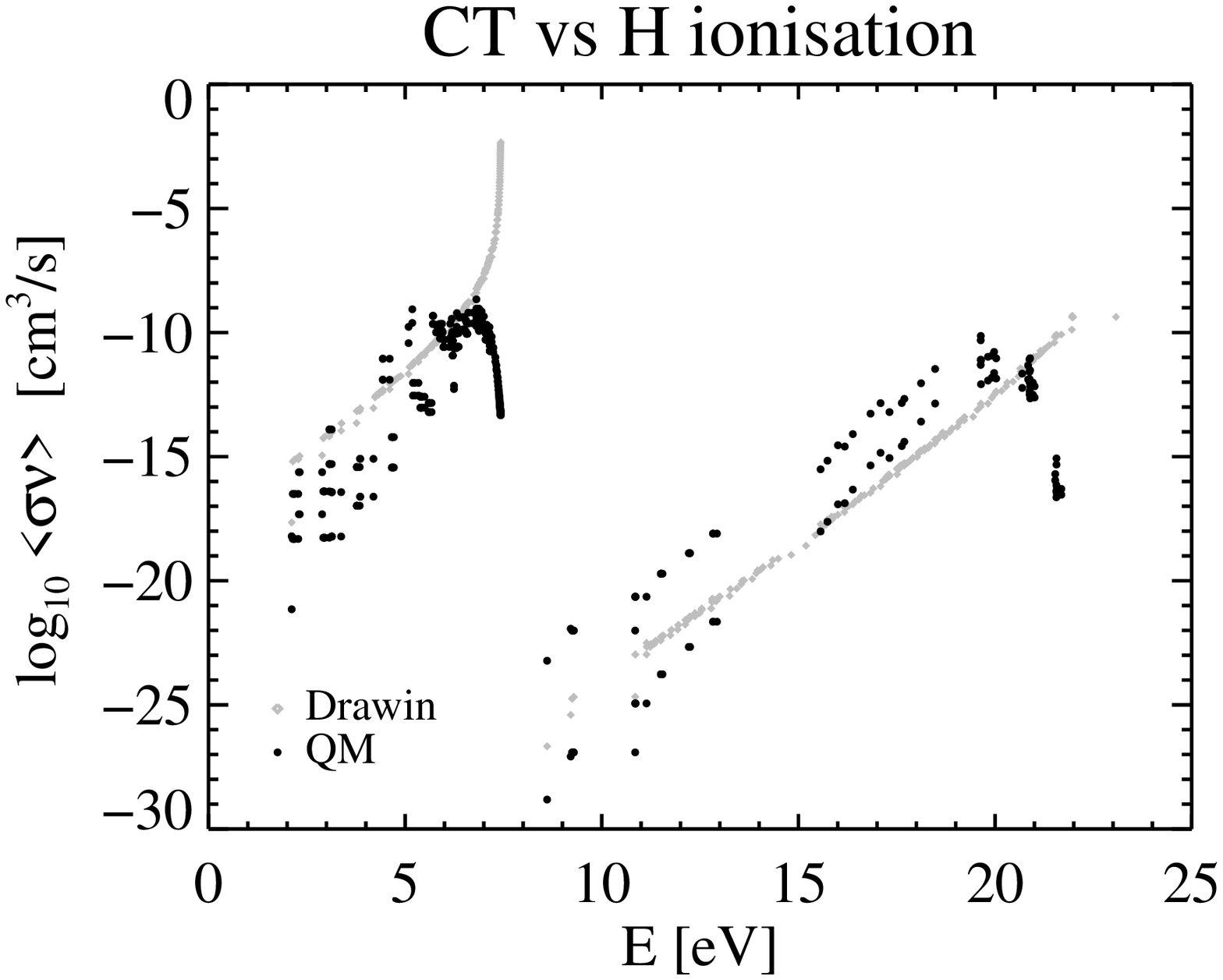}
\caption{Excitation (top panel) and ion-pair formation (bottom panel) rate coefficients for \mni~and \mnii~as used in this work compared to the Drawin's formulae \citep{Drawin1968,Drawin1969,Steenbock1984}.}
\label{fig:Hcol}
\end{figure}
\begin{figure*}[!ht]
\includegraphics[width=\linewidth]{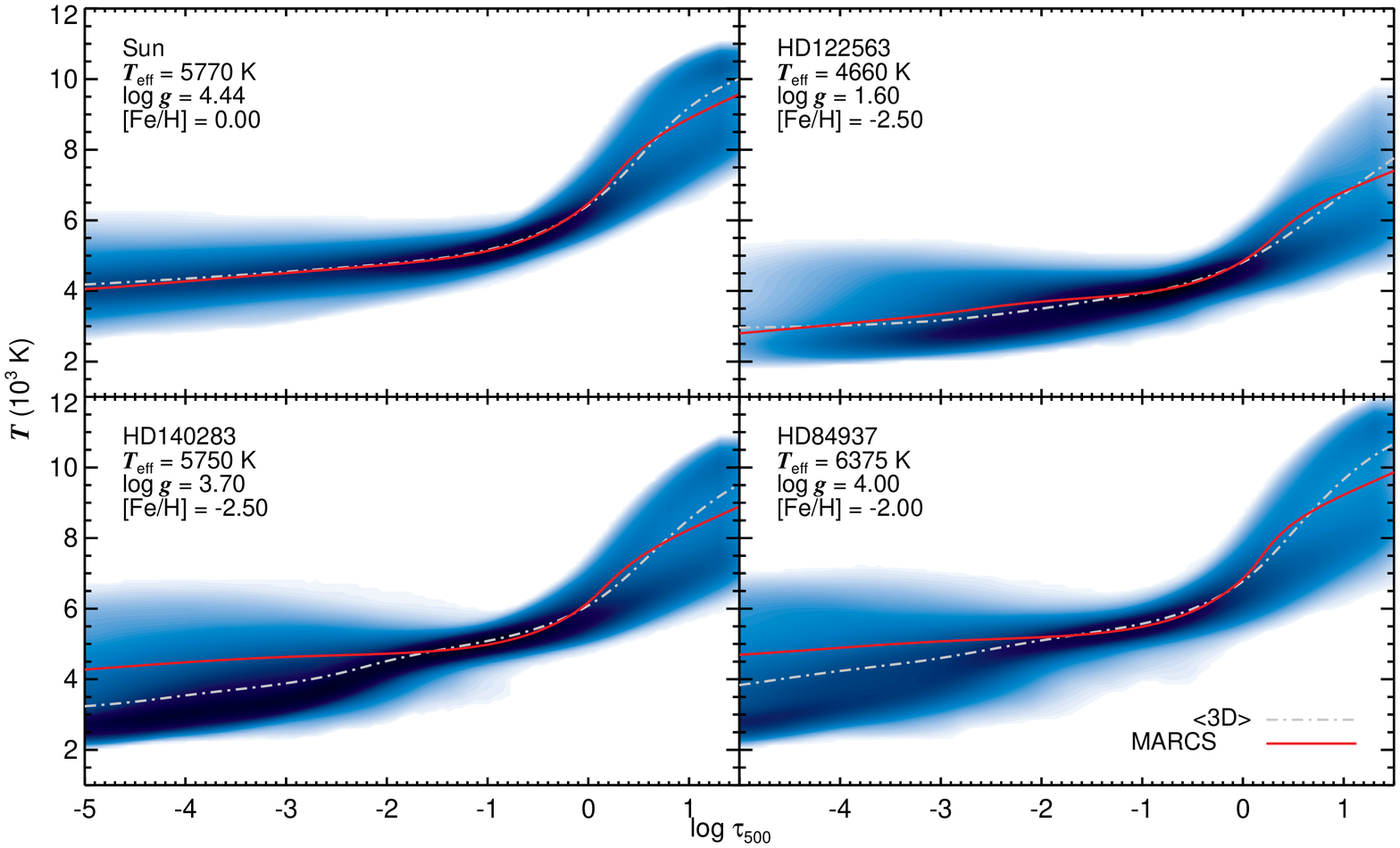}
\caption{3D, 1D, and \tda\ temperature structures as a function of the Rosseland optical depth for the four benchmark stars. The blue shaded regions indicate the kinetic temperature distributions in the representative snapshot from the 3D convection simulations. The stellar parameters are given in the inset. We note that \tda\ models are provided only to illustrate the difference between the average structure of the 1D and 3D models, however, the \tda\ are not used in our abundance analysis.}
\label{fig:tstruct}
\end{figure*}

In Fig. \ref{fig:Hcol} we compare the new H collision rates with the data computed using the \citet{Steenbock1984} formulation of the classical theory developed by \citet{Drawin1968,Drawin1969}. The classical formalism does not cover the mutual neutralisation and ion-pair formation processes. The differences in the H excitation rates for the individual energy levels amount up to $7$ orders of magnitude in both directions. The overall distributions as a function of energy difference have similar shapes, with the largest rate coefficients for the transitions between nearby energy levels. The rate coefficients for the charge transfer reactions are also qualitatively similar to the Drawin's bound-free recipe, which describes collisional ionisation, but quantitatively there are major differences of up to $5$ orders of magnitude. For the neutral species, the quantum-mechanical charge transfer rates are typically lower, whereas for the singly-ionised species these are larger than the Drawin. The Kaulakys rates are typically very low for the low-energy levels, but gradually increase closer to the ionisation threshold and thus somewhat compensate the downturn in the quantum-mechanical data, leading to higher collisional thermalisation. In sect. \ref{sec:abcor} we briefly report on how this impacts the line profiles and the NLTE abundance corrections.
\subsection{Model atmospheres}
\label{sec:atmos}

As in our previous papers, we use MARCS \citep{Gustafsson2008} and MAFAGS-OS \citep{Grupp2004a,Grupp2004b} model atmosphere grids. These are 1D LTE model atmospheres, with certain differences regarding the treatment of convective energy transport (mixing length), opacity, and the solar abundance mixture. The depth discretisation and the vertical extent of the models are also slightly different, as MAFAGS-OS usually cover the range from $-6$ to $+2$ in $\opd$, whereas the MARCS models sample the Rosseland optical depths from $-4$ to $+2$. Nonetheless, the thermodynamic structures of the models for the given input parameters are very similar \citep{Bergemann2012, Bergemann2017a}. 

The 3D model atmospheres are taken from the \textsc{stagger} model atmosphere grid \citep{Collet2011, Magic2013} computed with the \textsc{stagger} code \citep{Nordlund1995}. A 3D model consists of a series of computational boxes that represent a time series, which are referred to as snapshots. These snapshots are selected from a larger time series of snapshots that are produced from the \textsc{stagger} code and are selected at a time when they have reached dynamical and thermal relaxation. For our purposes -- and for the sake of time -- we have chosen to work with five snapshots. Importantly, and unlike an equivalent 1D model, 3D models provide $x$, $y$, and $z$ velocity fields for every voxel meaning that post-processing spectrum synthesis codes provide parameter-free description of Doppler broadening, including asymmetric line profiles, which trace these gas flows at each voxel.

Figure~\ref{fig:tstruct} depicts the 3D temperature structures (blue 2D histogram) in a representative snapshot for four benchmark stars, along with the 1D MARCS (red solid line) and \tda\ (dashed gray line) profiles. The average temperature of the full 3D model and the 1D hydrostatic model are fairly different in the outermost regions of the atmosphere, as seen by comparing the 1D hydrostatic with the \tda\ models. In particular, in the outer layers of metal-poor models the 1D hydrostatic models are significantly, up to $500$ K, hotter compared to the 3D structures \citep[see also][]{Bergemann2017a}. Also in deeper regions of the models - where the continuum usually forms - the models diverge. This is mostly due to the treatment of convection between the 1D and 3D model atmospheres.

The average temperature structure of the 1D model of the metal-poor RGB star HD 122563 is not very different from its \tda\ counterpart. Our adopted MARCS models are taken from \citet[][their Fig. 1]{Bergemann2012}. We explore the impact of adopted 1D model in Sec. \ref{bmk3d}, by performing the abundance analysis with a MAFAGS-OS suit of models. We also note that line formation is not only sensitive to the mean T($\tau$) and P($\tau$) structure, but also to the horizontal inhomogeneities. This star is a particular case, where the latter play a significant role in the abundance analysis.

The scope of this paper is limited to the analysis of a small sample 3D models, including that of the Sun, a typical dwarf and a typical giant (Table \ref{3dmodels}). We also include tailored 3D models computed for the parameters of the benchmark metal-poor stars HD 122563, HD 140283, and HD 84937 (Sect. \ref{sec:obs}). To make the NLTE radiative transfer problem computationally tractable, we have to resample the full 3D model cubes onto a less fine, yet equidistant, grid in horizontal coordinates. However, we test the effect of the resolution of the cube for radiative transfer in section \ref{3dnlte} and find virtually no differences in the resulting atomic number densities and line profiles for horizontal resolutions of (x,y,z) $= 30,30,230$ and the original cubes ((x,y,z) $= 240,240,230$). Hence, the former is taken to be the default resolution for most of the analysis presented in this work. 
\begin{table}
\begin{minipage}{\linewidth}
\renewcommand{\footnoterule}{} 
\caption{Parameters of 3D convective and 1D hydrostatic model atmospheres. In the 1D LTE radiative transfer calculations with MARCS models, we assume $\Vmic = 1$.}
\label{3dmodels}     
\begin{center}
\begin{tabular}{l l c c c c}
\noalign{\smallskip}\hline\noalign{\smallskip} No. & Name & $\teff$ & logg & $\feh$ &   \\
  & & K & dex & dex & \\
\noalign{\smallskip}\hline\noalign{\smallskip}
1 & Sun         & 5777  & 4.4  &  ~~0.0   &   \\
2 & HD 84937    & 6400  & 4.0  & $-$2.0   &   \\
3 & HD 140283   & 5750  & 3.7  & $-$2.5   &   \\
4 & HD 122563   & 4600  & 1.6  & $-$2.5   &   \\
\noalign{\smallskip}\hline\noalign{\smallskip}
%
5 & Sub-solar metallicity dwarf & 6500  & 4.0  & $-$1.0   &   \\
6 & Metal-poor dwarf & 6500  & 4.0  & $-$2.0   &   \\
%
7 & Sub-solar metallicity giant & 4500  & 2.0  & $-$1.0   &   \\
8 & Metal-poor giant & 4500  & 2.0  & $-$2.0   &   \\
\noalign{\smallskip}\hline\noalign{\smallskip}
\end{tabular}
\end{center}
\end{minipage}
\end{table}
We solve the 3D NLTE radiative transfer problem for a set of snapshots for each of the 3D model atmospheres listed in Table \ref{3dmodels}. These snapshots are extracted at regular time intervals from the full simulation that covers roughly two convective turnover timescales \citep{Collet2011,Magic2013}. The convergence criterion, that is the maximum relative correction in the population numbers, max$|\delta{N}/N|$, is set to 10$^{-3}$ that is fully sufficient according to our experience with 1D NLTE radiative transfer. 
\subsection{Statistical equilibrium}

We use two different codes to compute the SE of Mn. One is MULTI2.3 \citep{Carlsson1992}, the other code is DETAIL (Butler \& Giddings 1985). The codes solve the equations of radiative transfer and SE assuming a 1D geometry. The assumption of trace elements is used, that is, the element that is modelled in SE is assumed to have no effect on the model atmosphere. This is a good assumption for Mn as it is not an electron donor nor does it have a high impact on the overall opacity. Both codes adopt the accelerated lambda iteration (ALI) technique and the operator acting on the source function.

The basic differences between the codes are described in \citet{Bergemann2012}. The tests described in the following sections will be performed imposing the same input conditions (LTE populations) and the same model atmospheres, in order to maximise the consistency. The main difference between the codes are in the handling of thermodynamic parameters and of the background opacities. In particular, DETAIL takes the partial pressures and partition functions from the input model atmosphere, whereas MULTI2.3 includes a package to compute these parameters given the input $T(\tau)$ and $P_{e}(\tau)$ structures as a function of optical depth or column mass. 

In order to maximise the homogeneity of the analysis, we have also computed background opacity tables for MULTI2.3 using the updated linelists from DETAIL \citep{Bergemann2015}. The MARCS \citep{Gustafsson2008} and Turbospectrum \citep{Plez2012} codes were used to generate a table of opacities for a set of temperature and pressure points at more than $10^5$ wavelengths, Mn\footnote{Note that also Ba lines were omitted, in order to use these opacity tables for the 2nd paper in the series by Gallagher et al. (in prep).} lines being omitted. This table is then interpolated by MULTI2.3 to produce detailed line background opacities.

All calculations with MULTI2.3 are carried out by solving simultaneously the intensity at all angles, using the Feautrier method with all scattering terms included consistently (the ISCAT option set to 1). This is important when scattering in the background opacity is significant, as is the case in the Wien's regime. We have tested the line formation disabling this option, but found that this has a very strong effect on the blue and UV lines of \mni~and \mnii, significantly over-estimating the line abundances, because of reduced continuum intensities.

\multitd\ is an MPI-parallelised, domain-decomposed NLTE radiative transfer code that solves the equations of radiative transfer in 3D geometry using the accelerated lambda iteration (ALI) method. The formal solution of radiative transfer is done via the short characteristics method \citep{Kunasz1988} that solves the integral form of the radiative transfer equation across one subdomain per time step. The \citet{Carlson1963} quadrature A4 is employed to compute the angle-averaged radiation field in the SE solution. The approximate operator is constructed using the formulation of \citet{Rybicki1991,Rybicki1992}, where only the diagonal of the full $\Lambda$ operator is used \citep[for discussion of this approximation, see, e.g.][]{Bjorgen2017}. 
\begin{figure}[!ht]
\includegraphics[width=0.5\textwidth, angle=0]{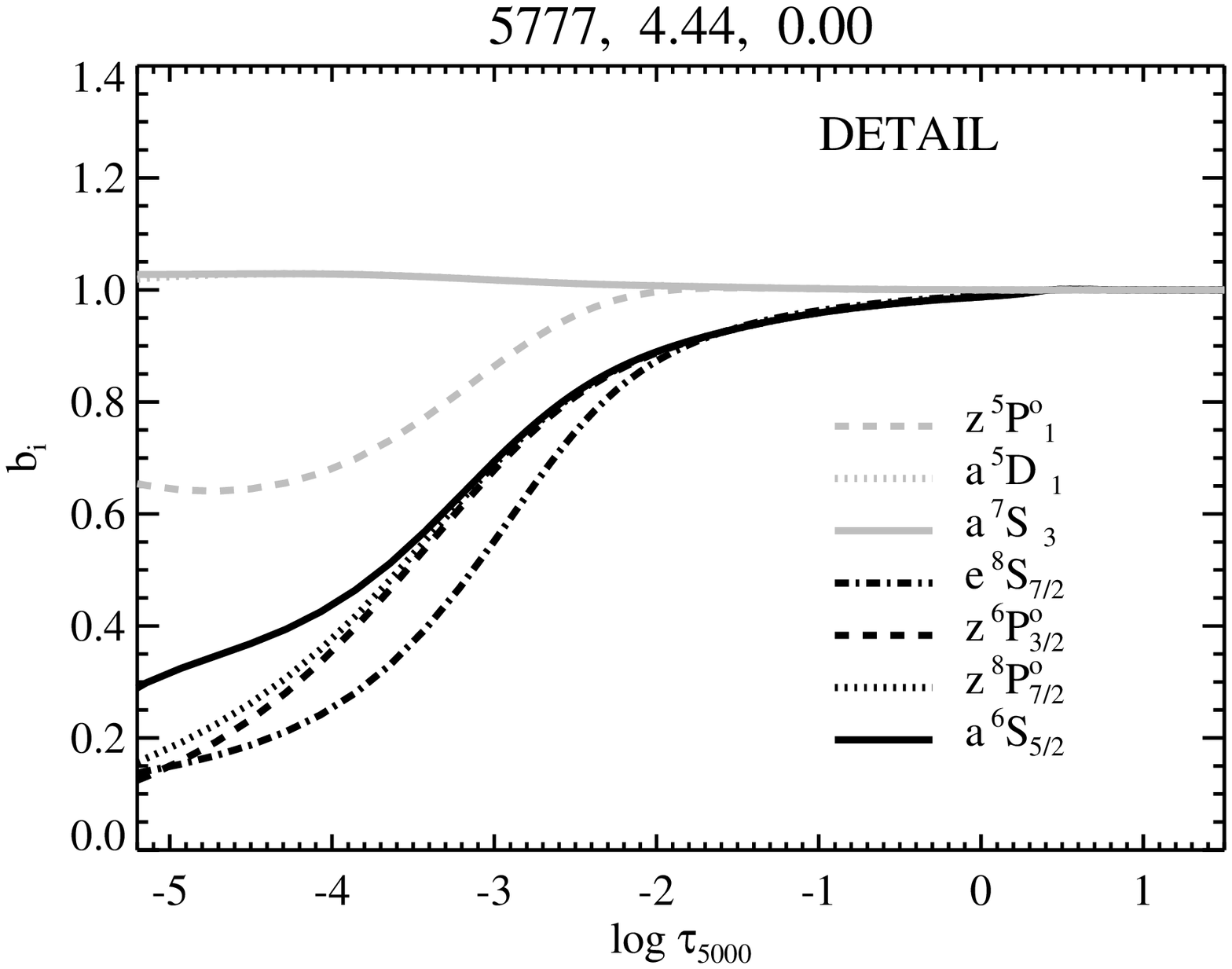}
\includegraphics[width=0.5\textwidth, angle=0]{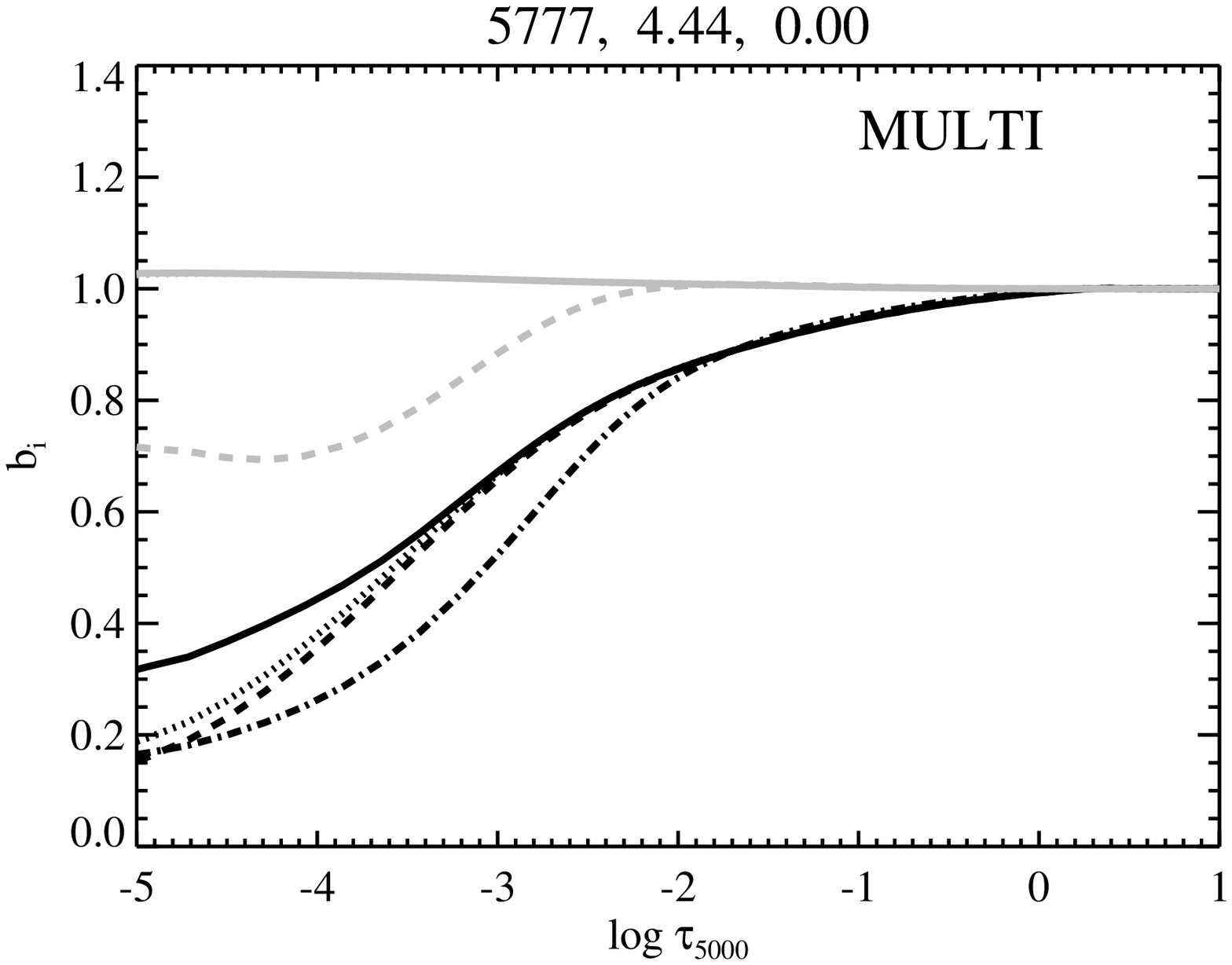}
\caption{Mn departure coefficients for the solar MARCS model atmosphere as a function of the optical depth at 5000 \AA, computed using DETAIL (top panel) and MULTI (bottom panels) codes. The surface parameters of the Sun, ($\teff$, $\log g$, and $\feh$) are given in the figure title.}
\label{fig:depsmulti}
\end{figure}

\multitd\ will accept three types of 3D model atmosphere formats, including the commonly used Bifrost \citep{Gudiksen2011} and \textsc{stagger} models \citep{Magic2013}. The code will also accept any 3D model providing the temperature, density, electron number density and $x$, $y$, and $z$ velocity fields are supplied on a cartesian grid that is both horizontally periodic and equidistantly spaced. The code can compute radiative transfer using the $1.5$D approximation, which treats each column of grid points in a model as a separate plane-parallel atmosphere, or using full 3D radiative transfer. The rate equations are assumed to be time-independent and the advection term is not included. For more information about the code, we refer the refer to \citet{Leenaarts2012} and \citet{Bjorgen2017}.%
%
%
\section{Results}\label{sec:results}
\subsection{1D NLTE}
\subsubsection{Departures from LTE}
\begin{figure}[!ht]
\includegraphics[width=0.5\textwidth, angle=0]
{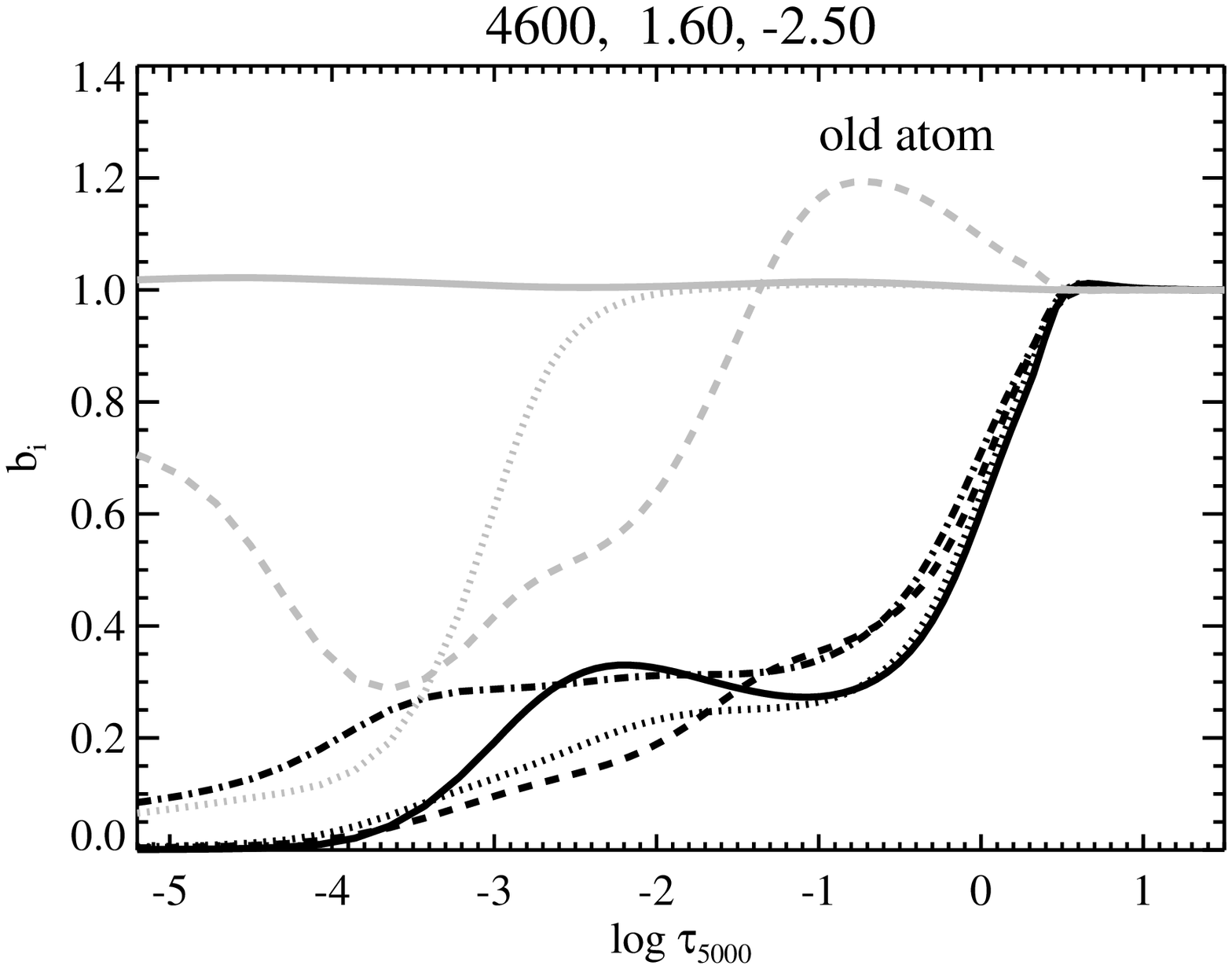}
\includegraphics[width=0.5\textwidth, angle=0]
{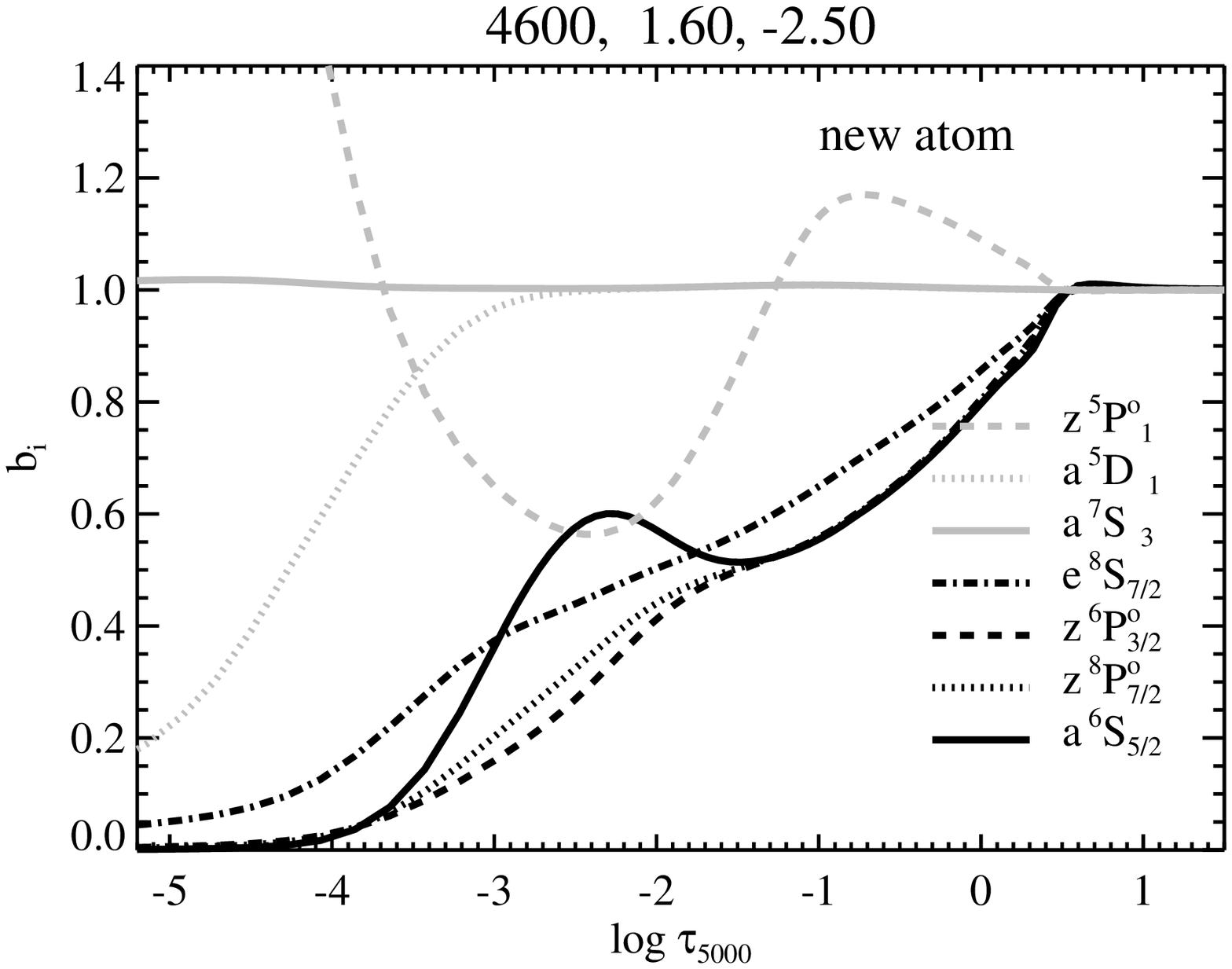}
\includegraphics[width=0.5\textwidth, angle=0]
{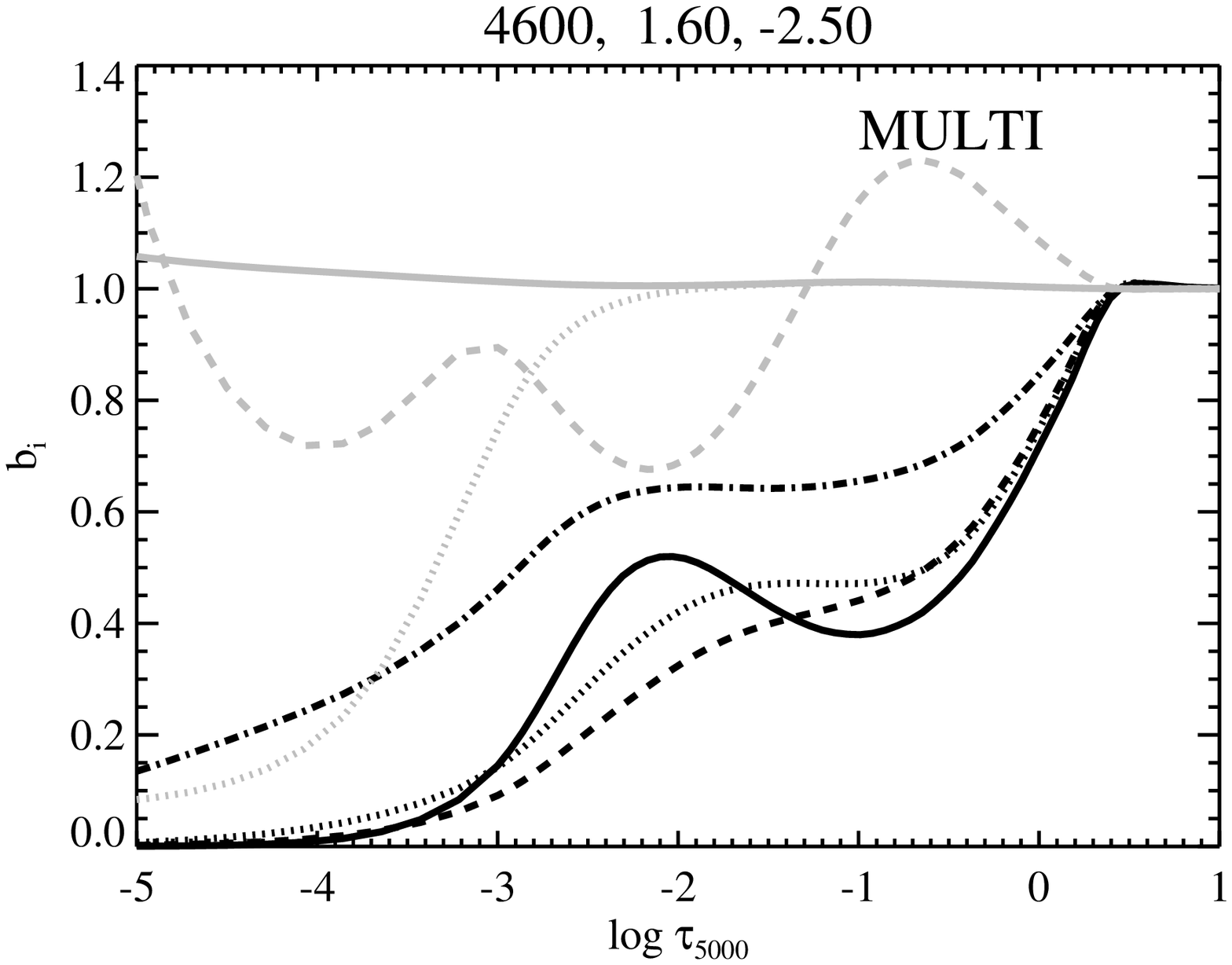}
\caption{Mn departure coefficients for the model atmosphere of a metal-poor red giant computed using the old Mn atom from \citet{bergemann2007} (top) and the new atom from this paper (middle panel: DETAIL, bottom panel: MULTI). The stellar parameters ($\teff$, $\log g$, and $\feh$) are given in the figure titles.}
\label{fig:depshd122}
\end{figure}
This work does not deal extensively with the properties of statistical equilibrium of Mn nor with the details of line formation, as this has been discussed in great detail in our previous work \citep{bergemann2007}. Also \citet{Bergemann2008} address the details of line transfer in metal-poor stars. It suffices to remind the reader that Mn, similar to other Fe-group elements, is a photo-ionisation dominated ion. Simply stated, the large photo-ioinisation cross-sections of \mni\ energy levels imply significant over-ionisation in the atmospheres of cool stars in the more general SE case compared to LTE. \mni\ becomes significantly over-ionised (compared to LTE) in metal-poor or in hotter stellar atmospheres due to their strong UV radiation fields. The effect of the radiation field is furthermore amplified in the atmospheres of giants owing to their lower densities and lesser efficiency of thermalising collisions.

This is reflected in the behaviour of \mni~level departure coefficients, $b_i$, which describe the ratio between NLTE and LTE atomic number densities. Departures from LTE take place in the line formation layers. Fig. \ref{fig:depsmulti} shows that for \mni, this ratio is typically $\lesssim 1$, as in NLTE the fraction of atoms in a given energy state is less than that predicted by the Saha-Boltzmann formulae. The departure coefficients of the ionic levels, \mnii, are very close to unity for the lower-lying energy levels, but deviate from thermal for the levels of higher excitation energy, $E \gtrsim 1$ eV. 
\begin{figure*}[!ht]
\includegraphics[width=1\textwidth, angle=0]{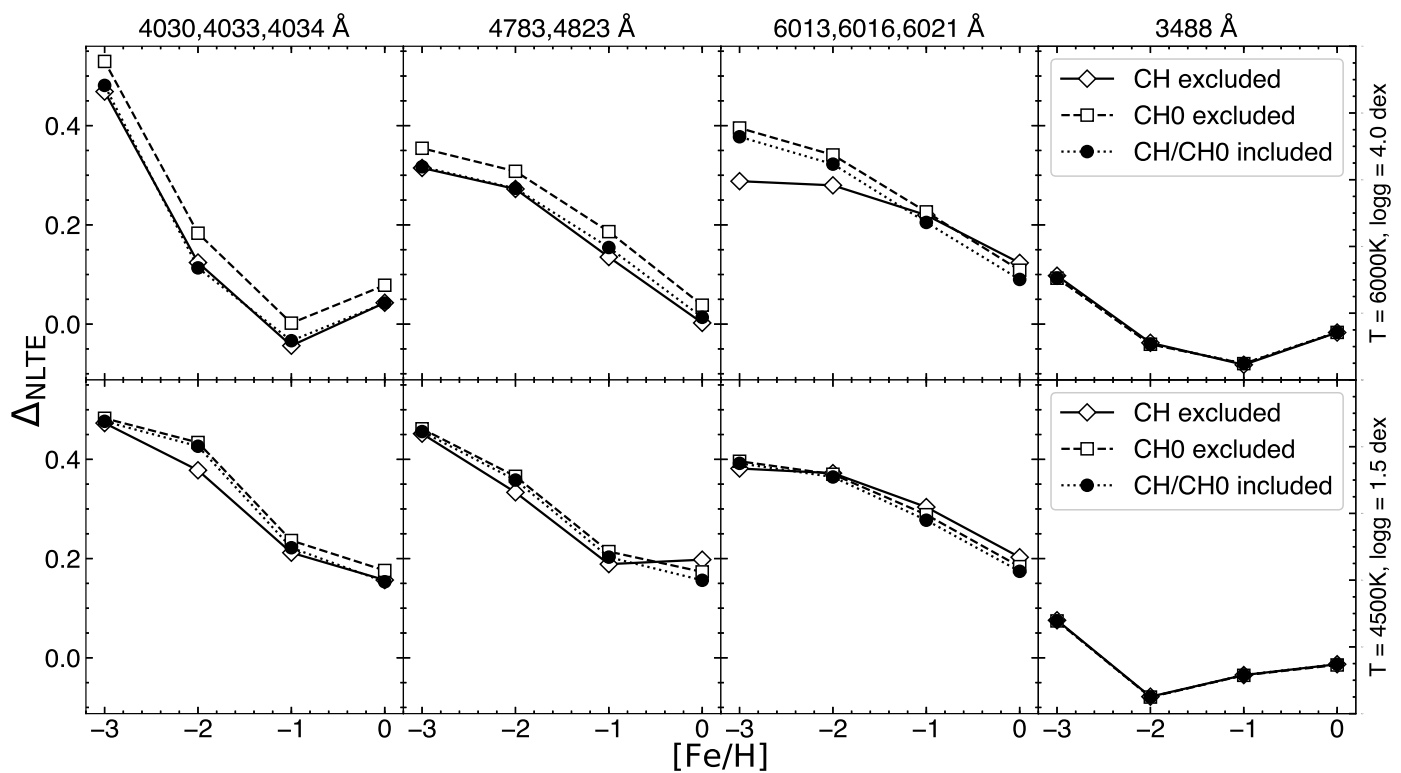}
\caption{NLTE abundance corrections for Mn I (4030,4033,4034, 4783,4823, 6013,6016,6021 \AA) and Mn II (3488 \AA) lines computed for a small grid of MARCS model atmospheres representative of dwarfs -- $T_{\rm eff} = 6000, \log g = 4.0$ (top panels), and red giants -- $T_{\rm eff} = 4500, \log g = 1.5$ (bottom panels) for a range of metallicities from 0 to $-3$ dex. Different curves represent the corrections derived using the model atoms with reduced complexity: (a) CH excluded - ignoring the excitation processes by collisions with H atoms, (b) CH$0$ excluded - ignoring the charge transfer reactions, and (c) CH$/$CH$0$ included - with excitation and CT rates from the quantum-mechanical calculations. None of the three models (a-c) includes the Kaulakys recipe.}
\label{fig:nlteabu}
\end{figure*}

Figure \ref{fig:depsmulti} shows that the departures from NLTE in the line formation region, $-2 \lesssim \opd \lesssim 0$, are moderate in the solar atmosphere, but only slightly smaller than our previous estimates in \citet{bergemann2007}. The differences with respect to the latter study are caused by the use of new quantum-mechanical photo-ionisation rates and H collision rates, as well as the implementation of fine structure for most of the \mni~levels. Collisions with electrons are not important in the physical conditions of the solar atmosphere. On the other hand, inelastic collisions with H atoms have a non-negligible effect on the atomic level populations and significantly contribute to the overall thermalisation of the system. This effect is not linear and may increase or decrease the departures from LTE for individual energy levels, and hence spectral lines, depending on the $\teff$, $\logg$ and metallicity of a star. 

The comparison of the $b_i$ profiles computed using DETAIL and MULTI2.3 (Fig. \ref{fig:depsmulti}) suggests that the codes are consistent, given the same input conditions, such as the model atom, line opacities, and model atmospheres. MULTI2.3 predicts slightly larger departures from LTE compared to DETAIL. This has also been shown in our earlier paper for Fe \citep{Bergemann2012}, and is likely related to continuum opacities and/or the numerical implementation of the coupled SE and radiative transfer equations.

The behaviour of departure coefficients is very different in the model atmosphere of a metal-poor red giant star (Fig. \ref{fig:depshd122}). All \mni~levels show a stronger under-population compared to the solar model, implying larger differences between LTE and NLTE abundances in metal-poor stars. 
The energy levels of the \mnii\ lines are also affected by NLTE. In particular, the levels of the upper term \Mn{z}{5}{P}{o}{} experience overpopulation caused by the radiative pumping in nine strong near-UV lines of \mnii\ multiplet Nr. 31 (\Mn{a}{5}{D}{}{} - \Mn{z}{5}{P}{o}{}) in the deeper layers. However, the levels of \Mn{z}{5}{P}{o}{} become under-populated at $\opd \sim -1.5$, as these lines progressively become optically thin. Consequently, one would expect significant NLTE effects on the formation of \mnii~lines, largely driven by the changes in the line source function itself.
Comparing the departure coefficients computed using the old atom from \citet{Bergemann2008} (Fig. \ref{fig:depshd122}, top panel) and the new atom in this work (Fig. \ref{fig:depshd122}, middle panel), we find substantial differences. The influence of new quantum-mechanical collisions with hydrogen is admittedly greater in the new atom, despite the larger photo-ionisation cross-sections. On the other hand, contrasting the results obtained with the DETAIL code with MULTI2.3 (Fig. \ref{fig:depshd122}, middle and bottom panels) confirms that, similar to the Sun (Fig. \ref{fig:depsmulti}), the two codes produce quantitatively similar outputs. In the outer layers, the departures are slightly different that could be related to the differences in the outer boundary conditions.
%
%
%
\subsubsection{NLTE abundance corrections}{\label{sec:abcor}}
\begin{figure*}[!ht]
\includegraphics[width=1\textwidth, angle=0]{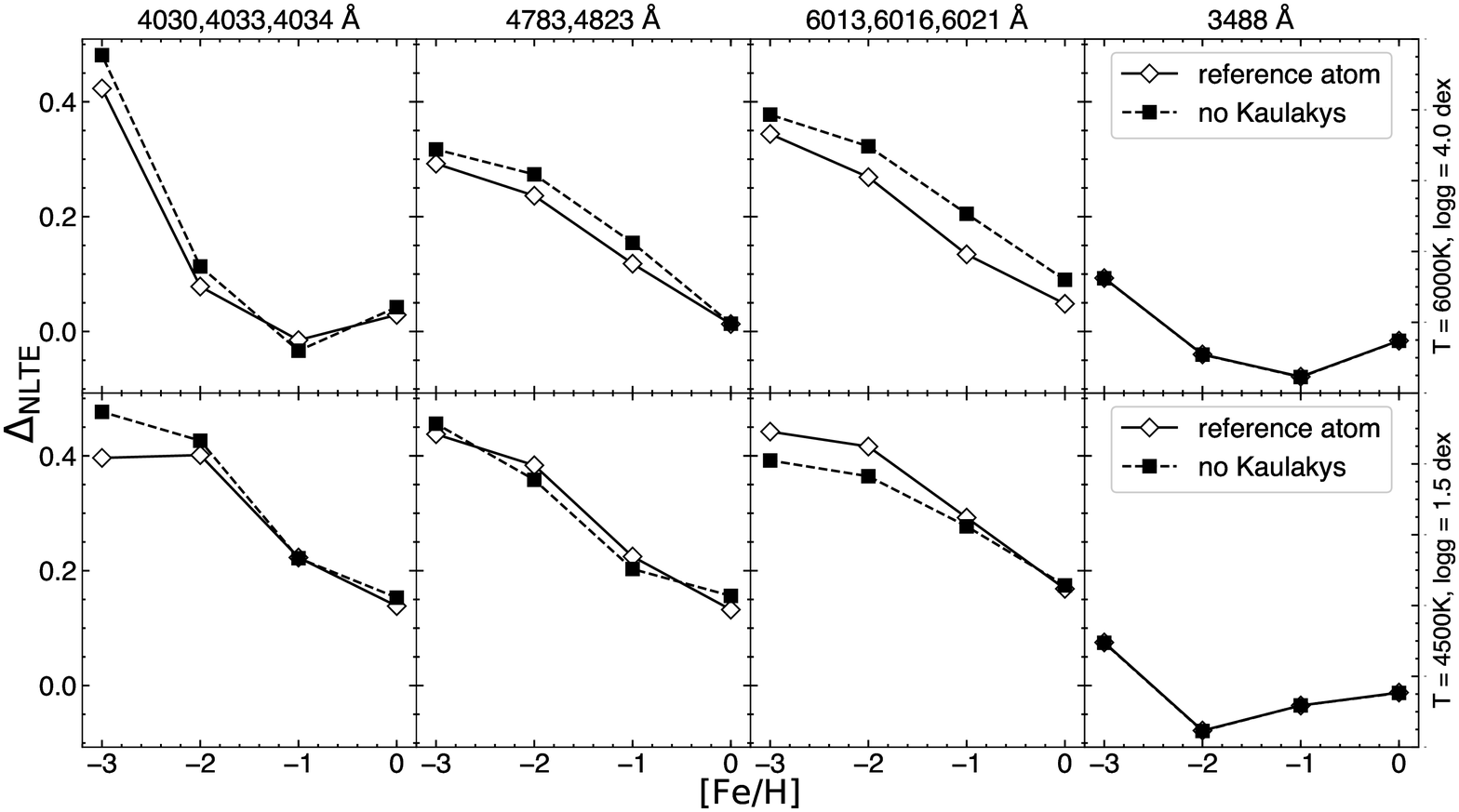}
\caption{NLTE abundance corrections for Mn I and Mn II lines computed for a small grid of MARCS model atmospheres representative of dwarfs -- $T_{\rm eff} = 6000, \log g = 4.0$ (top panels), and red giants -- $T_{\rm eff} = 4500, \log g = 1.5$ (bottom panels) for a range of metallicities from 0 to $-3$ dex. Different curves represent the corrections derived using two model atoms, one with and the other without collisions derived using the Kaulakys recipe.}
\label{fig:nlteabu2}
\end{figure*}

A NLTE abundance correction is the quantity that is commonly used in stellar abundance studies to correct the abundances derived under the assumption of LTE. The abundance correction are defined as $\Delta = A(\rm{NLTE}) - A(\rm{LTE})$, i.e., the difference in abundance required to fit a given spectral line assuming 1D LTE or 1D NLTE. We also employ this concept in our 3D NLTE analysis in Sec. \ref{3dnlte}.

Figure \ref{fig:nlteabu} illustrates the NLTE abundance correction for several metal-poor models in the metallicity range $-3$ to $0$. We only plot selected lines to illustrate the key results: the 3488 \AA\ line of \mnii, multiplet 4 (4030, 4033, 4034 \AA), multiplet 18 (4783, 4823 \AA), and multiplet 32 (6013, 6016, 6021 \AA) of \mni. The behaviour of abundance corrections \textit{within a given multiplet} is very similar, hence the lines are grouped by multiplets. The correction is not tailored to any particular star and is computed assuming the reference NLTE [Mn$/$Fe] of zero. We also explored the abundance corrections computed using LTE [Mn$/$Fe] abundance of $-0.5$ to $-0.8$, as it would be typically measured assuming LTE in metal-poor stars and found no significant differences in the corrections. The individual curves represent three possible scenarios, which differ in the completeness of the model atom. 

In Figure \ref{fig:nlteabu}, we illustrate the sensitivity of the corrections to the quantum-mechanical H data. The CH case corresponds to the model atom, which is  devoid of quantum-mechanical H collisional excitation processes (but including charge transfer). The CH0 model lacks the charge transfer processes, but includes the collisional excitation. These two cases are compared to the model, which has both excitation and charge transfer (CH/CH0 included). None of the three cases presented in this figure includes the \citet{Kaulakys1985} collision rates. 

The H collisions clearly have a different impact on the line formation in the atmospheres of giants and dwarfs. Based on the NLTE abundance corrections in the figures, it appears that for dwarfs, H collisions serve as a thermalising agent, decreasing the difference between NLTE and LTE. In the atmospheres of red giants, the effect is somewhat counter-intuitive: the lines of multiplet $32$ (6013-6021 \AA\ triplet) have smaller NLTE corrections, when H collisions are excluded. In fact, this is the indirect effect of over-ionisation, which is more efficiently transferred to higher-excitation states by collisions with H. On the other hand, the lines of multiplet $18$ (4783, 4823\AA) behave as expected from the simple  considerations of increased rate of collisional thermalisation. Fig. \ref{fig:nlteabu} also suggests that charge transfer (CT) reactions are more important in the atmospheres of dwarfs. Neglecting CT fully typically leads to abundance corrections over-estimated by $0.05$ dex for the models with [Fe$/$H] $\gtrapprox -3$. 

Figure \ref{fig:nlteabu2} illustrates the influence of collision rates computed using the \citet{Kaulakys1985} recipe. The reference atom includes the \citet{Kaulakys1985} collisions in addition to the data from \citet{Belyaev_2017}, and we compare this model with the model that is devoid of the Kaulakys data. The differences with our reference model atom are modest, and do not exceed $0.05$ dex for the dominant part of the parameter space. The effects that are possibly most significant occur when we neglect \citet{Kaulakys1985} collisions. This leads to slightly over-estimated NLTE abundance corrections for the multiplet $32$ \mni~lines (6013 - 6021 \AA\ triplet) in the RGB models. On the other hand, the NLTE abundance corrections for these very high-excitation lines are systematically under-estimated in the model of a dwarf.

The general picture is that the NLTE abundance corrections for the \mni~lines are positive and increase with decreasing metallicity, supporting our earlier study of Mn~\citep{bergemann2007,Bergemann2008} and of other similar ions \citep[e.g. Fe, ][]{Bergemann2012, Lind2012}. The corrections are slightly larger for the RGB model, especially at lower [Fe$/$H]. The higher-excitation lines, such as those of multiplets 18 and 32 are more sensitive to NLTE. Their NLTE abundance corrections typically increase  with decreasing metallicity, but this trend slightly flattens below [Fe/H] $\sim -2$. Another noteworthy feature of these diagrams is that fact that the \mnii~lines are also not immune to NLTE. It has been often assumed in the literature that the lines of ionic species do not show NLTE effects. The few strong excited lines of \mnii~at $1.85$ eV show the classical NLTE effect of photon loss. This effect is small, but shows in the atmospheres of dwarfs and giants (Fig. \ref{fig:nlteabu2}). It implies that \textit{lower} abundances would be obtained from \mnii~line especially for the metal-poor stars with [Fe$/$H] $= -2$. At lower metallicity, the \mnii~lines become weak enough and radiative pumping effects dominate that leads to positive NLTE abundance corrections.

It is interesting, and it possibly presents the main difference with respect to our earlier study, that the strong resonance triplet of \mni~at 4030-4034 \AA\ and the excited lines show very similar NLTE abundance corrections. The NLTE corrections  exceed just about $0.4$ dex in the atmospheres of RGB stars with [Fe/H] $=-3$. This is important, as LTE abundances derived from the resonance \mni~lines are known to be significantly lower compared to high-excitation \mni~features \citep{Bonifacio2009, Sneden2016}. In the 1D NLTE analysis, there is no room for differentially larger NLTE corrections for the 4030-4034 \AA\ triplet lines, compared to the high-excitation features. In our previous work, a higher degree of over-ionisation in \mni, and, in particular, over-ionisation from the ground state, was achieved by employing a tailored $S_H$ scaling factor to the Drawin collisional (excitation and ionisation) Mn I + H I rates. As a consequence, the NLTE abundance corrections for the resonance triplet lines were significantly higher. 

Below we show that 3D NLTE calculations suggest substantial differential effects between \mni~lines of different excitation potential, which help to improve the excitation balance (Sect. \ref{3dnlte}), effectively providing the physical basis for the effect, which is mimicked by using inefficient H collisions in 1D models.
\begin{figure*}
\hbox{
\includegraphics[width=0.5\textwidth, angle=0]{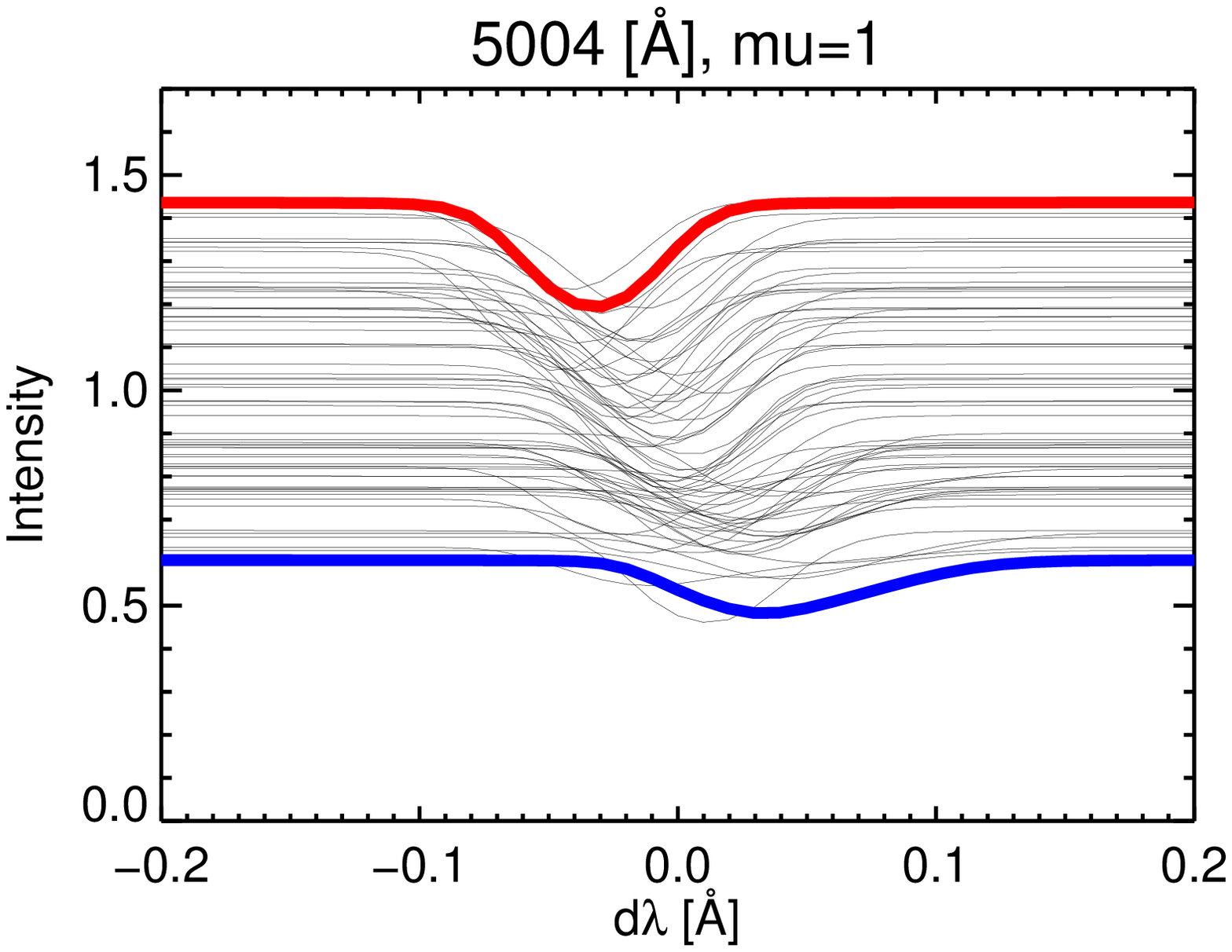}
\includegraphics[width=0.5\textwidth, angle=0]{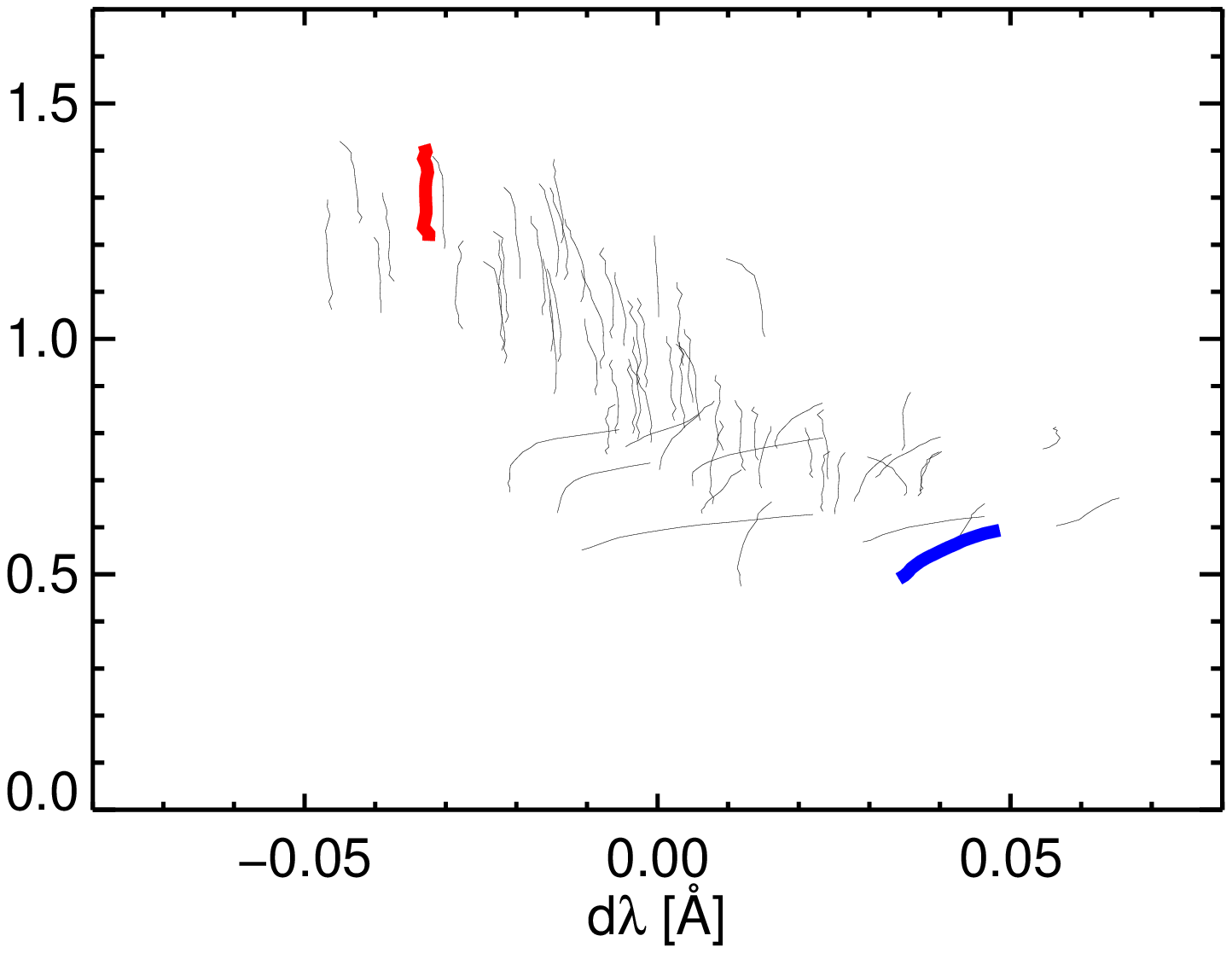}}
\hbox{
\includegraphics[width=0.5\textwidth, angle=0]{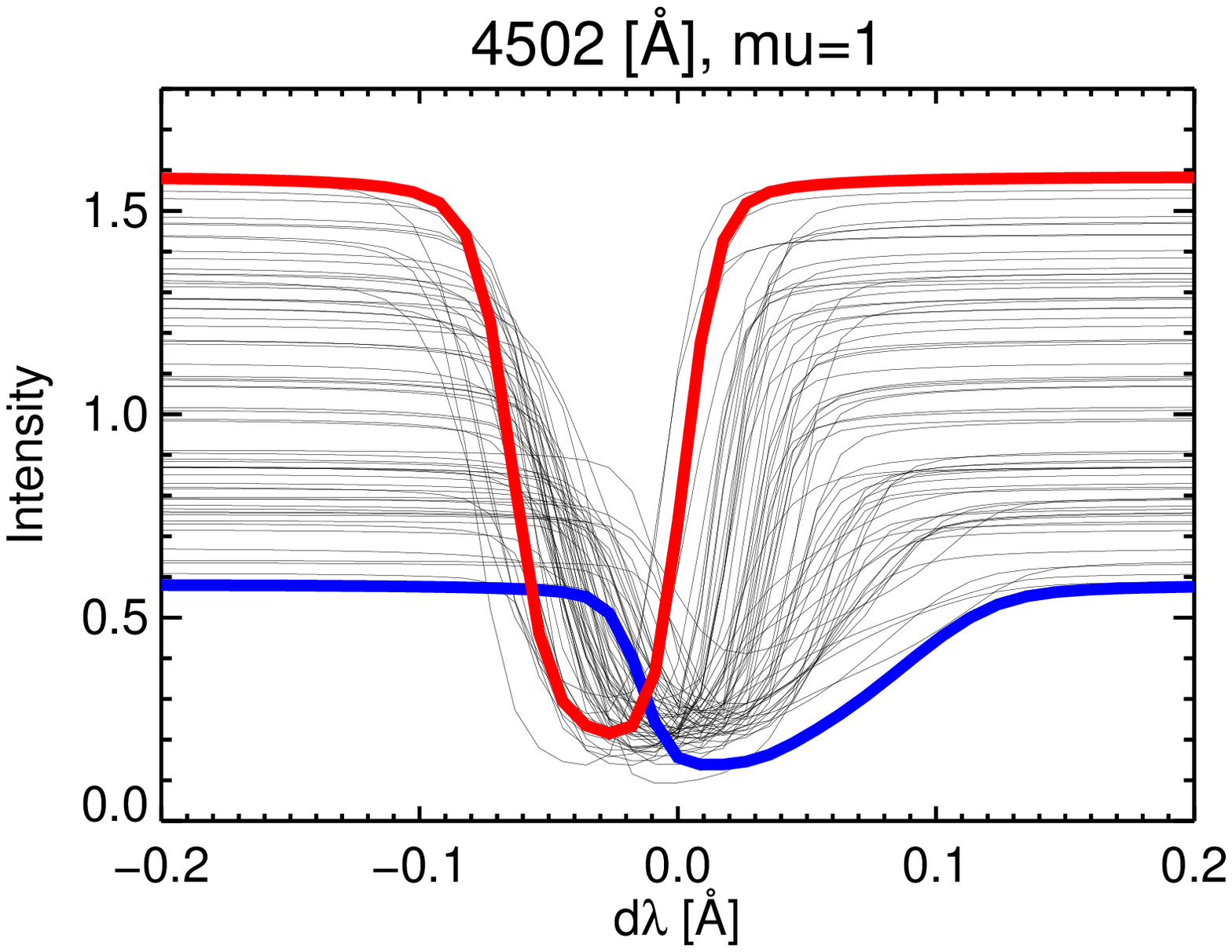}
\includegraphics[width=0.5\textwidth, angle=0]{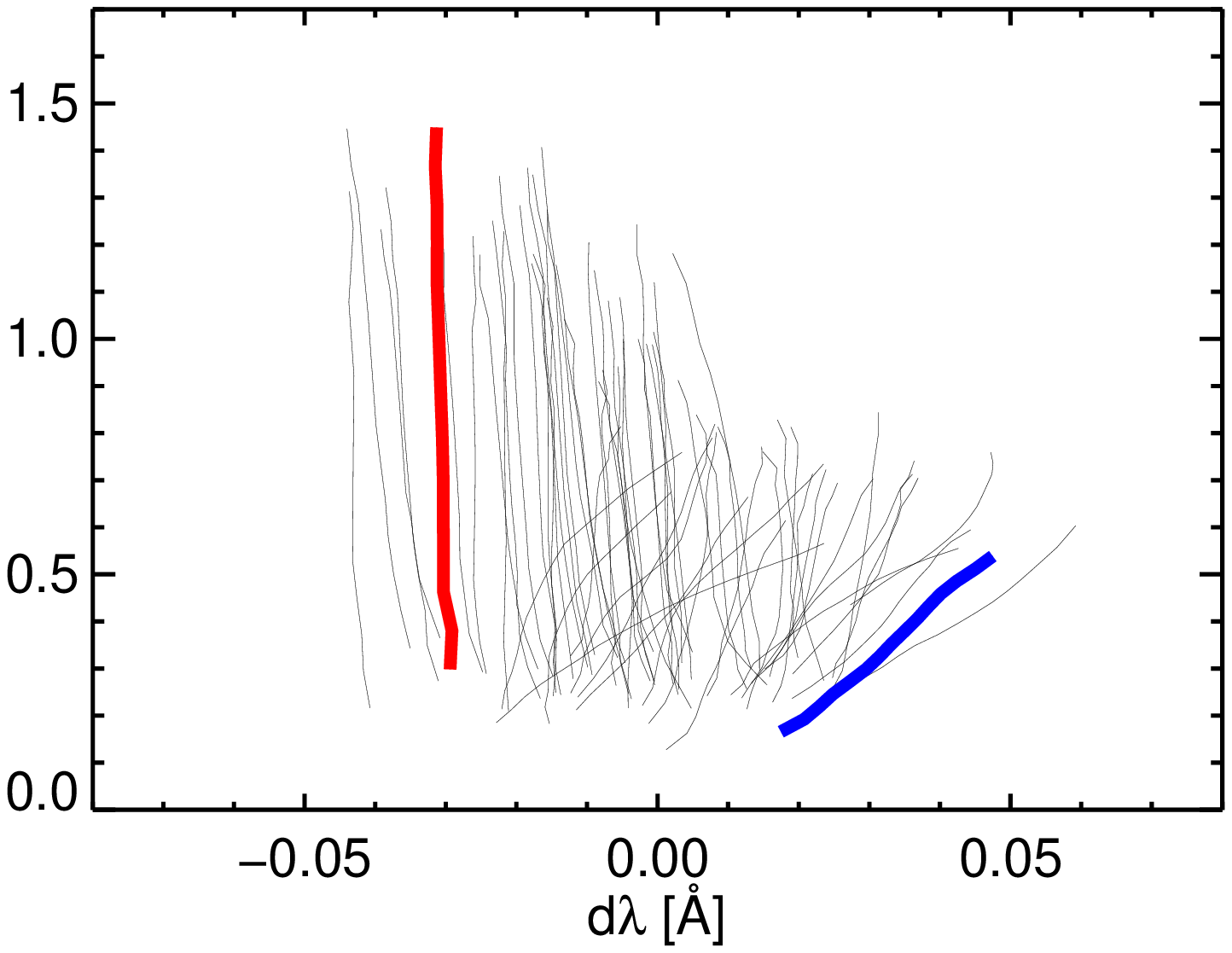}}
\caption{Spatially resolved line profiles (left panels) and their bisectors (right panels) for the solar disk centre in the model snapshot $020$. The solid red and blue curves indicate the profiles extracted from the granule, respectively, inter-granular lane, which are marked with red and blue boxes in Fig. \ref{fig:int5432}.}
\label{fig:bisec}
\end{figure*}
%
%
\subsection{3D NLTE}{\label{3dnlte}}

We begin the 3D NLTE analysis with a brief account of line formation in 3D inhomogeneous atmospheres, with the emphasis on the solar model. This is useful in order to understand the key differences between 1D and 3D abundances that will be the subject of Sections \ref{sun3d} and \ref{bmk3d}. 

Section \ref{photon} will deal with the properties of line formation in full 3D solar simulation cubes, that we refer to as \textit{photon kinematics}. In Section \ref{testcase}, we discuss simplified radiative transfer models and explore how these impact the line profiles compared to the full 3D NLTE solution. The results of the 3D NLTE solar abundance analysis are presented in Section \ref{sun3d}. In Section \ref{sec:mps3d}, we discuss the impact of 3D NLTE on the line profiles, their equivalent widths (EW), and on the abundance diagnostic for four metal-poor models and compare them with the results of LTE and NLTE calculations with 1D hydrostatic models, as well as with the predictions of 3D LTE models. Finally, in section \ref{bmk3d}, we use 3D convective models of the benchmark metal-poor stars HD 84937, HD 140383, and HD 122563 to derive 3D NLTE abundances.

\begin{figure}[!ht]
\includegraphics[width=\linewidth]{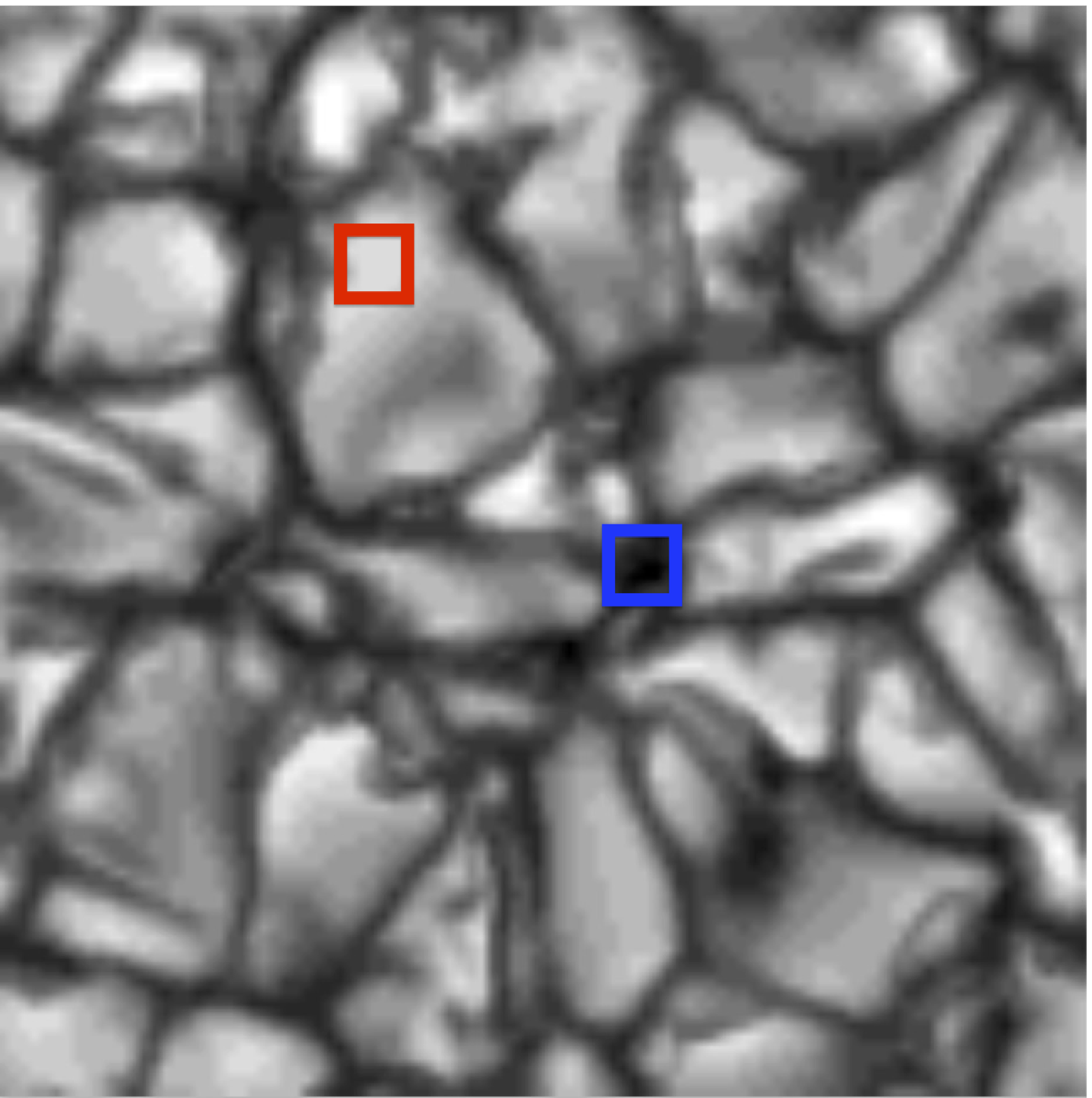}
\caption{A monochromatic image of solar granulation at the wavelength of $5394$ \AA~ at the solar disk centre.}
\label{fig:int5432}
\end{figure}
%
%
%
\subsubsection{Photon kinematics}{\label{photon}}
Figure \ref{fig:bisec} illustrates spatially-resolved NLTE intensity profiles of two \mni~lines in the solar model at the disk centre. All profiles are normalised to the average continuum intensity for the corresponding spectral line in the snapshot. The lines were chosen such that the effect of the HFS is minimal, in order to isolate the effect of granular motions on the profiles. The profiles are taken for every fourth point along each horizontal coordinate in the simulation domain (i.e. for $8 \times 8  =  64$ points out of 900), to not overload the figure. The bisectors for each line component are shown in the right-hand side panels. In addition, the solid red and blue curves indicate the profiles extracted from the granule, respectively, inter-granular lane, which are marked with red and blue boxes in Fig. \ref{fig:int5432}.

\begin{figure*}
\hbox{
\includegraphics[width=0.5\textwidth, angle=0]{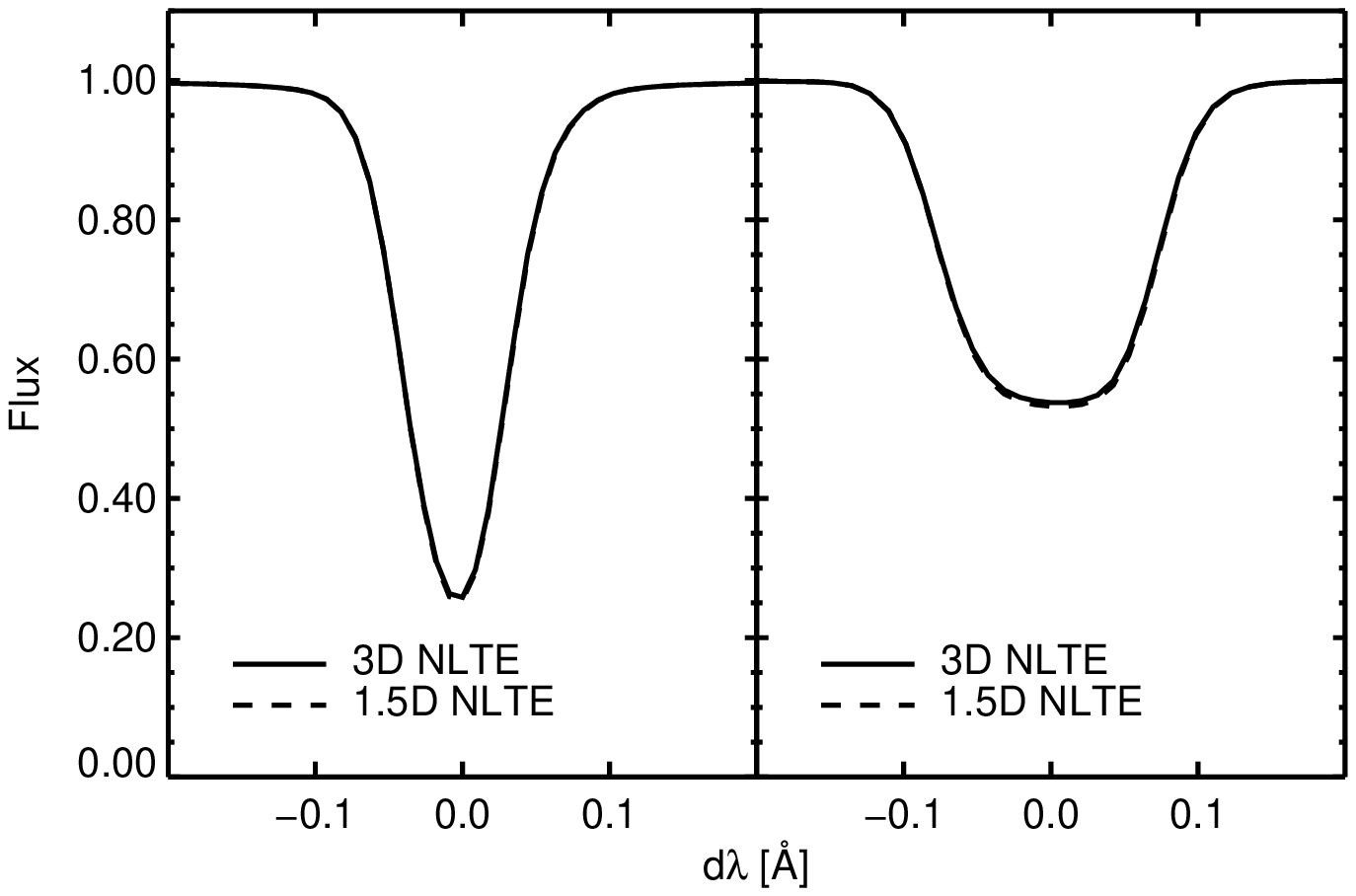}
\includegraphics[width=0.5\textwidth, angle=0]{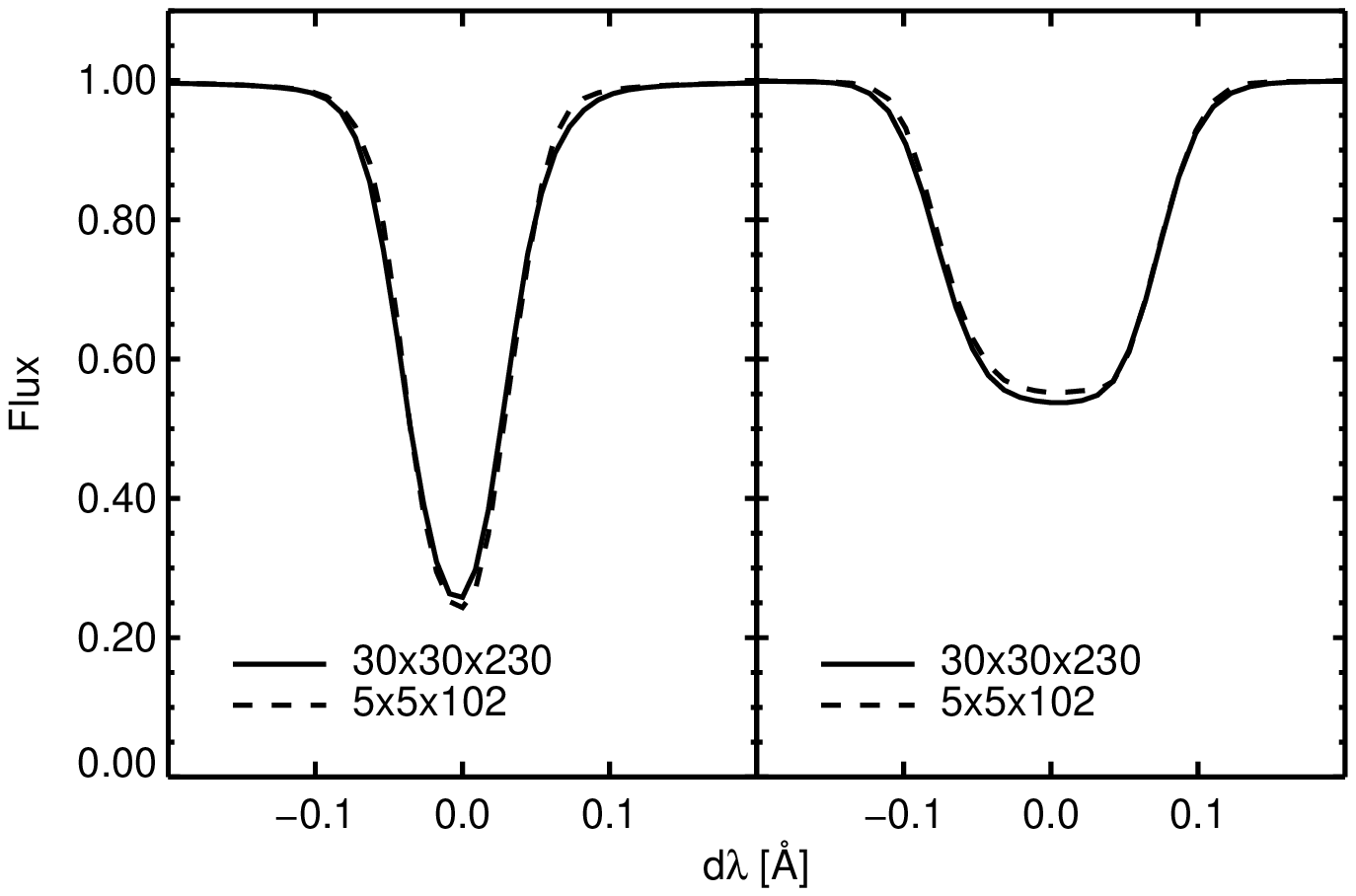}}
\hbox{
\includegraphics[width=0.5\textwidth, angle=0]{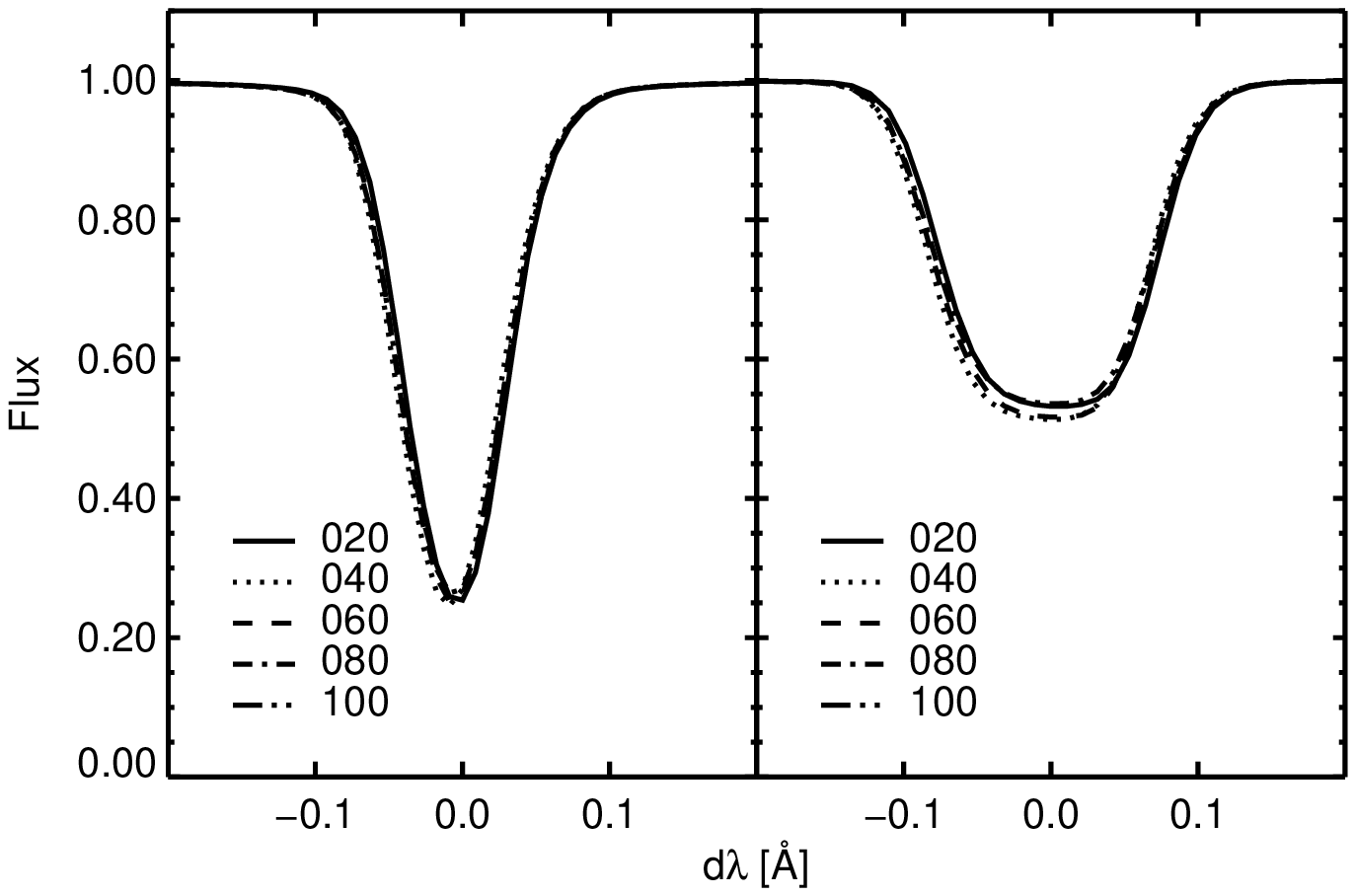}
\includegraphics[width=0.5\textwidth, angle=0]{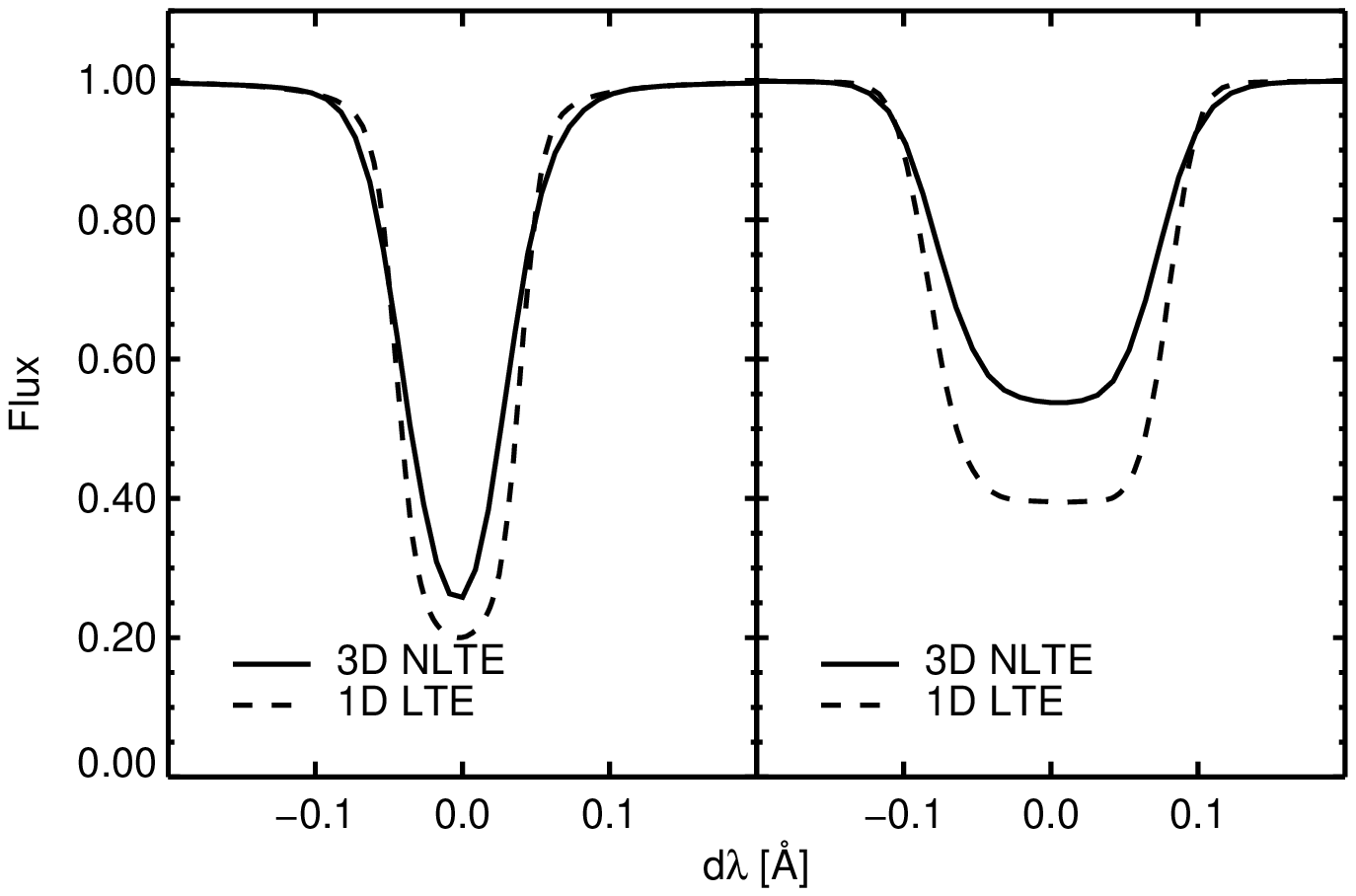}}
\caption{A comparison of the theoretical profiles for the \mni~lines at 4502 and 5394 \AA\ for the solar snapshot 020. Top left panel: 1.5D NLTE versus full 3D NLTE. Top right panel: Line profiles computed with the 3D data cubes scaled down to the resolution of (x,y,z) $= 30,30,230$ (solid line) and $5,5,102$ (dashed line). Bottom left panel: full 3D NLTE for the five solar snapshots. Bottom right panel: 3D NLTE profiles versus 1D LTE profiles.}
\label{fig:test3}
\end{figure*}

Overall, the behaviour of the lines is very similar to that described earlier by \citet[][e.g. their Fig. 6 for a Sun-like star $\alpha$ Cen A]{Dravins1990a}. The weaker high-excitation \mni~line at 5004 \AA~shows a strong anti-correlation between the depth of the line core and the line shift (the right-hand panel). This is, in fact, the weakest unblended solar \mni~line with the EW of only 13 m\AA. The line profiles with the strongest blue-shift and the highest intensity contrast form above the granules, where the upwards streaming motions of hotter material are characterised by higher velocities, and higher granular temperatures account for the brighter background continuum. The dominant non-LTE effect of over-ionisation leads to the brightening in the line core. The lack of any pronounced curvature in the bisectors of the blue-shifted components suggests that there is little vertical variation of velocity field in the upflows.

The \mni~line at 4502 \AA, which is stronger but has the same lower excitation potential as the 5004 \AA\ line, also shows a very broad distribution of line shifts. This line is close to saturation, as is evidenced by the broad rectangular inner core in the bluest components. Similar to the $5004$ line, the bisectors of the blue-shifted components, which form above granules, are typically $l$-shaped, i.e., the line profiles are nearly symmetric. Although the blue-shifted components are very strong, and are characterised by very extended (vertically) line formation regions, this again suggests that the vertical variation of velocity fields in the granules is small. On the other hand, the bisectors of the profiles that are forming above the inter-granular lanes tend to approximate a $c$-shape. These line components are highly asymmetric, their cores are very broad, and tend be skewed to the red. \citet{Dravins1990a} found that this is a characteristic feature of the lines that form across inter-granular regions with a larger vertical velocity gradient with depth.
%
%
%
\subsubsection{3D test cases}{\label{testcase}}
Calculations of 3D NLTE radiative transfer are very computationally expensive. Hence, we explored whether a simplified treatment of NLTE radiative transfer with 3D simulations offers a suitable alternative to full 3D NLTE calculations. 

In particular, we illustrate and discuss the results of calculations obtained using more compact  model atmosphere cubes and different radiative transfer solvers (1.5D versus full 3D). We also compare the flux profiles obtained using different solar snapshots. In what follows, we limit the discussion to two \mni~lines only. One of them is a strong resonance line at 5394 \AA, where the effects of NLTE and 3D convection are most pronounced. The second line is that at 4502 \AA, discussed in the previous section. This line is least affected by the HFS, hence its shape is a good test case to explore the effects of NLTE and 3D convective flows.

Figure \ref{fig:test3} (top left panel) illustrates the impact of the radiative transfer solver, full 3D NLTE vs $1.5$D, that is, column-by-column solution, ignoring non-vertical radiative transfer, using the reference model atom. The difference is negligible and it amounts to less than $0.5\%$ in the flux level, or $< 0.01$ dex in abundance, that is far less than the uncertainty incurred by the other sources of error. In the case of metal-poor stars, the differences are also modest and do not exceed a few percent in EW. Since the computational time of 3D NLTE scales as $O(n^{4/3})$, where n is the number of grid points, that is the number of iterations is proportional to the number of grid points in the z direction, it is a reasonable approximation to use $1.5$D NLTE calculations, at least for the stars with parameters not too different from the models tested in this work. 
\begin{figure*}[!ht]
\includegraphics[width=1\textwidth, angle=0]{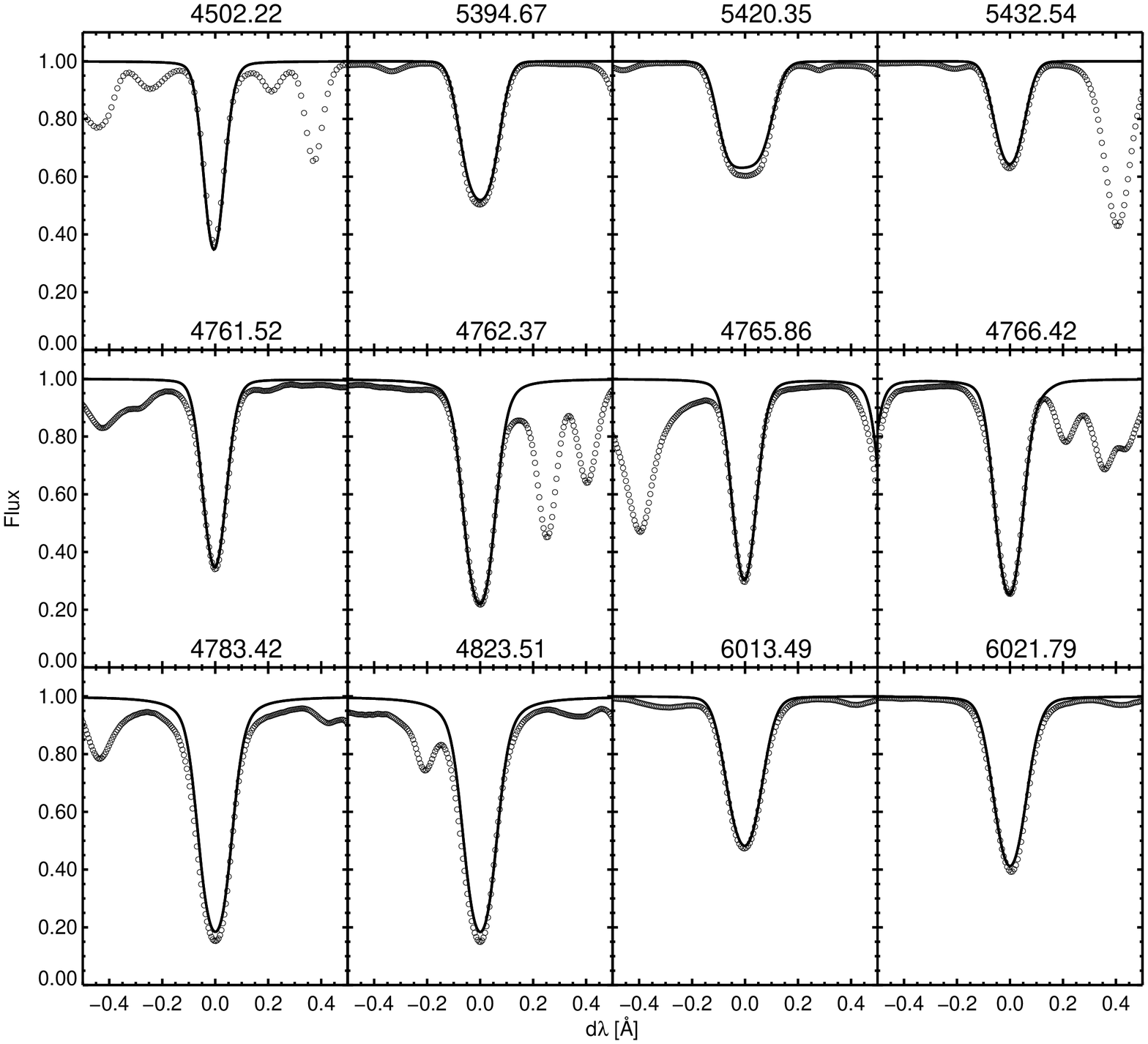}
\caption{Comparison of the 3D NLTE line profiles for selected \mni~lines with the observed solar flux spectrum. All model profiles were generated using the same Mn abundance of $5.47$ dex.}
\label{fig:prof_new}
\end{figure*}

Figure \ref{fig:test3} (top right panel) illustrates the impact of the model atmosphere resolution. We employ three models: one resampled to the resolution of 5x5x102, the other with the resolution of 30x30x230 (our default resolution), as well as the maximum resolution of 100x100x230 that we can afford with the computing time available to us. The comparison of the medium- and maximum resolution simulation can be found in the Appendix. The effects are visible in the line profiles, but they only slightly change the shape of the line and have a very modest effect on the line equivalent width. This effect comes mostly through the impact of velocities on the line opacity, $\kappa (x, \nu \cdot (1 - \nu(x) / c))$ as is best seen in the lines, which are not very sensitive to HFS, such as the $4502$ \AA~line (left panel in Fig. \ref{fig:test3}). Their shapes are closely approximated by a Gaussian, and the effects of velocity fields are easily distinguished. In particular, the red wing of the line appears to be slightly skewed, i.e. darker, in the model with higher resolution. There is also an implicit effect of velocities on the line source function, as it depends on the radiation field at the line frequencies, which is non-local and connected to the velocity distribution at all spatial points along the beam.

We also find (Fig. \ref{fig:test3}, bottom left panel) that the flux profiles computed using different (regularly-spaced in time) snapshots do not show significant variations. The 5394 \AA~line changes its depth by $\sim 3\%$. The position of the line centre changes by a few m\AA. The 4502 \AA~line is almost insensitive to the temporal effects. This suggests that the statistical properties of velocity fields, temperature, and pressure inhomogeneities in each snapshot are representative of some average distribution values for the relaxed simulation given the values of $\teff$, $\log g$, and $\feh$. The temporal variability of spatially resolved profiles has been examined in detail in \citet[][their Sect.5]{Dravins1990c, Dravins1990b}. They found that the characteristic evolution of line shapes and depths lacks a unique timescale, that was a manifestation of the lack of a unique scale in the spatial power spectra. Hence, the  ensemble averages of spatially resolved line profiles are very similar in different snapshots, as we will also demonstrate in Sect. \ref{sec:mps3d} for other models. 

Finally, Fig. \ref{fig:test3} (bottom right panel) illustrates the line profiles computed with 1D LTE and 3D NLTE. This is a major difference that amounts up to $0.2$ dex in the abundance (see next section). Similar to the earlier studies for ions with similarly complex electronic structure and ionisation potential \citep[e.g.][for Fe I]{Amarsi2016, Lind2017}, we find that the effects of 3D NLTE are to weaken the line compared to 1D LTE. \mni~is a photoionisation-dominated ion, that is the NLTE effects are driven by strong non-local radiation field across multiple ionisation channels of the ion in the UV and in the optical. Of all \mni~lines, the resonance lines at 5394 and 5432 \AA~are most sensitive to this effect, as the lines are comparatively weak in the solar spectrum, and their NLTE effect is mainly caused by the change in the line opacity that scales in the first order as the ratio of the NLTE to LTE number densities of the lower energy level of the transition. The higher-excitation lines, such as that at 4502 \AA~show more modest NLTE effects in the model atmosphere of the Sun, both in 1D and in 3D. 

The results of our test calculations indicate that vertical radiative transfer in NLTE with full 3D data cubes at the resolution of (x,y,z) $=30,30,230$ is the most suitable approach and we employ this method in our calculations of Mn abundances for the Sun (Sect. 4.2.3) and the benchmark stars (Sect. 4.4). The approach allows affordable 3D NLTE calculations within reasonable timescales and provides line profiles consistent with the original resolution of the 3D cubes, as well as, with the full 3D radiative transfer calculations.
%
%
%
\subsubsection{Solar Mn abundance}{\label{sun3d}}
The Sun is a metal-rich star, and as such, the spectral lines in most cases appear to suffer from blending and/or strong sensitivity to damping or turbulence. As a matter of fact, most \mni~lines are not very useful, being either blended or too strong, with a few exceptions. Nonetheless, we start with the analysis of all \mni~lines that are typically used in the analysis of FGK stars, and later assume a more conservative approach that takes into account only the most reliable features.

 The LTE analysis of the solar \mni~ lines is carried out using the SIU spectrum synthesis code (Reetz 1999). SIU allows interactive spectral fitting, which is very convenient to test for the presence and effects of blends and asymmetries within a line. This is a very accurate approach, especially when applied to ultra-high-resolution spectra, such as the solar atlases, and has been commonly exploited in the literature.

Figure \ref{fig:prof_new} presents some examples of 3D NLTE line profiles for the solar spectrum. Note that in this plot, we have not adjusted the Mn abundances, but assumed the abundance of $5.47$ dex, as recommended by earlier studies of the solar Mn abundance. Most of the \mni~lines are very broad, with the width of $0.2-0.3$ \AA, which is the consequence of the HFS. It is obvious that the available atomic and HFS data describe the line profiles remarkably well, both in terms of the asymmetries and the multi-component structure. The latter is particularly prominent in the cores of the weak resonance or ground-state lines, such as the 5394 and 5516 \AA~line. The core of the line at 5420 \AA\ is slightly too weak, compared with the observations. This could be due to its sensitivity to activity \citep{Wise2018}. Note that also the 5394, 5432 and 6013 lines could be affected \citep{Wise2018}, and in particular, the 5394 \AA\ line has a long history of research owing to its variation with the global solar activity cycle \citep{Danilovic2005, Livingston2007}. Some studies suggest that the line is not as sensitive to granular motions as other features, because of its wide HFS, and is more sensitive to the intergranular magnetic concentrations, similar to the CH G-band \citet{Vitas2009}. According to our 3D NLTE results, however, the 5394 \AA\ line is sensitive to convection and shows a significant difference between modelling with 1D hydrostatic and 3D convective models. Overall, the agreement of the model 3D NLTE line profiles with the observed solar data suggests that the atomic data quality and the physical quality of spectral models are sufficient to satisfactorily describe the properties of Mn lines in the solar spectrum.

\begin{figure}[!ht]
\includegraphics[width=0.5\textwidth, angle=0]{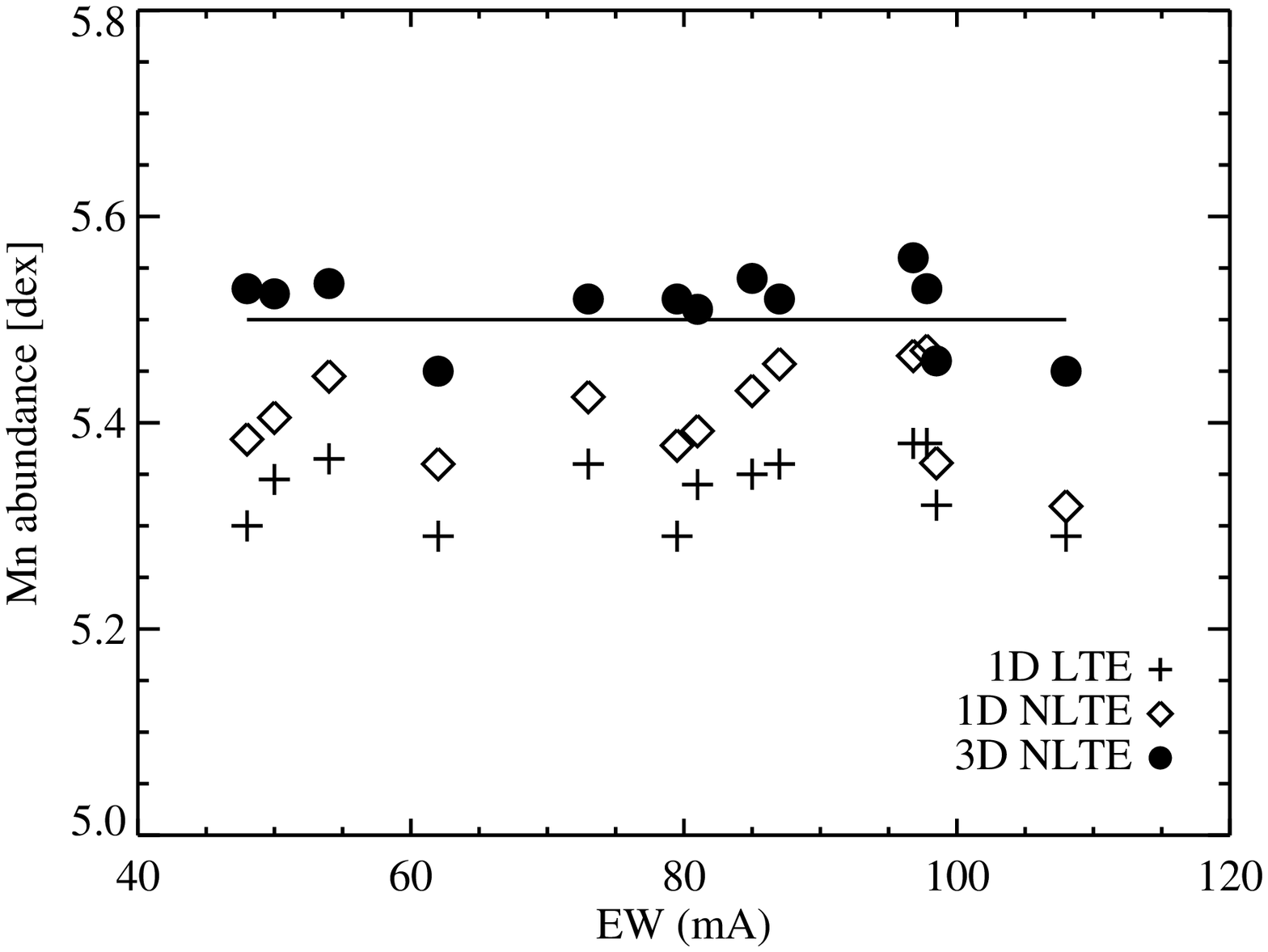}
\includegraphics[width=0.5\textwidth, angle=0]{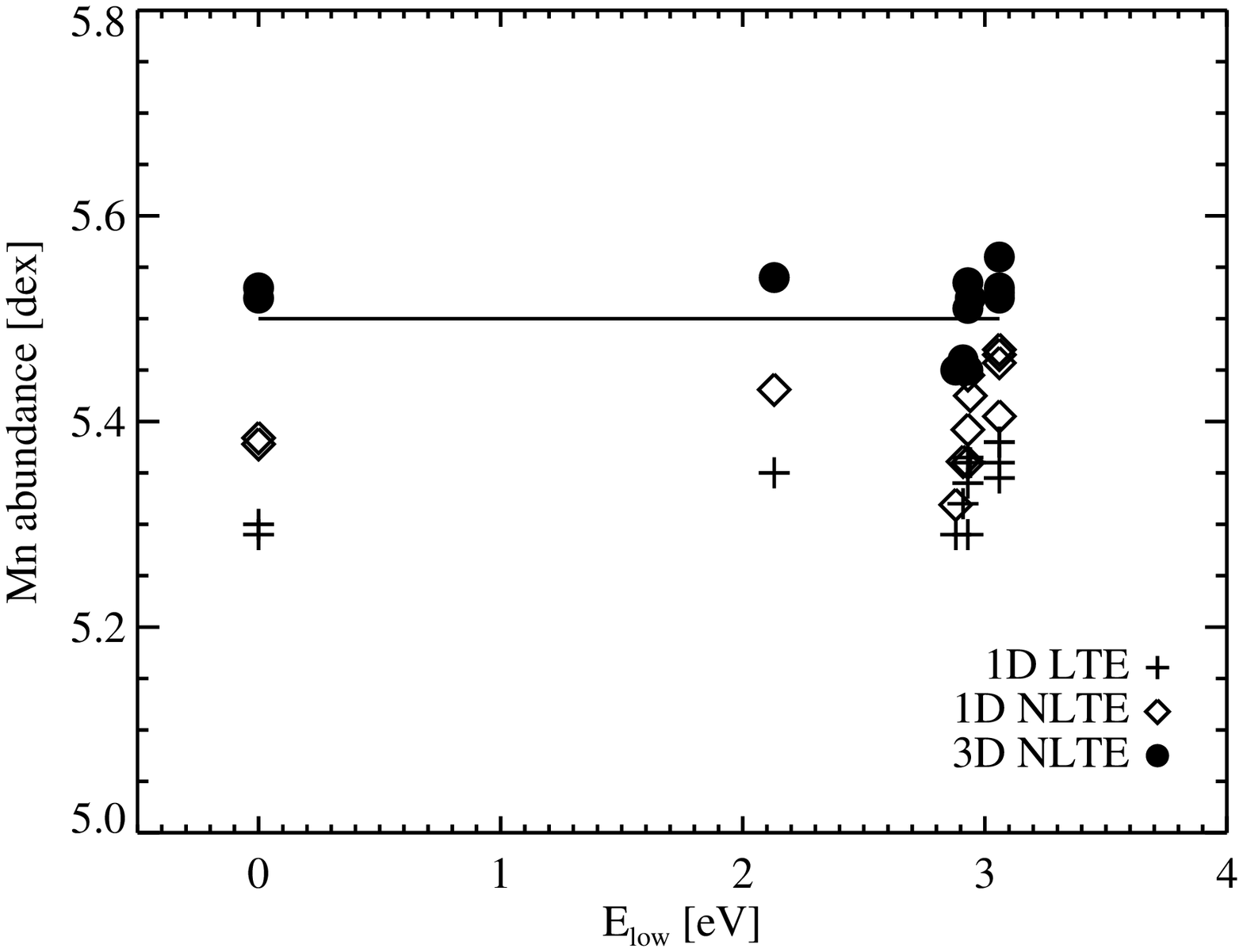}
\caption{Mn abundances determined from the individual \mni~lines in the solar flux spectrum using different models: 1D LTE, 1D NLTE, and 3D NLTE. The solid line denotes the Mn abundances measured in CI chondrites, $5.50 \pm 0.03$ dex \citep{Lodders2003}.}
\label{fig:sun}
\end{figure}

The abundances determined from the \mni~lines that we view as reliable photospheric abundance diagnostics are shown in Figure \ref{fig:sun}. The major challenge in the analysis of the solar spectrum is the contamination of line profiles by blends. These are present in the wings of the strong \mni~lines, such as the $4761$ \AA, $4762$ \AA, $4765$ \AA, $4766$ \AA, but also the lines of multiplet 18 ($4783$ and $4823$ \AA). Besides, there are no quantum-mechanical data for elastic collisions with H atoms for the two latter lines \citep{Barklem2000}, hence, we have to resort to the standard Uns\"old formalism that carries an additional source of an uncertainty. We disregard such lines when computing the solar abundance of Mn. The 1D NLTE and 3D NLTE Mn abundances are derived by applying the corrections computed using MULTI2.3 and MULTI3D for the individual spectral lines.

In 1D LTE, the low-excitation potential and/or very strong lines give a lower abundance of Mn compared to the high-EP (weaker) lines. The difference is not large, but significant given the very quality of the observed data and well-constrained fundamental parameters of the Sun. The effects on NLTE in the solar atmosphere are moderate and change the average solar abundance by only $\sim 0.02$ to $0.07$ dex. NLTE alone does not help to improve the solar excitation balance. More important is the combined influence of convection and NLTE. In the case of NLTE calculations with the 3D convective solar model, the effect of over-ionisation is amplified owing to stronger background radiation fields in the granules. The 3D NLTE - 1D LTE difference is equivalent to $\sim 0.22$ dex in abundance for the resonance lines, but $0.15$ dex for the higher excitation lines of other multiplets. 

Our solar abundances of Mn are $5.34 \pm 0.04$ dex in LTE, $5.41 \pm 0.05$ dex in NLTE, and $5.52 \pm 0.04$ in 3D NLTE. The 1D NLTE and 3D NLTE values are consistent with the meteoritic abundance of Mn $5.50 \pm 0.03$ dex reported by \citet{Lodders2003}. Our 1D LTE result is $\sim 0.1$ dex higher compared to the previous estimate reported in \citet{bergemann2007}, $5.23 \pm 0.1$ dex. On the other hand, it matches very well our LTE estimate obtained using the MAFAGS solar model, the revised transition probabilities, and the SIU spectrum synthesis code in \citet{Blackwell-Whitehead2007}, $5.33 \pm 0.1$ dex. The new 1D NLTE values are also higher than the estimates in \citet{bergemann2007}. In both cases, 1D LTE and 1D NLTE, the difference is due to the revision of the transition probabilities and the NLTE model atom. 
\subsubsection{3D NLTE effects in metal-poor model atmospheres}{\label{sec:mps3d}}
\begin{figure*}[!ht]
\hbox{
\includegraphics[width=0.5\textwidth, angle=0]{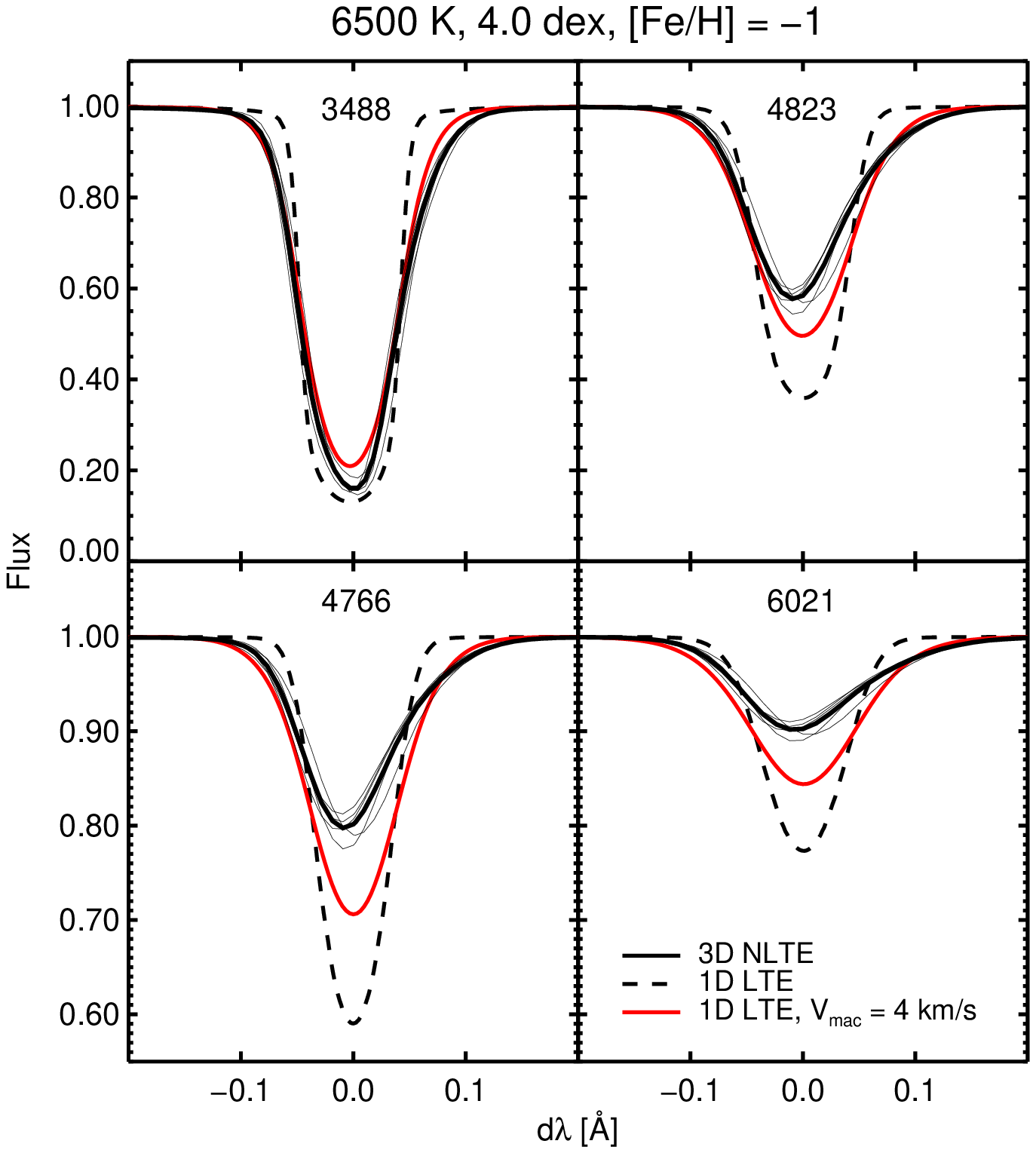}
\includegraphics[width=0.5\textwidth, angle=0]{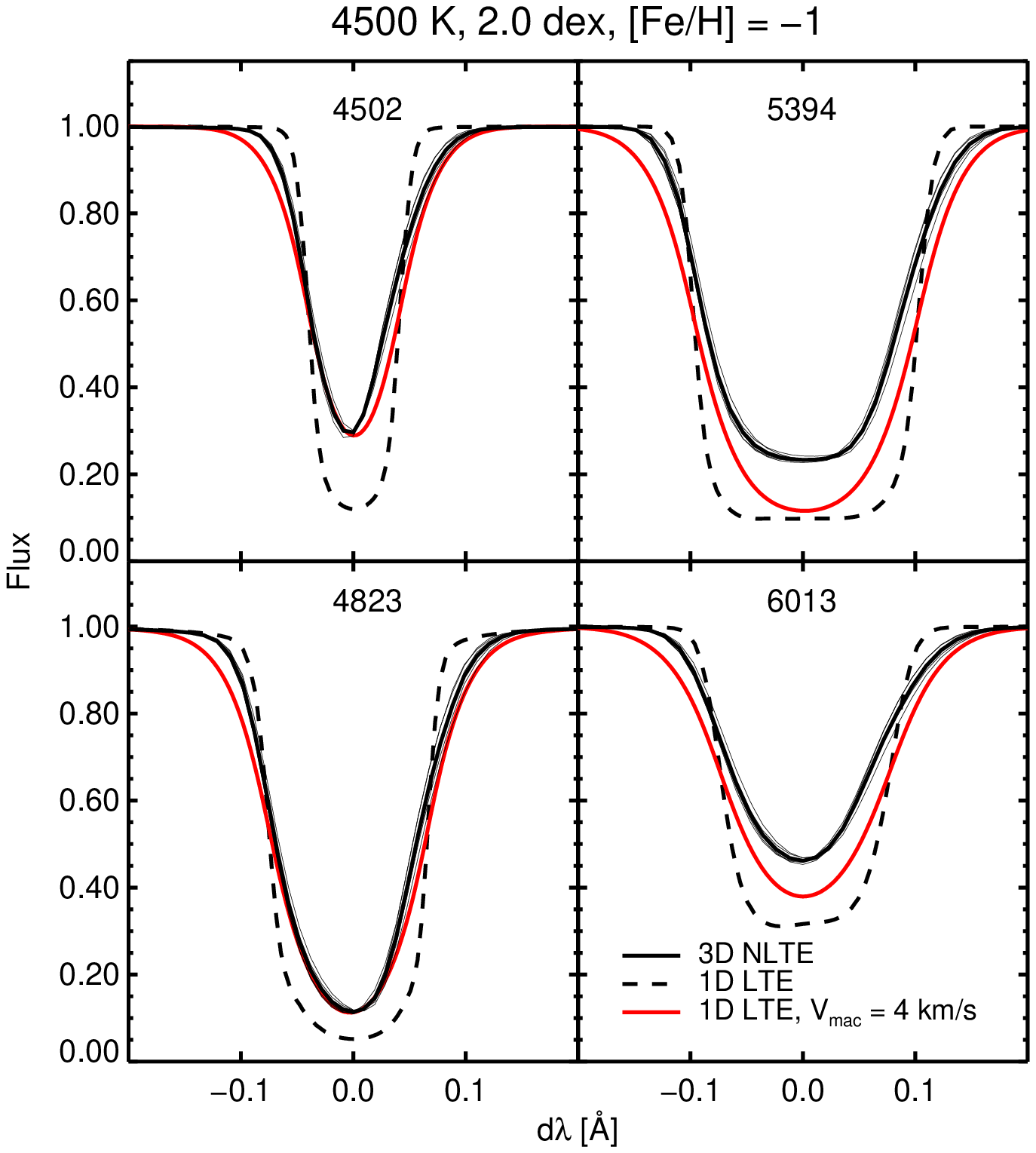}}
\hbox{
\includegraphics[width=0.5\textwidth, angle=0]{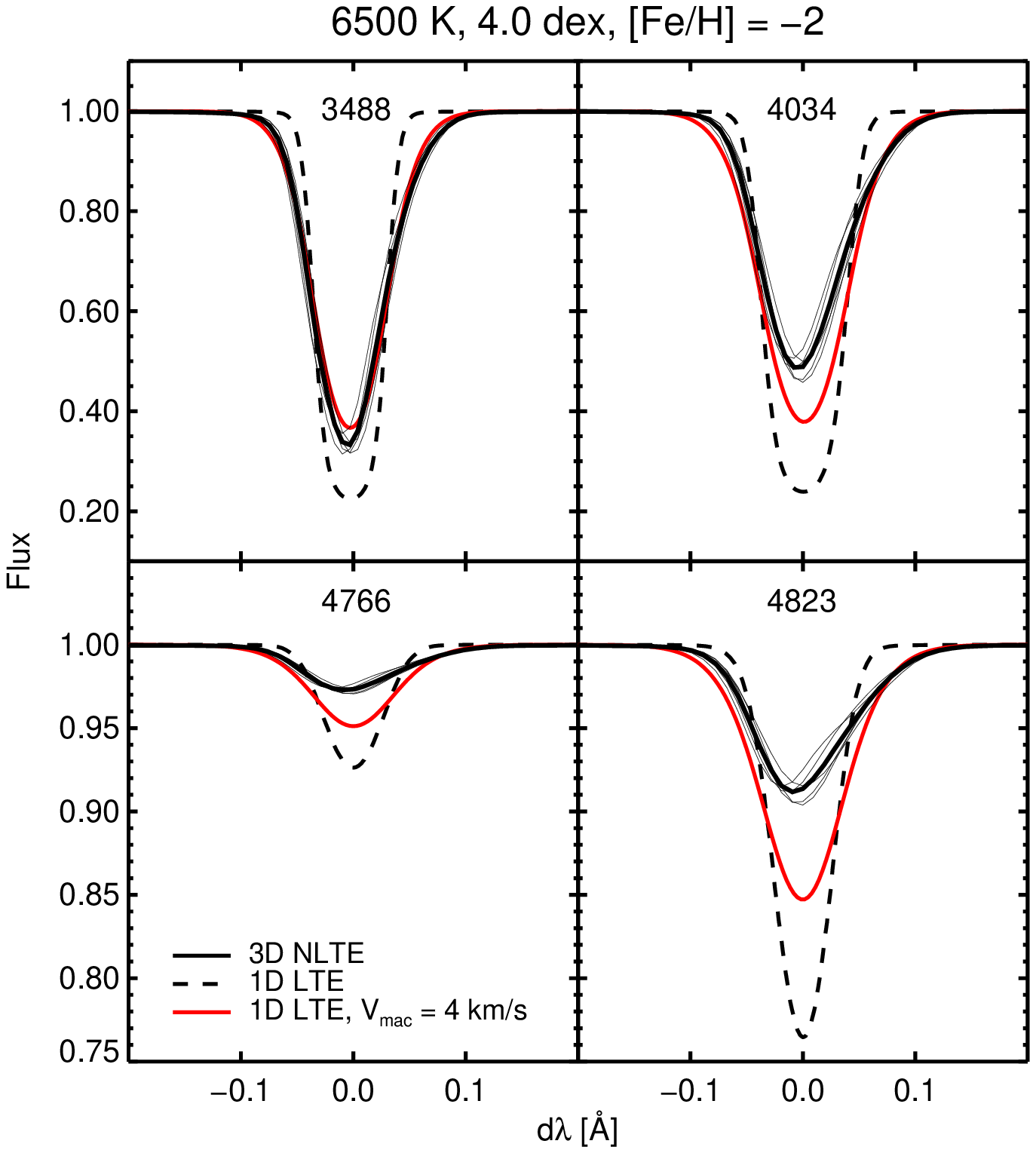}
\includegraphics[width=0.5\textwidth, angle=0]{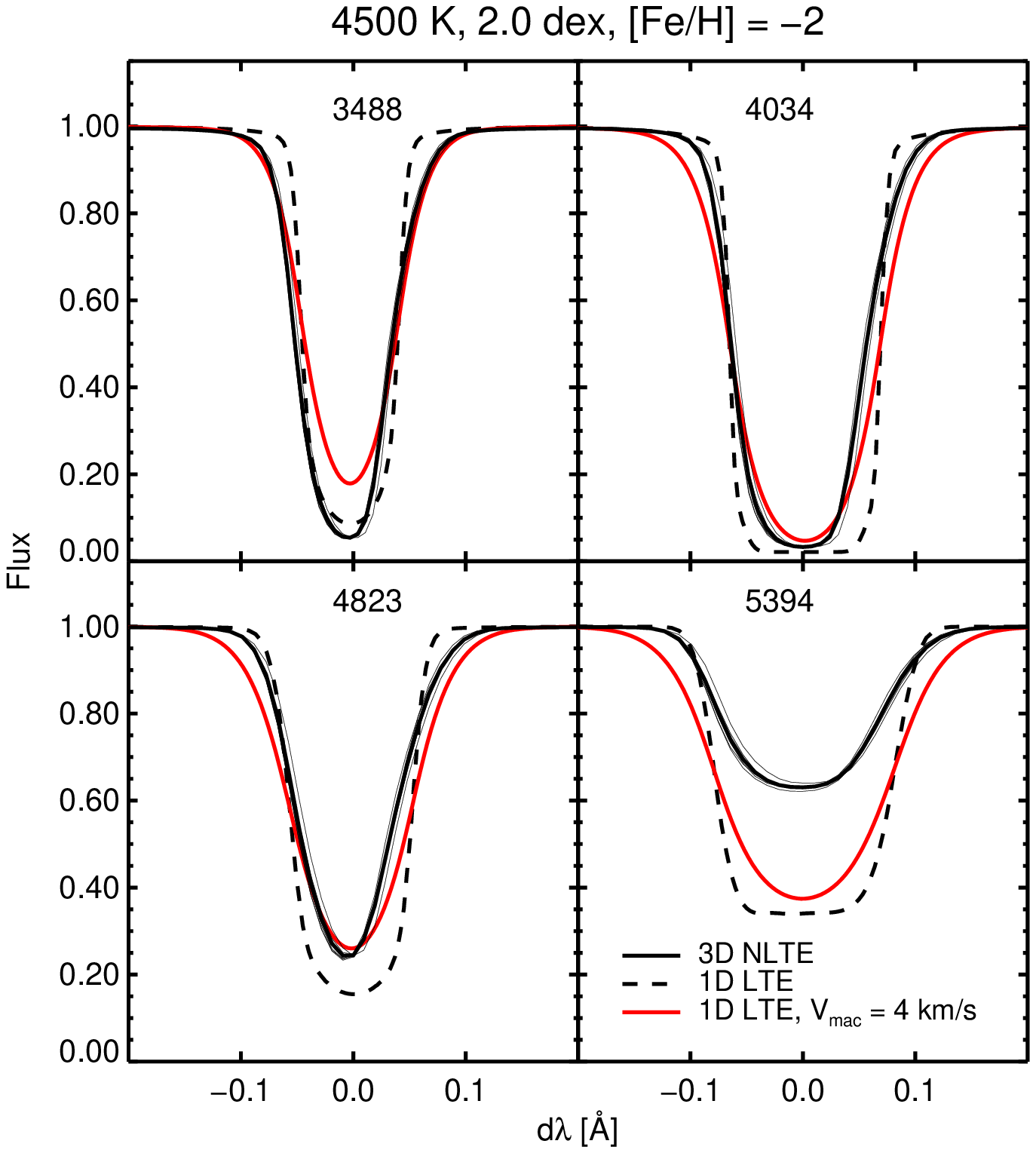}}
\caption{Line profiles of selected \mni~lines in the models of a Turn-off and an RGB star with metallicities $\feh =-1$ (top) and $\feh=-2$ (bottom panels), respectively. Thick solid curves correspond to the averaged profiles from five snapshots taken at equidistant time steps. Thin lines are the profiles from individual snapshots. 1D LTE profiles are computed assuming $\Vmic = 1$ km/s. The 1D LTE profiles would be stronger, if larger  $\Vmic$ values were assumed. Note that different spectral lines of \mni~and \mnii~are shown in the  panels.}
\label{fig:prof3dnlte}
\end{figure*}
%
%
\begin{figure*}[!ht]
\hbox{
\includegraphics[width=0.5\textwidth, angle=0]{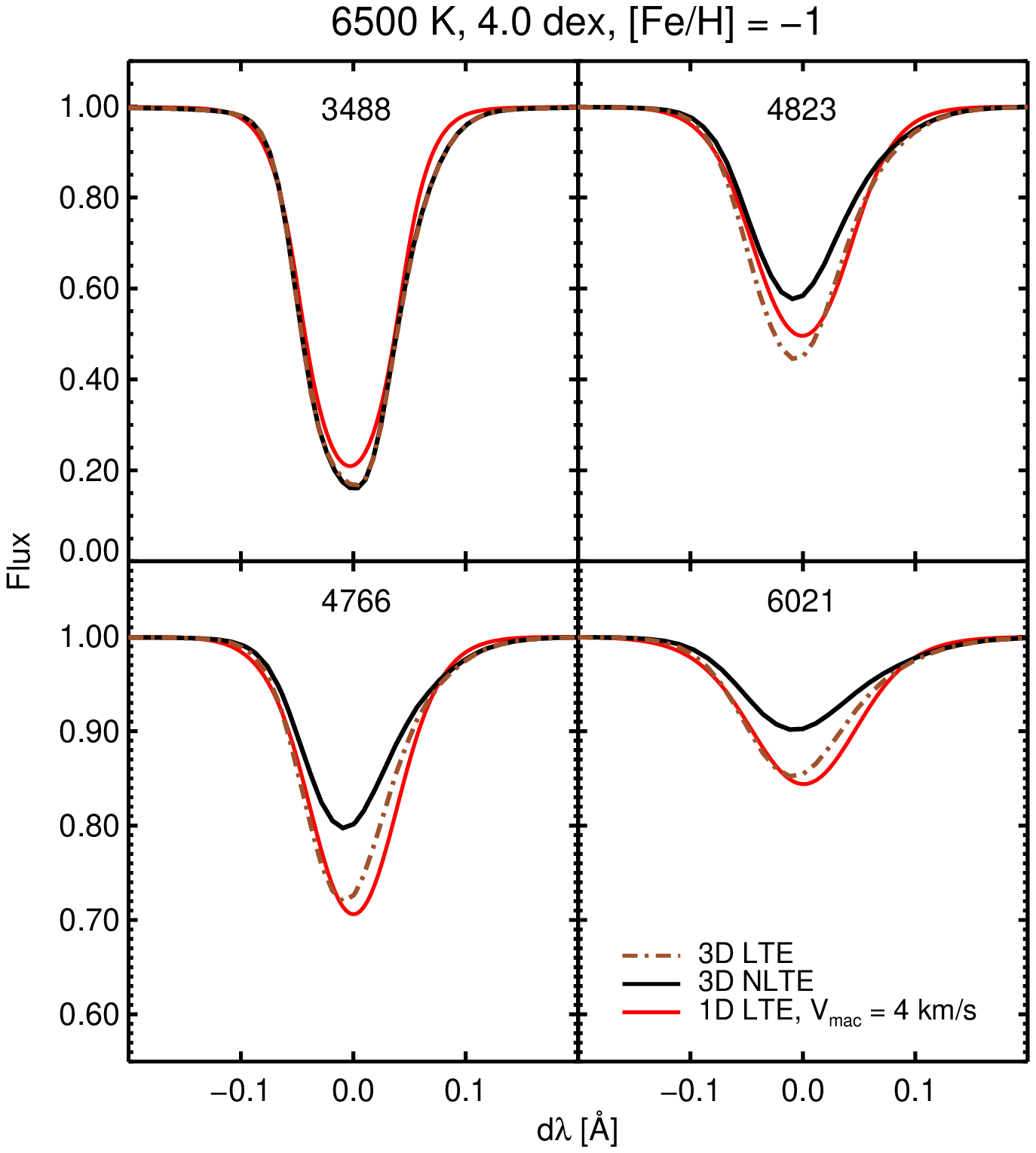}
\includegraphics[width=0.5\textwidth, angle=0]{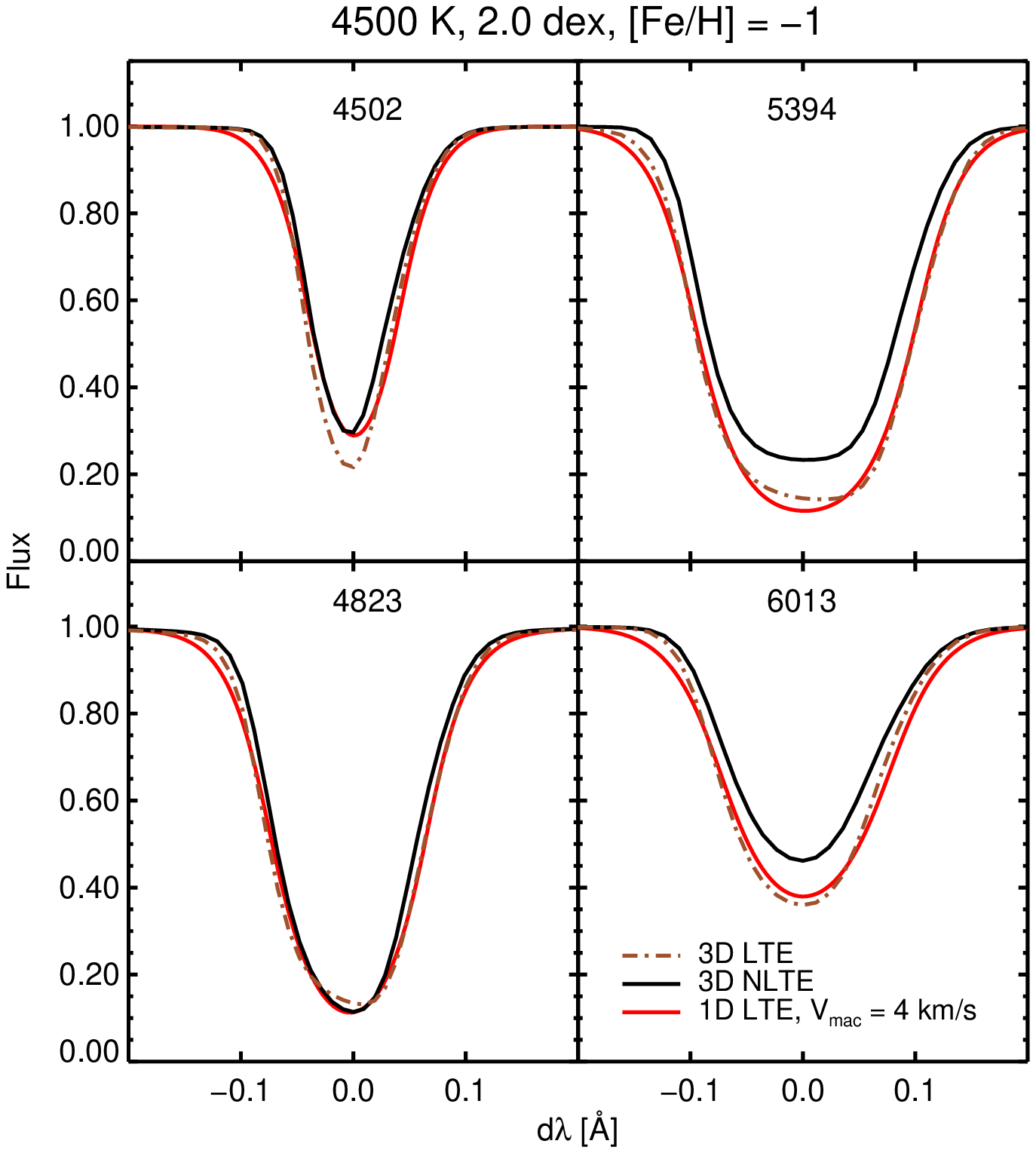}}
\hbox{
\includegraphics[width=0.5\textwidth, angle=0]{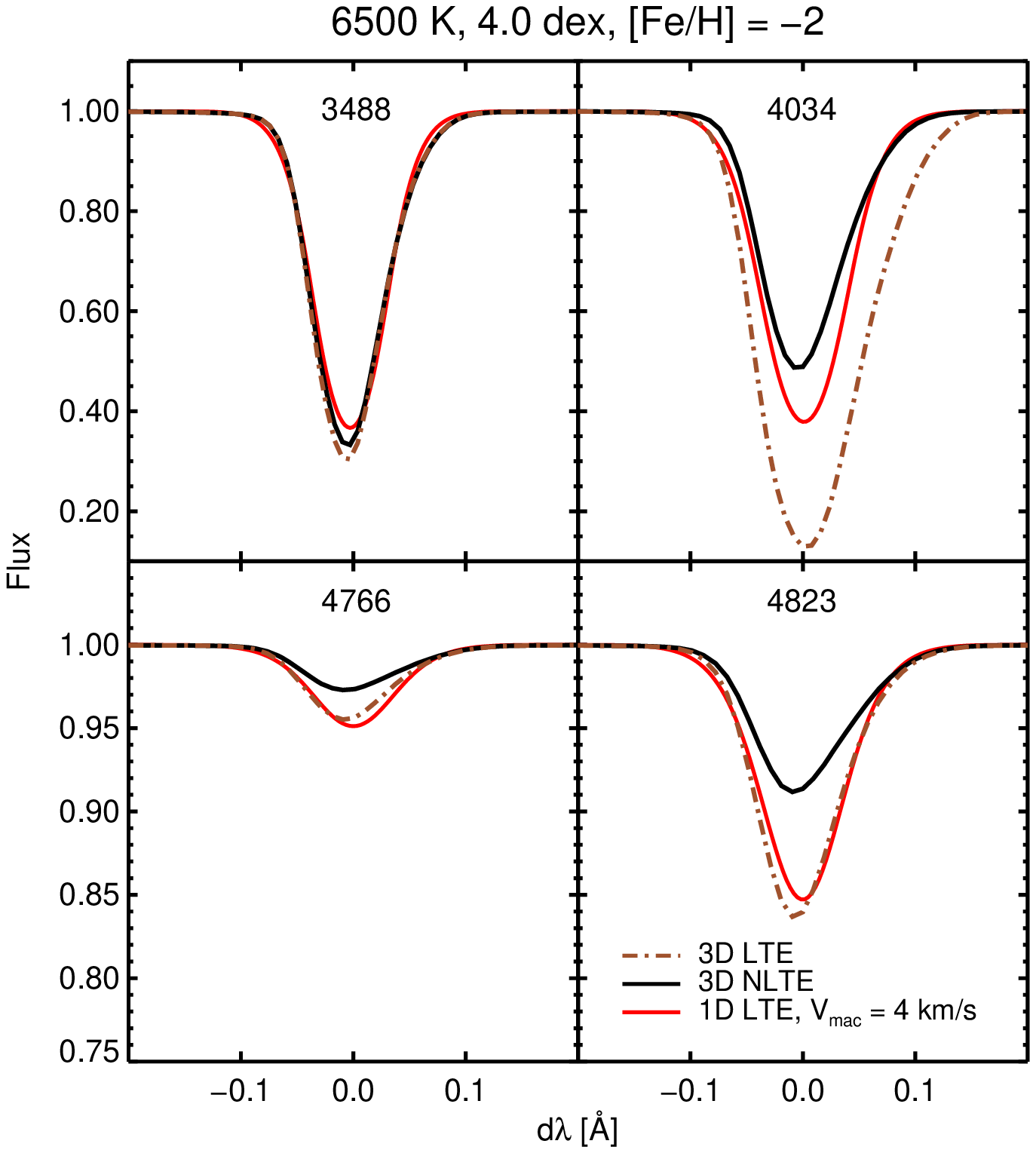}
\includegraphics[width=0.5\textwidth, angle=0]{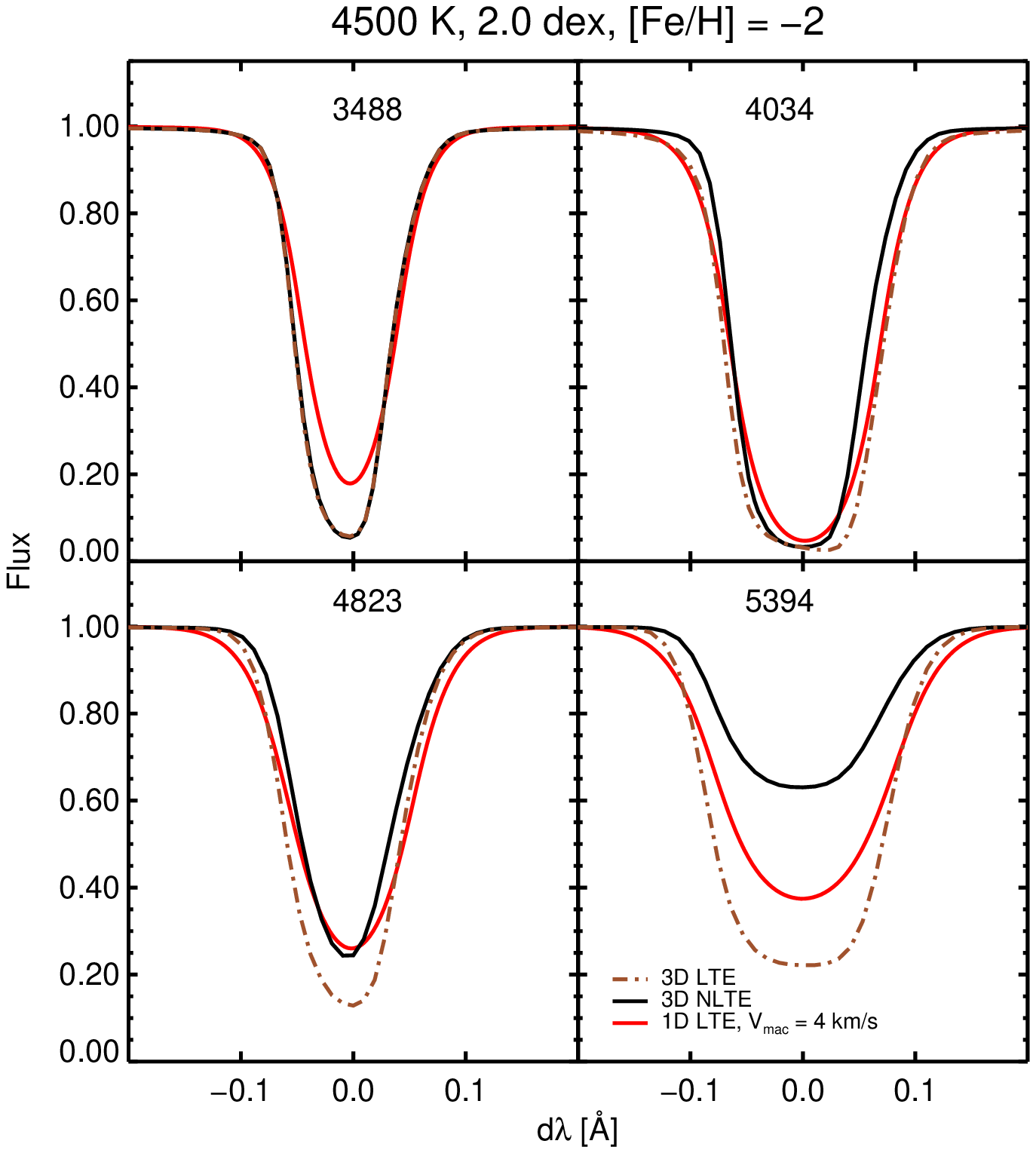}}
\caption{Same as Fig. \ref{fig:prof3dnlte}, but thick brown dot-dashed curves now denote the  profiles from 3D LTE calculations for the same five snapshots.}
\label{fig:prof3dlte}
\end{figure*}

Our analysis of line formation with 1D models suggests that the majority of Mn lines are sensitive to NLTE. Also the calculations for the 3D hydrodynamical model of the Sun indicate a strong impact of convection on the line profiles, which does not cancel the NLTE effects, but rather amplifies them for the most spectral features of \mni. 

In this section, we present the analysis of line formation in several model atmospheres of metal-poor stars. We explore the models with parameters with $\teff = 6500$ K, $\log g = 4$ dex, $\feh = -1$ and $\teff$ = 4500 K, $\log g = 2$, $\feh = -1$, as well as their metal-poor counterparts with $\feh = -2$.  For simplicity and didactic clarity, all calculations are performed assuming a scaled solar Mn abundance. Deviations from this assumption may affect  the line formation, yet at this stage we are interested in the combined effects of 3D radiative transfer in convective models and NLTE and the impact on \mni~lines with different properties. We explore which lines are least sensitive to these effects, and employ them in the abundance analysis in the follow-up study of a larger stellar sample.

Our results for the four 3D models are illustrated in Fig. \ref{fig:prof3dnlte}. We also overplot the predictions of 1D LTE modelling computed using the same Mn model atom and the same Mn abundance. Similar to the solar model, the 3D NLTE line profiles are much weaker than the 1D LTE counterparts. This is the consequence of over-ionisation, which is stronger in metal-poor conditions due to reduced line blanketing and stronger UV radiation fluxes. The topology of convection in FGK stars is set by mass conservation. The hotter rising granules occupy the dominant ($\sim2/3$) part of the stellar surface, whereas the cooler downdrafts are confined to narrow inter-granular lanes \citep{stein1989,nordlund2009}, although the area  subtended by the down-flows increases with depth. Hence, over-ionisation in the granules is the dominant NLTE effect that defines the appearance of the spatially-integrated line profiles. In contrast to the Sun, where the effects of convective shifts were marginal, the integrated flux profiles in the metal-poor models show strong asymmetries, with typical blue-shifted cores and skewed wings. This is best seen by comparing the symmetric 1D profiles, convolved with an arbitrary macro-turbulence and the 3D profiles, which do not assume any ad-hoc broadening. Also the line profiles from individual snapshots are remarkably similar to their temporarily-averaged counterparts, although a slightly more pronounced temporal variation is seen in the NLTE line profiles computed using the 3D models of dwarfs (Fig. \ref{fig:prof3dnlte}). 

The differences between 1D LTE and 3D NLTE profiles are very large, with only a few notable exceptions. These exceptions are the blue lines of \mnii, such as the 3488 \AA\ and 3497 \AA\ lines. For dwarfs, these low-excitation \mnii~features are nearly invariant to the changes in the model atmosphere structure, and they indeed show very similar line shapes and equivalent widths, regardless of whether one adopts LTE or NLTE radiative transfer. Also a standard choice of the micro-turbulence $\Vmic =1$ km/s appears to be satisfactory for a dwarf model, producing a line profile, which closely resembles the full 3D NLTE profile of the \mnii~3488 \AA\ line even in the model with $\feh = -2$. For the computational complexity of 3D NLTE modelling, we are unable to explore a larger parameter space of 3D models in this work, but the behaviour of the models with metallicity suggests that this conclusion may hold also for other stars on the main-sequence, as long as $\feh$ is not in the extremely metal-poor domain.

Yet, this fortunate result does not hold for the atmospheres of giants, neither does it generally hold for the \mni~lines. The 3488 \AA\ line of \mnii~ suffers from non-negligible deviations from NLTE in the RGB models, being significantly stronger in 3D NLTE. On the  other hand, the 3D NLTE abundances of all lines of \mni~are significantly higher compared to 1D LTE. This is best illustrated by exploring the differences between 1D LTE and 3D NLTE, respectively, 1D NLTE and 3D NLTE abundances, derived from the line of a given equivalent width. This quantity is typically referred to as a 3D NLTE abundance correction (3D NLTE - 1D LTE). 

\begin{figure}
\center
\includegraphics[width=0.4\textwidth, angle=0]{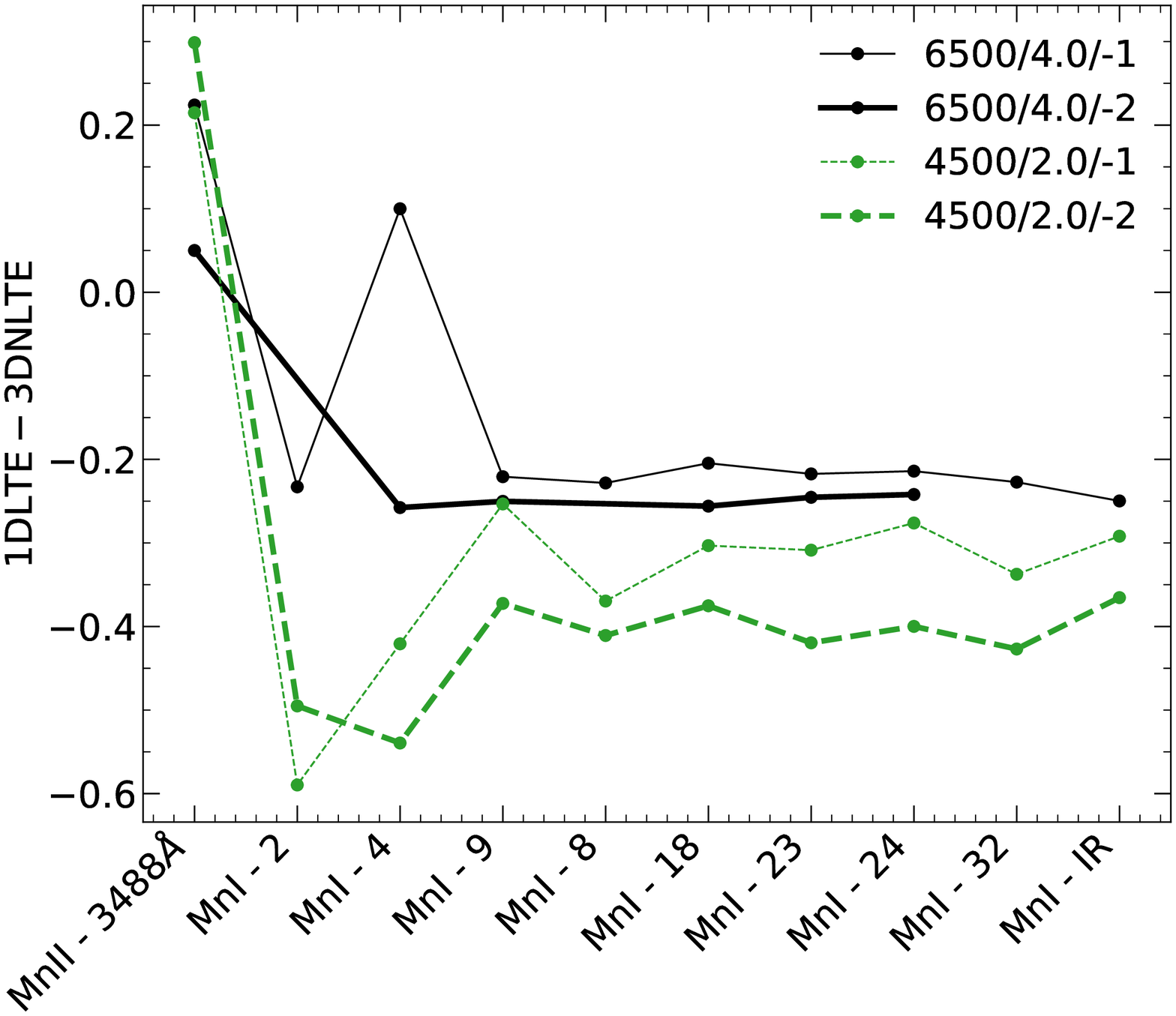}
\includegraphics[width=0.4\textwidth, angle=0]{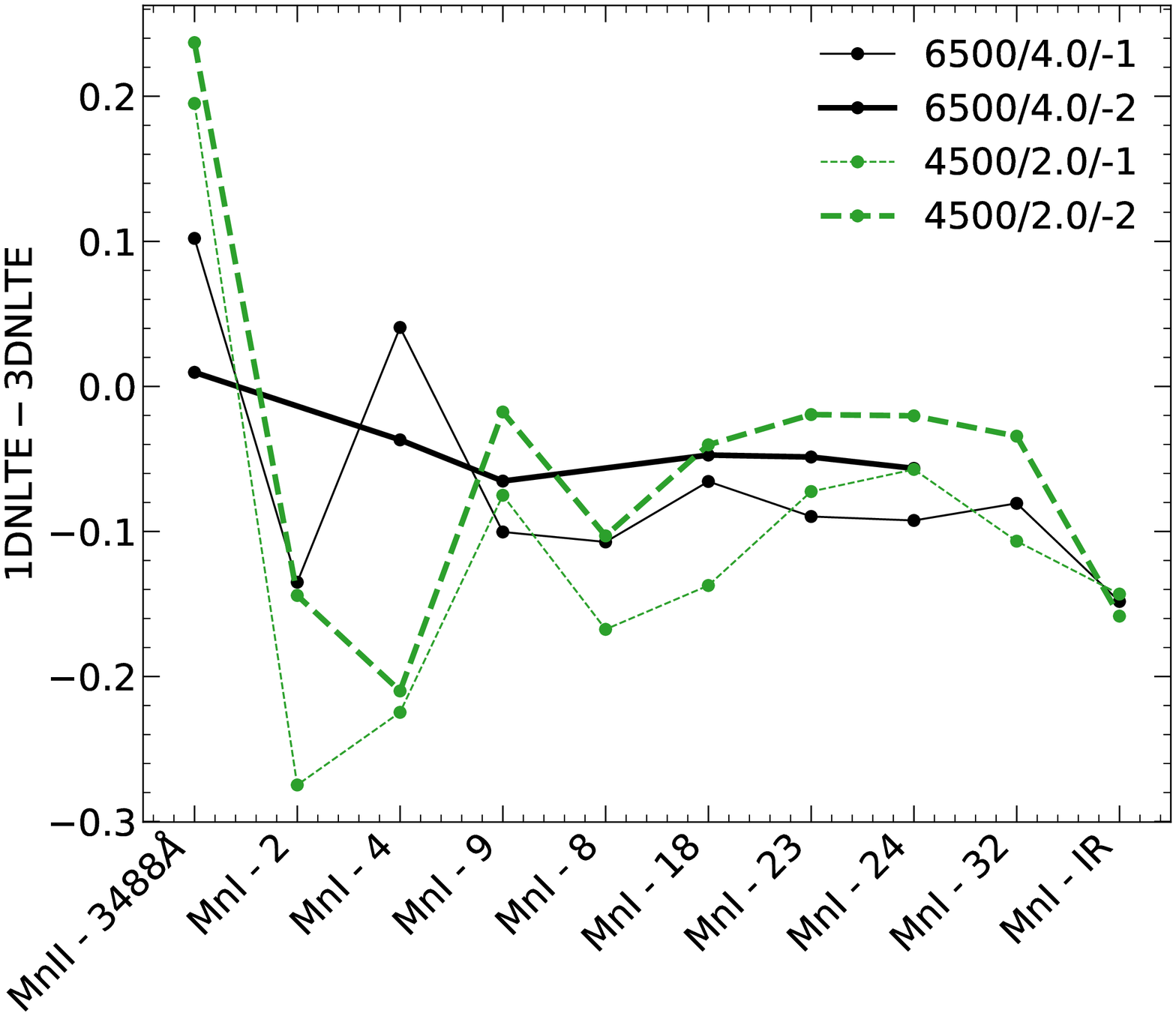}
\caption{3D NLTE abundance corrections for selected models.}
\label{fig:nlte3dcor}
\end{figure}

Figure \ref{fig:nlte3dcor} (top panel) shows that the 3D NLTE abundance corrections amount to $0.25$ dex for most \mni~lines in the atmospheres of dwarfs, but in the models of RGB stars the (3D NLTE - 1D LTE) abundance differences tend to be significantly larger. In all cases, 1D analysis tends to substantially under-estimate the Mn abundance. In particular, for the lines of multiples 18 and 32 (e.g. 4823 and 6016 \AA) we find the abundance corrections of $0.3$ to $0.4$ dex in metal-poor models, whereas the blue resonance lines at 4030-4034 \AA\ show even larger 3D NLTE corrections of $\sim 0.5$ dex at $\feh = -2$. This is a very interesting result. On the one hand, our results  suggest that the large abundance discrepancies and excitation imbalance seen in the analysis of RGB stars \citep{Bonifacio2009} may indeed be accounted for by 3D NLTE. On the other hand, we also find that 3D NLTE effects do not significantly impact the excitation balance in the atmospheres of very metal-poor dwarfs, as 3D NLTE corrections are similar for the lines of all \mni~multiplets. This is why this effect may go un-detected in 1D LTE studies of stars with similar parameters (e.g the study of HD 84937 by \citealt{Sneden2016}, or \citealt{Mishenina2015}). Also the 3D NLTE corrections for the \mnii~lines are very small in the metal-poor atmospheres of dwarfs, and do not exceed $-0.05$ dex at [Fe/H] $= -2$.

Figure \ref{fig:nlte3dcor} also suggests that, whereas 1D LTE always under-estimates abundances for the \mni~lines, there are a few \mni~lines that can be reliably modelled with 1D NLTE. These are the lines of multiplets $9$ (e.g. 4055, 4070, 4082 \AA), $23$ (4761, 4762, 4765, 4766 \AA), $24$ (e.g. 4436, 4451, 4498, 4502 \AA), and 32 (6013, 6016, 6021 \AA). For these features, the differences between 1D NLTE and 3D NLTE are marginal, and do not exceed $0.1$ dex in the lowest-metallicity ([Fe/H] =$-2$) models of both dwarf and RGB stars. The lines of other multiplets show a very significant systematic deviations from 1D NLTE, caused by the impact of convective inhomogeneities.  Also for dwarfs, the impact of convection is modest and indeed, most optical \mni~lines, including the resonance \mni~triplet at 403 nm, would be suitable for abundance diagnostic, leading to a modest bias of $\sim_{-0.15}^{+0.05}$ dex. The exact value would also depend on the abundance of Mn itself, but also on the assumed value of micro-turbulence in the 1D models, so our estimates of 3D NLTE abundance corrections may not be directly applicable to the 1D LTE measurements in the observed stars. Rather we recommend to employ those lines, which according to out tests, show minimal impact of convection and/or NLTE.

The IR lines of \mni, e.g. those that are used in the H-band abundance analysis \citep{Shetrone2015}, show significant departures from 1D and LTE. However, it is remarkable that all three Mn lines that we tested in this work are extremely consistent what concerns the 1D NLTE-3D NLTE differences in dwarf and RGB models. In effect, the abundances for all these models are systematically under-estimated by $-0.15$ dex with respect to 1D NLTE, and the bias appears to be insensitive to metallicity of the model. It, hence, appears reasonable to apply a positive correction to all H-band Mn 1D NLTE abundance measurements to obtain physically unbiased results.
\subsection{3D LTE versus 3D NLTE}
Figure \ref{fig:prof3dlte} compares the line profiles computed using 3D LTE, 3D NLTE, and 1D LTE, the latter convolved with a macro-turbulence of 4 km$/$s. 

For the [Fe/H] $=-1$ dwarf model, the 3D LTE line profile of the \mnii~line at 3488 \AA\ closely resembles the equivalent 1D LTE profile. The \mni~lines at 4766 and 6021 \AA\ are slightly shallower in 3D LTE compared to 1D LTE, but deeper for the 4823 \AA~line. On other hand, all lines of \mni~are much fainter in 3D NLTE compared to 3D LTE, suggesting that a 3D LTE analysis would strongly under-estimate the Mn abundances. 

This tendency is qualitatively very similar in the dwarf model at [Fe/H] $=-2$. One notable exception is that the resonance line at 4034 \AA\ is now exceptionally strong in 3D LTE, much stronger than the profiles computed using 1D LTE and 3D NLTE, the behaviour that is known also for other chemical species. The major implication of this is that 3D LTE modelling is unlikely to solve the excitation imbalance in Mn, which is well-known from 1D LTE studies \citep[e.g.][]{Bonifacio2009,Sneden2016}. The unfortunate problem is that even in 1D LTE the resonance \mni~lines yield significantly lower abundances, relative to their higher excitation counterparts. The 3D LTE line profiles of the resonance lines are stronger than their 1D LTE counterparts computed assuming the same Mn abundance, meaning that the 3D LTE abundances of these lines must be reduced to recover the correct equivalent width. For the higher-excitation lines, the differences between 1D and 3D LTE at [Fe/H] $=-2$ are not large, because the lines are very weak, and form deep in the atmosphere, at $\opd > -2$, where the temperature inhomogeneities are not pronounced and the average structure of the 1D hydrostatic models closely resembles that of 3D models (see, e.g. Fig.6).

The behaviour is very similar in the 3D models of giants. 3D LTE line profiles are not too different from 1D LTE at [Fe/H] $=-1$, modulo the modest effect of convection on the line shapes, but they are always stronger compared to 3D NLTE. The differences are exacerbated at low metallicity, [Fe/H]$=-2$, where 3D LTE calculations greately over-estimate the strength of all \mni~lines compared to 3D NLTE. The UV \mnii~line at 3488 \AA\ shows very small NLTE effects in the metal-poor RGB model, in fact, its 3D LTE and 3D NLTE profiles are very similar. On the other hand, the 1D LTE assumption under-estimates the strength of the \mnii~line, hence, over-estimating Mn abundance derived from this feature.
\subsection{Benchmark metal-poor stars}\label{bmk3d}
\begin{figure*}
\center
\hbox{\includegraphics[width=0.5\textwidth, angle=0]{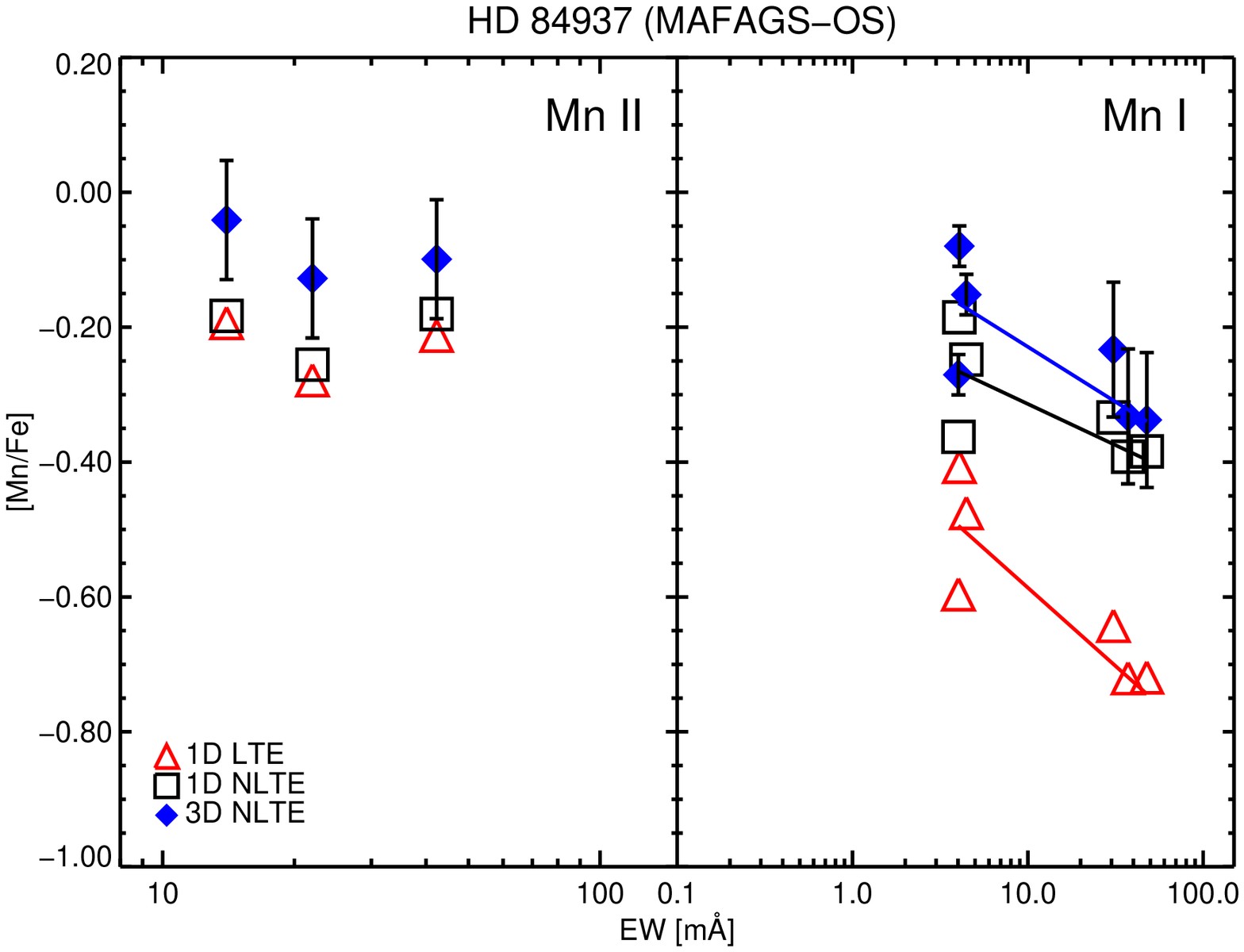}
\includegraphics[width=0.5\textwidth, angle=0]{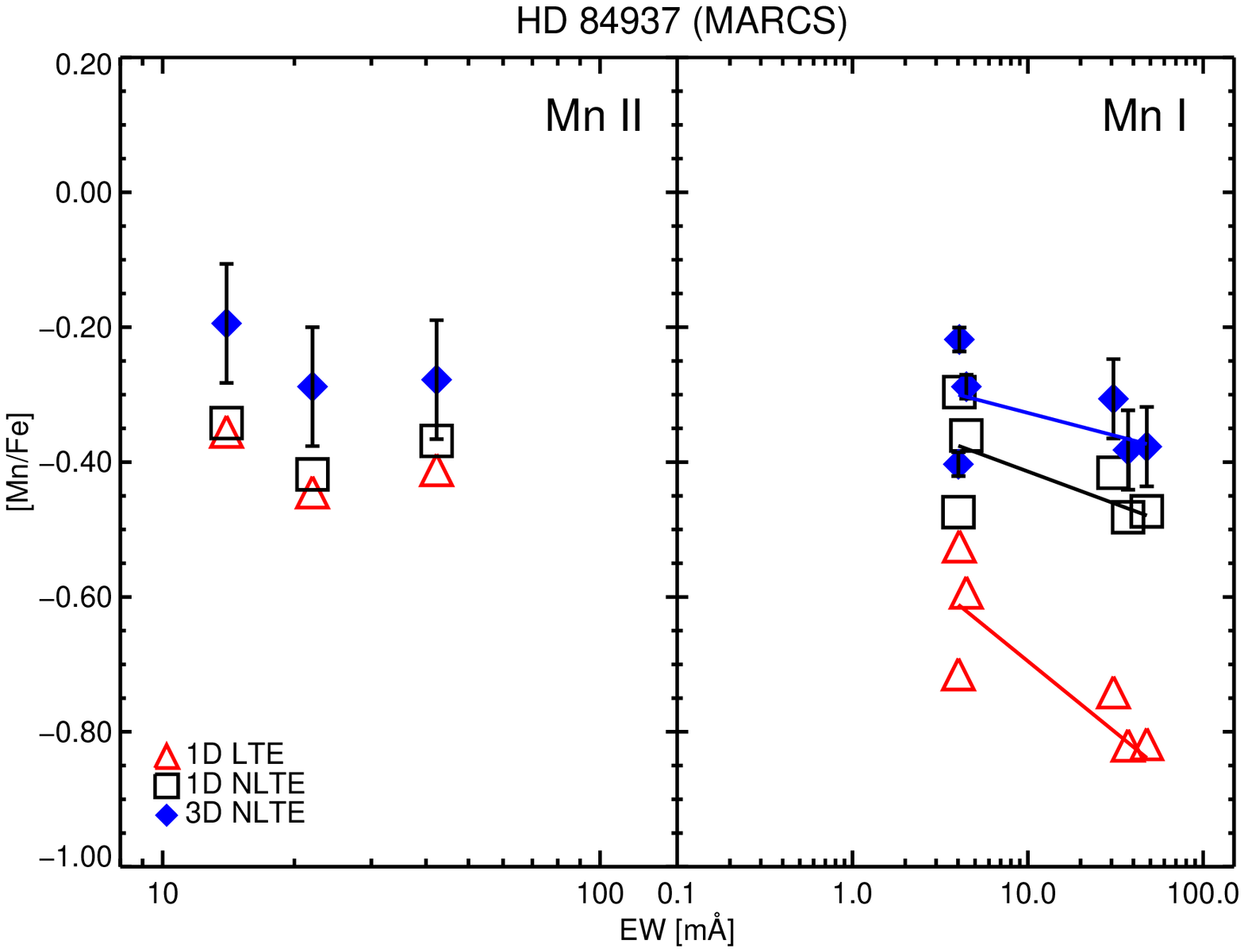}
}
\hbox{
\includegraphics[width=0.5\textwidth, angle=0]{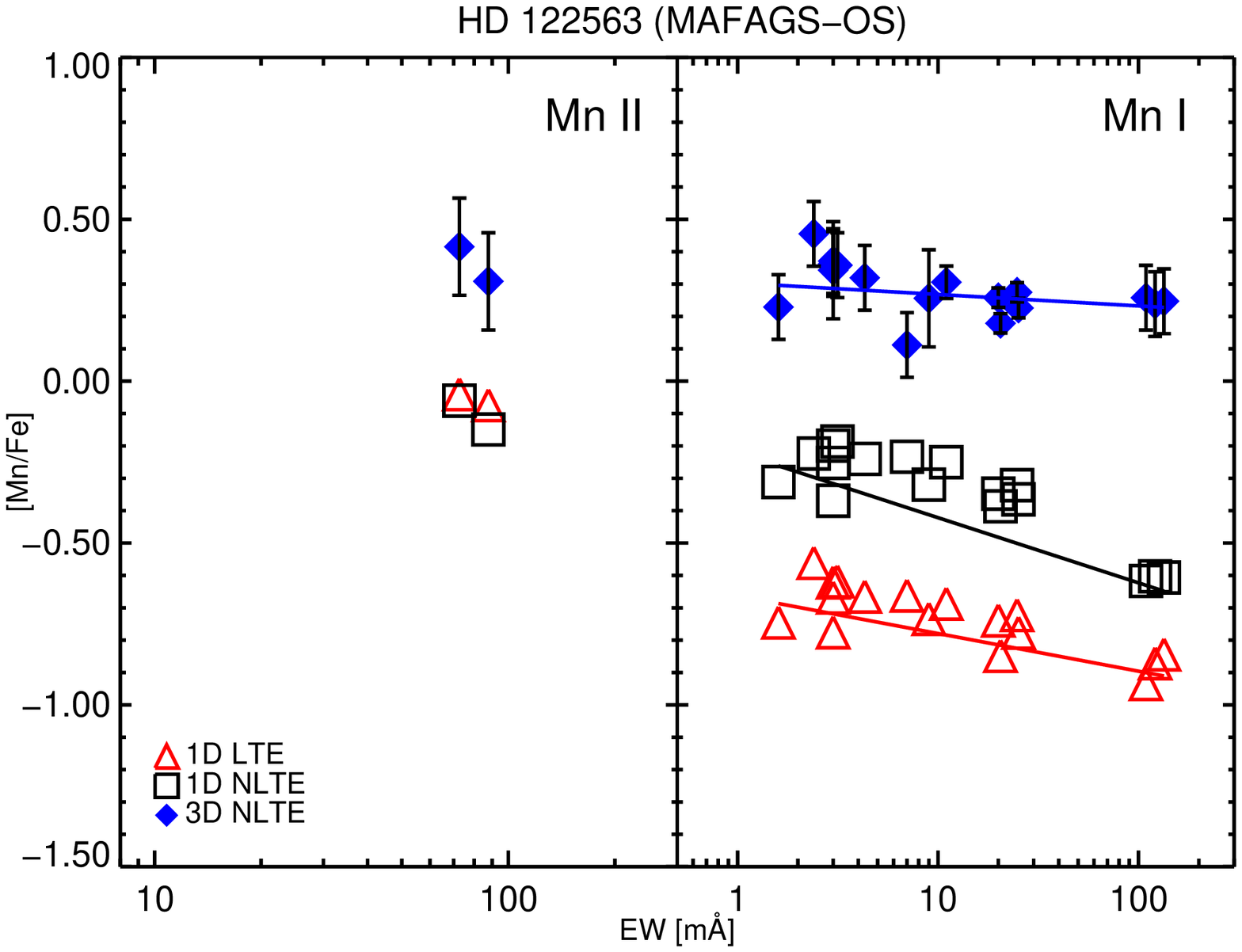}
\includegraphics[width=0.5\textwidth, angle=0]{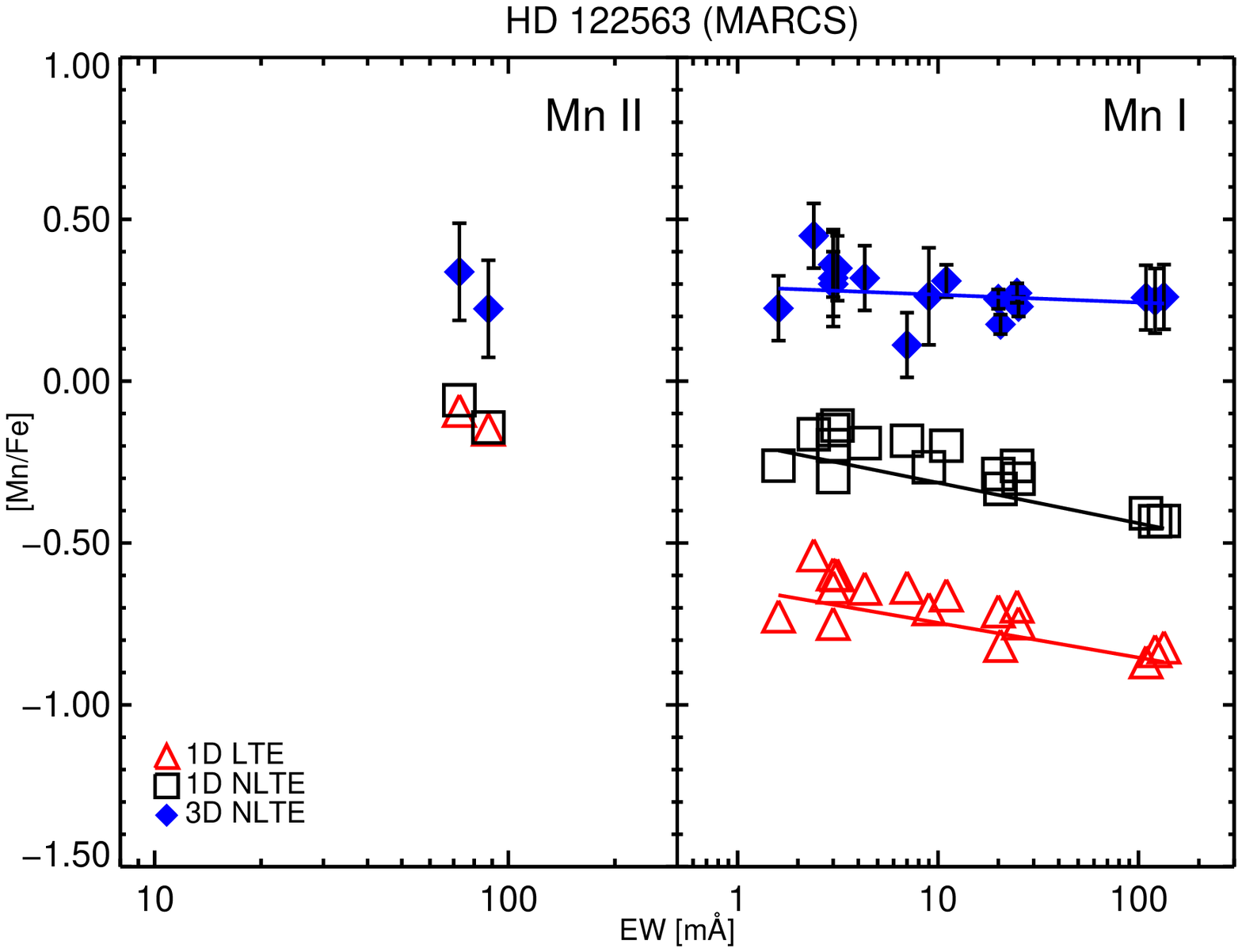}
}
\hbox{
\includegraphics[width=0.5\textwidth, angle=0]{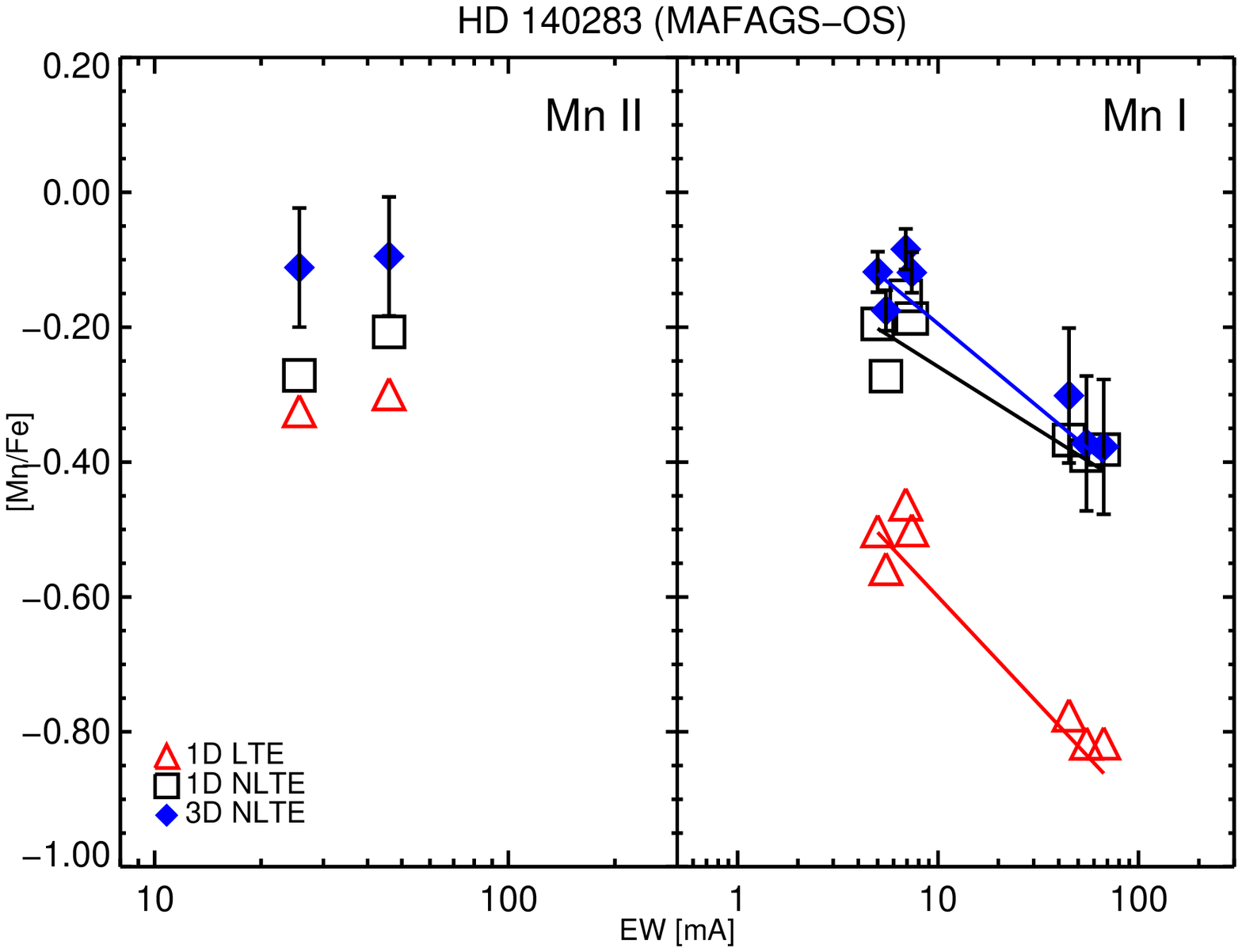}
\includegraphics[width=0.5\textwidth, angle=0]{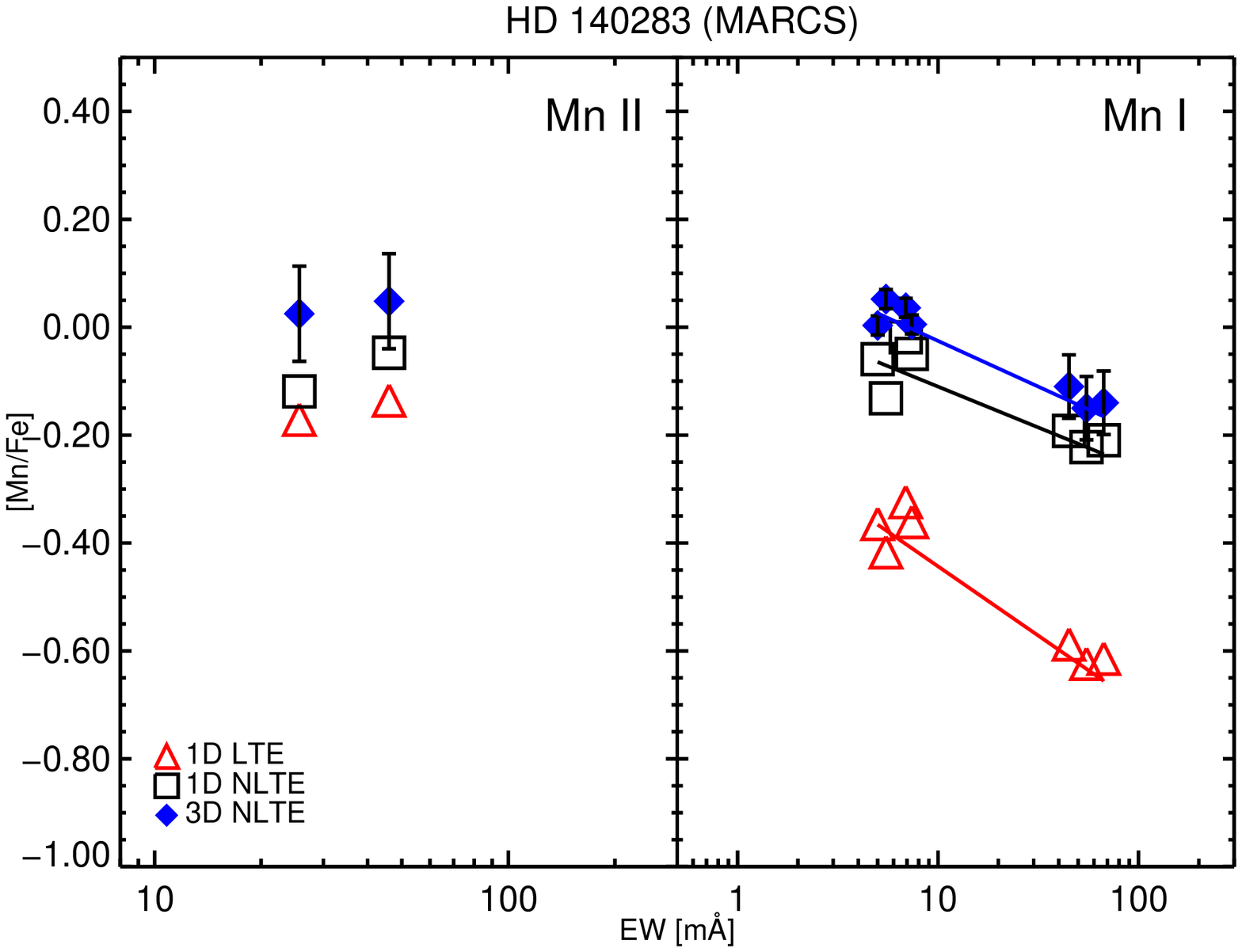}
}
\caption{Mn abundances in three metal-poor stars. Left panels: the abundances determined using MAFAGS-OS and 3D models. Right panels: the abundances determined using the MARCS and 3D models.}
\label{fig:abundstars}
\end{figure*}

The Mn abundances in the stellar spectra were computed by matching the observed equivalent widths to the grids of 1D LTE and 1D NLTE spectral lines computed using MULTI2.3. The equivalent widths were measured from the UVES-POP spectra of these stars using the SIU code. Unlike in our analysis of the Sun, we have chosen to not employ SIU for abundance measurements in metal-poor stars, as blending is not a problem anymore and a detailed spectrum synthesis is unnecessary. Also MULTI2.3 has an advantage in that it includes background scattering, which is essential for the blue and UV lines. 3D NLTE corrections were computed separately and applied to the 1D LTE abundances derived using the measured EWs.

The results for all three benchmark metal-poor stars are shown in Fig. \ref{fig:abundstars}. The error bars correspond to the uncertainties of the EW measurements. We show the abundances derived using the MAFAGS-OS models (left panels) and the MARCS models (right panels) in order to illustrate the impact of the 1D model atmospheres. Overall, the results obtained MAFAGS-OS and MARCS models are in agreement, especially for the model of the metal-poor RGB star HD 122563. The differences can generally be explained by the small differences in the input parameters of the models. In particular, the MAFAGS-OS models assume the following parameters: (a) HD 122563: $\teff = 4600$ K, $\log g = 1.60$ dex, [Fe/H] $= -2.5$ dex, (b) HD 140283: $\teff = 5773$ K, $\log g = 3.66$ dex, [Fe/H] $= -2.38$ dex, and (c) HD 84937: $\teff = 6350$ K, $\log g = 4.09$ dex, [Fe/H] $= -2.15$ dex, which can be compared to Table \ref{3dmodels}. This is most likely at the origin of the somewhat lower abundances that we get for HD 84937. It is also noteworthy that MARCS models lead to slightly more consistent abundances from \mni~lines for all three reference stars.

Regardless of the choice of 1D models, 1D LTE modelling reveals a significant ionisation imbalance, in that the \mni~lines give systematically lower abundances compared to \mnii~lines, confirming the results reported in the literature \citep{Johnson2002}. The difference is most pronounced for the RGB star HD 122563, for which the offset is close to $0.7$ dex. We also confirm the \mni~excitation imbalance in all three stars, with strong resonance lines of \mni~producing systematically lower abundances compared to higher excitation lines. For HD 122563, the offset between the lines of multiplet 4 (the 4030-4034 \AA\ triplet) and the lines of other multiplets is $\sim 0.2$ dex, whereas for the subgiant HD 140283 and turn-off star HD 84937 the offset is slightly larger, of the order $\sim 0.3$ dex.

The ionisation balance is significantly improved for all three stars in 1D NLTE. In particular, for HD 84937 and HD 140283, the abundances derived from the high-excitation lines are now consistent with the abundances derived from the \mnii~lines. For HD 122563, the ionisation balance in 1D NLTE is only partially improved, but there is still a differential effect of $-0.2$ dex for the high-excitation lines and $-0.5$ dex for the low-excitation lines.

The improvement is most striking in 3D NLTE. The 3D NLTE abundance corrections are very large and positive for both ionisation stages. For HD 122563, this brings \mni~and \mnii~lines into agreement. For HD 84937 and HD 140283, the 3D NLTE abundances are higher compared to 1D NLTE, however, the differential effect is not as large. It should be emphasised that stellar parameters of these three stars are well-constrained by independent methods, and the fact that ionisation balance is satisfied in 3D NLTE, without resorting to ad-hoc parameters, like micro-turbulence in 1D, is remarkable. For comparison, to explain the ionisation imbalance in HD122563 by the error in stellar parameters, its $\teff$ would have to be increased to $4900$ K (that would bring the resonance lines up by $\sim +0.7$ dex to be consistent with \mnii), or alternatively the $\Vmic$ increased to 3.5 km$/$s dex that would push the \mnii~abundances down by $\sim -0.7$ dex. Decreasing the $\log g$ might help to increase the abundance from the resonance lines, but a change to $\log g < 0.5$ would be necessary. All these changes in stellar parameters can be ruled out, given the robust constraints from asteroseismology and interferometry \citep{Creevey2012, Karovicova2018, Creevey2019}. Also micro-turbulence, despite being a free parameter, is relatively well-constrained in the literature \citep{Bergemann2012,Afsar2016}.

HD 84937 was recently analysed employing 1D LTE models by \citet{Sneden2016}. They adopted somewhat different stellar parameters for this metal-poor turn-off star: $\teff = 6300$ K, $\log g = 4.0$, $\feh = -2.15$, and $\Vmic = 1.5$ km/s. Their estimates of [Mn/Fe] for this star are $\sim -0.25$ for the Mn II lines and $\sim -0.3$ for the Mn I lines (their Fig. 7). On the other hand, they also tabulate the abundances derived by taking the ratios of the elements derived from the lines in the same ionisation stage (e.g. Mn I to Fe I and Mn II to Fe II). In the latter case, their values are $-0.42$ for the Mn II lines and $-0.46$ for the Mn I lines (their Table 5).
    
 Our 1D LTE estimates for HD 84937 are [Mn/Fe] $= -0.24$ dex, as derived from the \mnii~ lines, and [Mn/Fe] $= -0.47$ dex for the high-excitation \mni~lines. Similar to Sneden et al., we find that the \mni~triplet lines give the abundance that are $\sim 0.2$ dex lower compared to the high-excitation lines. Our abundances derived from the Mn I lines may appear to be lower, but this is likely the consequence of the adopted metallicity. Indeed, we adopt $\feh = -2.0$ dex in this work, consistent with our $<\rm{3D}>$ NLTE estimate in \citet{Bergemann2012}($\feh = -2.04$ to $-1.95$ dex, depending on the choice of the H scaling factor), which was recently confirmed from the full 3D NLTE Fe analysis by \citet{Amarsi2016}, who find [Fe/H] $= -1.96 \pm 0.02\rm{(stat)} \pm 0.04 \rm{(sys)}$ dex. On the other hand, \citet{Sneden2016} adopt a very low metallicity, $\feh = -2.15$ dex, that leads to significantly higher Mn abundances. It is not clear which of the estimates from \citet{Sneden2016} are to be given preference and which of their both methodological approaches is more consistent with our estimate. Different from that study, we do not consider lines with EWs $\lesssim 3$ m\AA\ in the UVES spectrum of HD 84937 as reliable, such as the lines at 4762,4765, 4766 \AA, and the red triplet at 6013-6021 \AA. This, combined with the choice of the atmospheric models, stellar parameters, and the methods to determine the abundances, may account for the somewhat different results in our study and in \citet{Sneden2016}. 
\section{Conclusions}\label{sec:conclusions}
We present the first 3D NLTE analysis of Mn line formation in inhomogeneous model atmospheres. We employ 3 different statistical equilibrium codes, two of them compute radiative transfer in 1D geometry (MULTI2.3 and DETAIL), whereas MULTI3D is used for detailed 3D NLTE calculations.

The NLTE model atom is assembled using the new atomic data for different processes. We use the R-matrix method to compute new photo-ionisation cross-sections for 84 terms of \mni, and employ them in place of hydrogenic cross-sections. We also compute new collision rates for 71 terms of Mn I interacting with H and the for the first excited state of \mnii~interacting with H. The latter are supplemented with the data for 19 \mni~levels and the ground state of \mnii~presented by \citet{Belyaev2017}. We also implement the collision rates computed using the scattering-length approximation according to \citet{Kaulakys1985, Barklem2016}. The new photo-ionisation cross-sections and the new rates of inelastic collisions represent the main difference with respect to our earlier results.

We confirm the earlier studies \citep{bergemann2007,Bergemann2008} that the NLTE effects in \mni~are driven by over-ionisation. The qualitative behaviour of the departure coefficients and the NLTE abundance corrections is very similar. All \mni~lines in the optical suffer from strong NLTE effects and display positive NLTE abundance corrections, which increase with decreasing metallicity of the model atmosphere. LTE modelling under-estimates Mn abundances, by $\sim -0.5$ dex in the models of metal-poor red giants, and to a lesser degree in the models of dwarfs. The NLTE abundance corrections are sensitive to the implementation of collision rates that affects the results at the level of $\sim 0.1$ dex. Different from \citet{Bergemann2008}, we find that the new model does not produce large NLTE effects for the resonance lines of \mni. We attribute this to the use of a tailored ad-hoc scaling factor to collisions in the previous study, that, however produced qualitatively similar results to our 3D NLTE calculations.

Our 3D modelling of solar Mn lines reveals unique features of line formation in the convective models, known from the previous studies \citep[e.g.][]{Dravins1981, Dravins1987, Dravins1990a, Dravins1990b, Nordlund1990}. The spatially-resolved Mn lines show pronounced asymmetries. For the weaker lines, we find a strong anti-correlation between the line core depth and the line shift. The strong lines show a very broad distribution of line shifts, associated with granular motions. The bluer components are rather symmetric, but red components of the spatially-resolved features are very asymmetric and broad, tracing the differences in the line formation in the granules and in the inter-granular lanes.

We perform 1D LTE, 3D LTE, 1D NLTE, and 3D NLTE calculations for a large set of Mn lines, including the \mnii~UV lines, but also the commonly-used optical lines, and the IR \mni~lines in the H-band. We find large differences between the four scenarios, which can be broadly summarised as follows:
\begin{itemize}
	\item All lines of \mni~are significantly weaker in 3D NLTE, compared to 1D LTE. As a consequence, 1D LTE under-estimates the Mn abundances by $\sim -0.6$ dex for the blue  resonance \mni~lines, but $-0.4$ dex for the optical lines in the RGB models with $\feh = -2$. The effect is smaller for the models of dwarfs: the systematic bias incurred by 1D LTE is $\sim -0.2$ dex and does not change substantially with $\feh$.
	\item The commonly-used blue lines of \mnii~ at 3488 and 3497 \AA\ are typically too weak in 1D LTE and 1D NLTE calculations, compared to 3D NLTE modelling. This effect is caused by the significant line scattering, and is greately amplified in the 3D inhomogeneous model atmospheres. Hence, 1D LTE and 1D NLTE typically over-estimate Mn abundance derived from the \mnii~3488 and 3497 \AA\ lines. 3D NLTE effects depend strongly on the atmospheric parameters of a star, and, in particular in metal-poor dwarf models with $\feh =-2.0$, the 3D NLTE results are close to 1D LTE.
	\item The impact of convection is modest for the high-excitation \mni~lines with lower excitation potential $> 2$ eV. In particular, the least affected lines are those belonging to multiplets 9, 23, 24, and 32. We recommend to use these lines for the abundance analysis in 1D NLTE.
	\item 3D LTE modelling substantially over-estimates the line strength of resonance \mni~lines, compared to 1D LTE and 3D NLTE. As a consequence, the excitation imbalance reported for \mni~lines across a broad metallicity range \citep{Bonifacio2009} will not be cured, but rather amplified using 3D LTE. 
	\item All IR H-band lines of \mni~suffer from strong systematic bias. 1D LTE under-estimates the abundance derived from these lines by $0.2$ to $0.35$ dex. On the other hand, the difference between 1D NLTE and 3D NLTE is only $-0.15$ dex, and this value does not depend on $\feh$, $\teff$, or $\logg$ of the model atmosphere. We suggest to employ this correction in 1D NLTE studies, in order to account for 3D effects.
\end{itemize}

We derive new 3D NLTE solar abundance of Mn, $5.52 \pm 0.03$ dex, which is consistent with the CI meteoritic abundance, $5.50 \pm 0.03$ \citep{Lodders2003}. The 1D LTE and 1D NLTE abundances are lower, $5.34 \pm 0.04$ dex in LTE, $5.41 \pm 0.05$ dex in NLTE. Our 3D NLTE value is slightly higher compared to \citet{Asplund2009} and \citet{Scott2015}, $5.42 \pm 0.04$ dex,  however, they did not compute full 3D NLTE, but applied the NLTE corrections computed using a 1D model. In particular, in the latter study, we employed a Mn atom with Drawin's collision rates and hydrogenic photo-ionisation, in contrast to the detailed quantum-mechanical data in this work.

Our results for the metal-poor benchmark stars offer a considerably different picture on Mn abundances at low metallicity, contrasting with earlier 1D LTE studies. We find that 3D NLTE abundances in HD 84937, HD 140283, and HD 122563 are significantly higher. For HD 122563, we obtain a perfect ionisation balance in 3D NLTE, that cannot be otherwise explained by the atomic data uncertainties and stellar parameters. The 3D NLTE Mn abundances derived from the \mni~lines are $\sim 0.5 - 0.7$ (HD 84937, HD 140283) to 1 dex (HD 122563) higher compared to 1D LTE results. Also the 3D NLTE abundance derived from the \mnii~lines are $\sim 0.15$ (HD 84937, HD 140283) to $0.4$ dex (HD 122563) larger compared to 1D LTE. Effects of this magnitude are also expected for other stars in this metallicity regime. We strongly recommend applying NLTE, and, if possible, 3D NLTE radiative transfer in the analysis of Mn lines. Alternatively, the \mni~lines of certain multiplets (see above) can be used as a relatively reliable (with a bias of $\sim 0.1$ dex) diagnostics with hydrostatic models.

In the next paper in the series (Eitner et al. in prep), we will apply the models developed in this work to a large sample of stars to explore the chemical evolution and nucleosynthesis of Mn in the Galaxy.
\bibliographystyle{aa}
\bibliography{references}
\begin{acknowledgements}
All calculations are run on MPCDF clusters Draco and Cobra. We thank Anish Amarsi for kindly providing us the H collision rates computed using the Kaulakys recipe. This study is supported by SFB 881 of the DFG (subprojects A05, A10) and by the Research Council of Norway through its Centres of Excellence scheme, project number 262622. SAY and AKB gratefully acknowledge support from the Ministry for Education and Science (Russian Federation), projects No. 3.5042.2017/6.7, 3.1738.2017/4.6. JL has received support through a grant from the Knut och och Alice Wallenberg foundation (2016.0019). Funding for the Stellar Astrophysics Centre is provided by The Danish National Research Foundation (Grant agreement no.: DNRF106). BP is partially supported by the CNES, Centre National d'Etudes Spatiales.
\end{acknowledgements}
\begin{appendix}

\section{Figures}
%
%
%
%
%
%
%
\begin{figure*}[!ht]
\hbox{
\includegraphics[width=0.5\textwidth, angle=0]{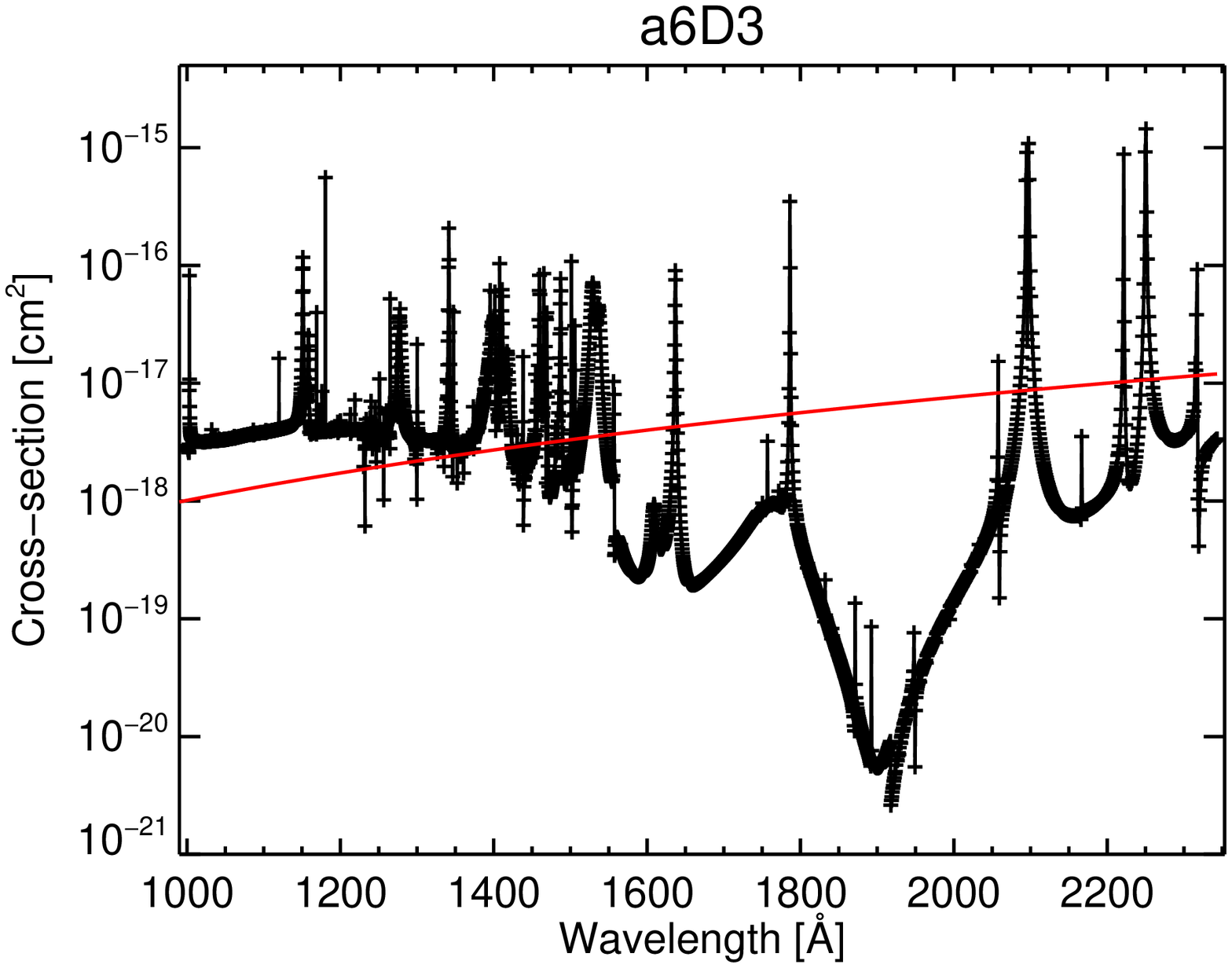} 
\includegraphics[width=0.5\textwidth, angle=0]{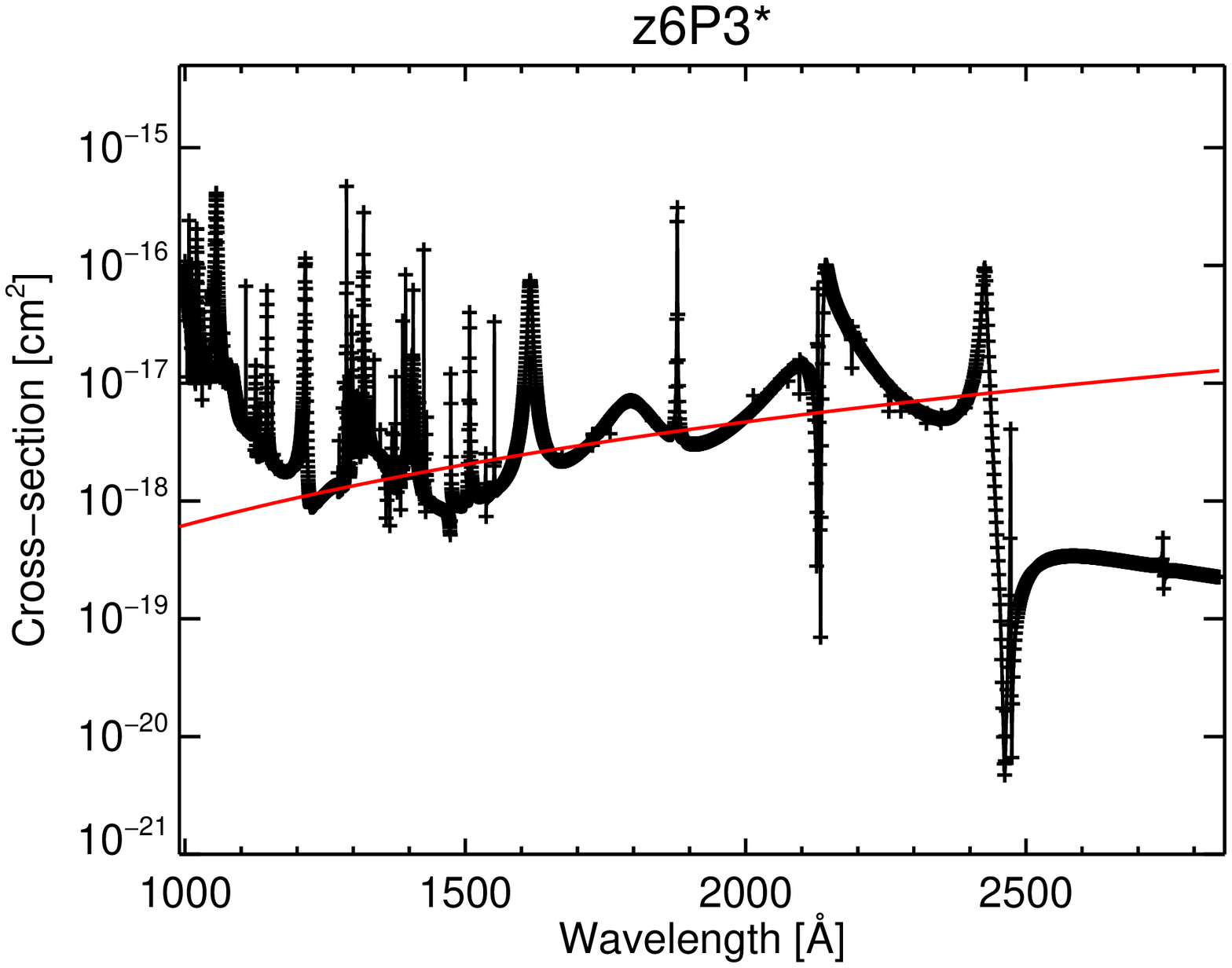}}
\hbox{
\includegraphics[width=0.5\textwidth, angle=0]{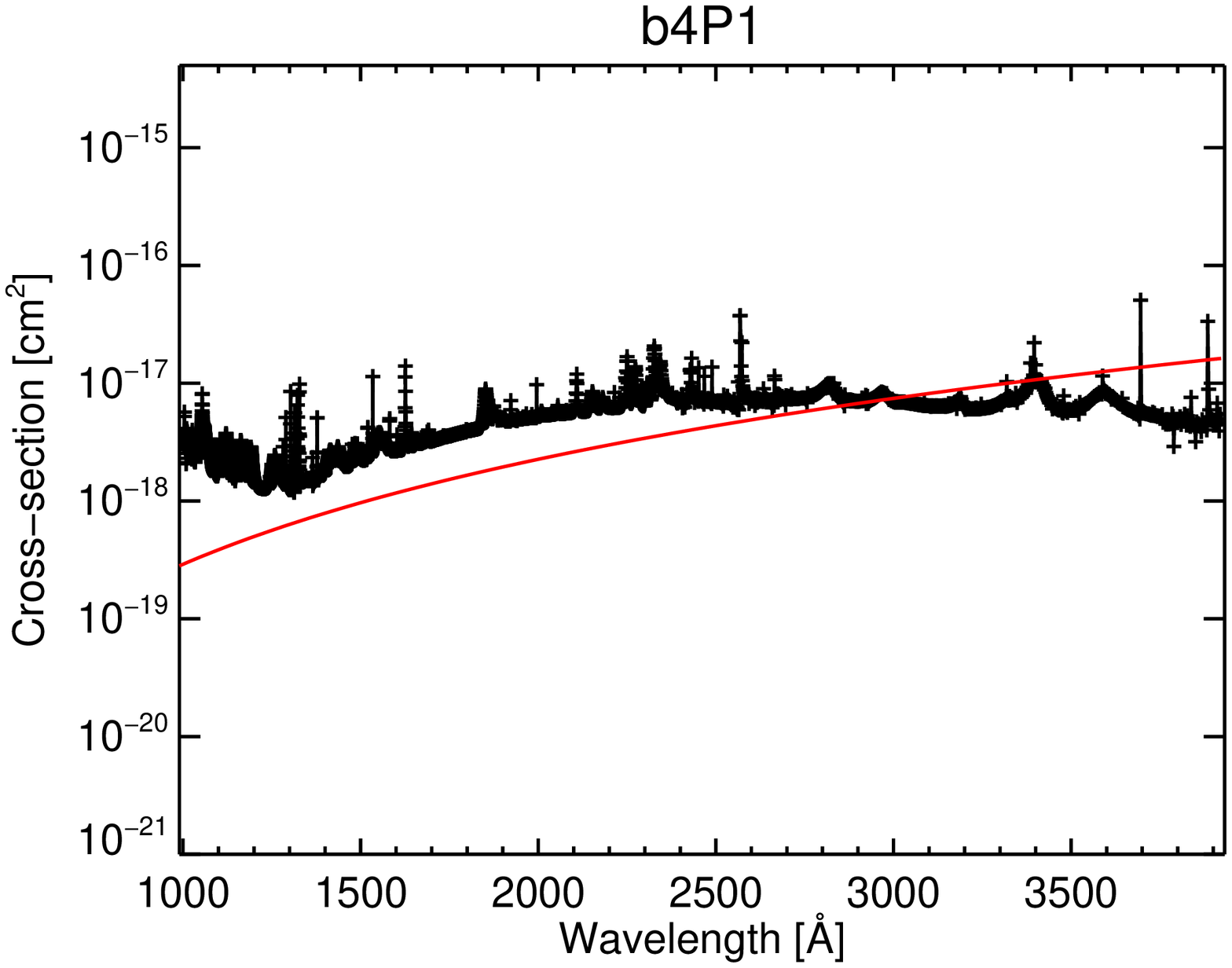} 
\includegraphics[width=0.5\textwidth, angle=0]{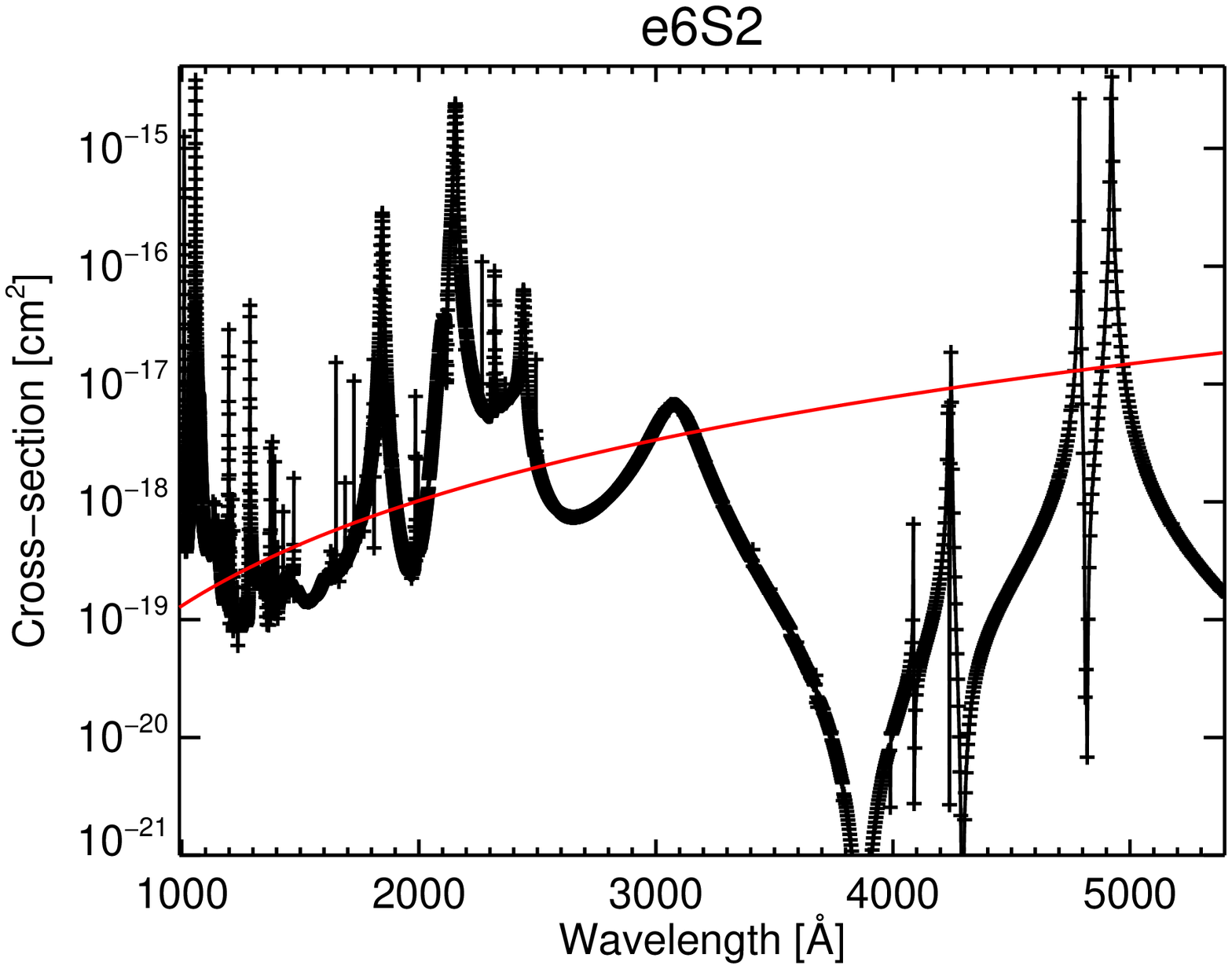}}
\hbox{
\includegraphics[width=0.5\textwidth, angle=0]{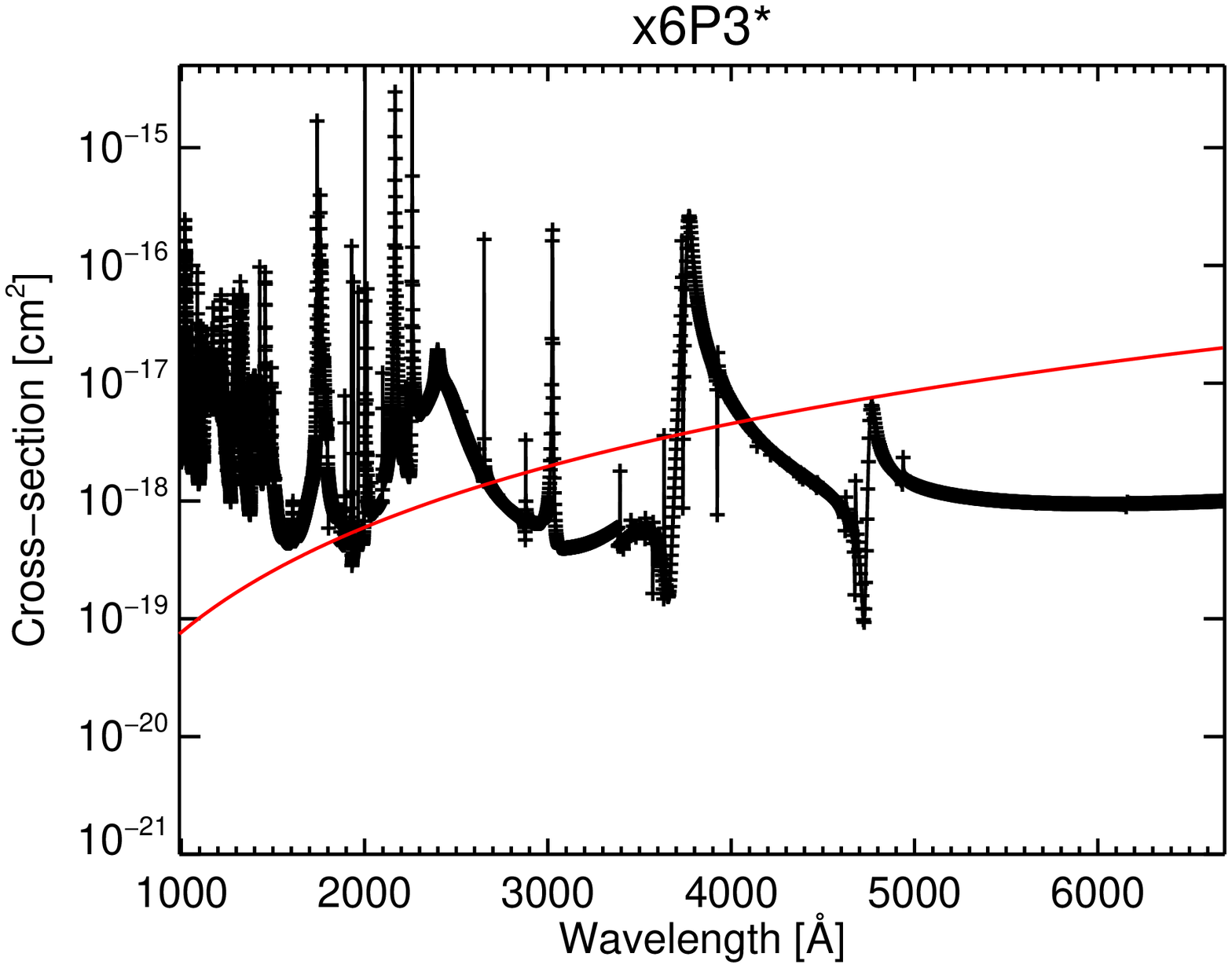} 
\includegraphics[width=0.5\textwidth, angle=0]{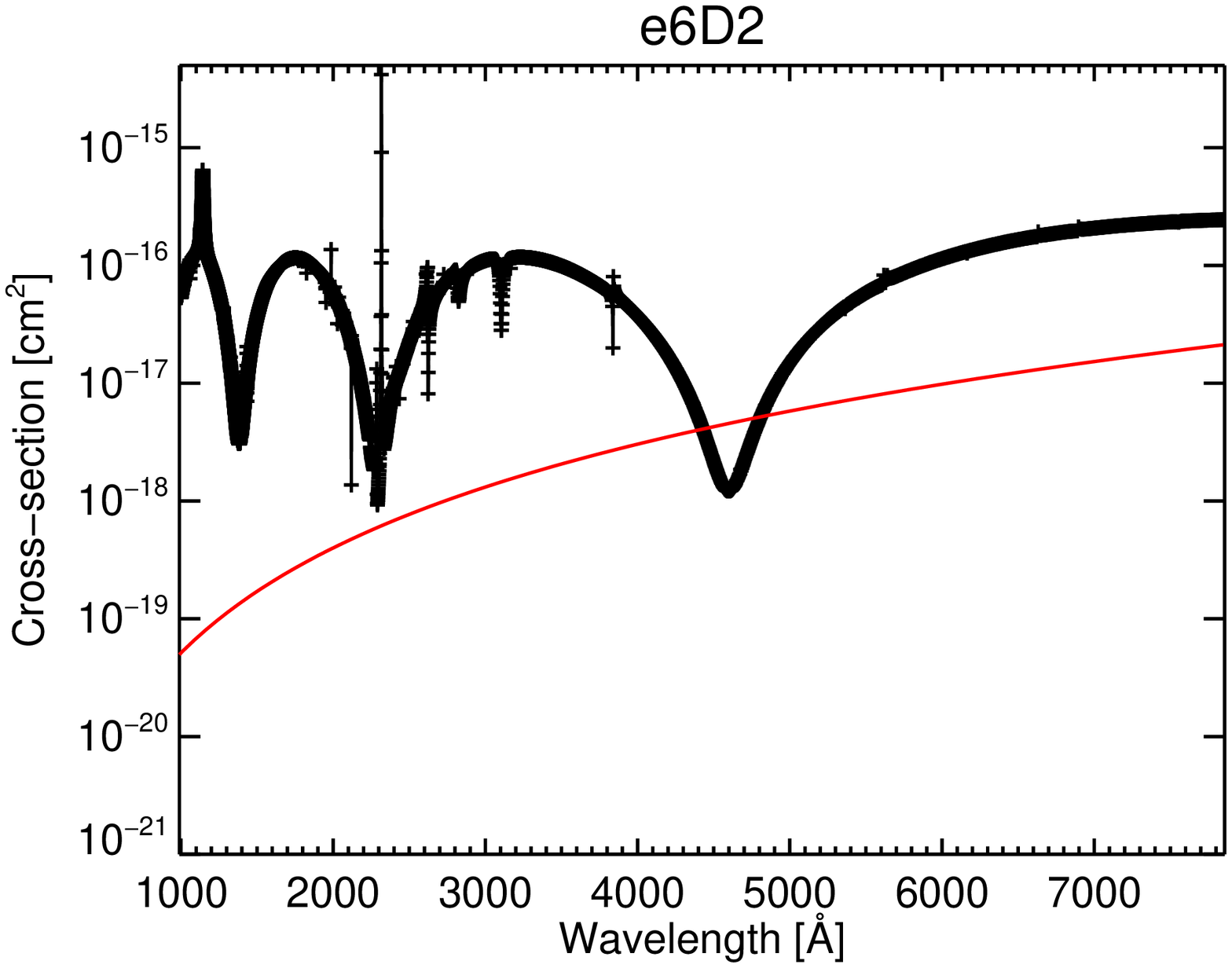}}
\caption{Photo-ionisation cross-sections for selected \mni~levels with quantum-mechanical data: top to bottom: a6D3, z6P3*, b4P1, e6S2, x6P3*, e6D2. Red lines illustrate the hydrogenic cross-sections computed using the effective principal quantum number.}
\label{fig:photo_app}
\end{figure*}

\newpage 

\begin{figure}[!ht]
\includegraphics[width=0.5\textwidth, angle=0]{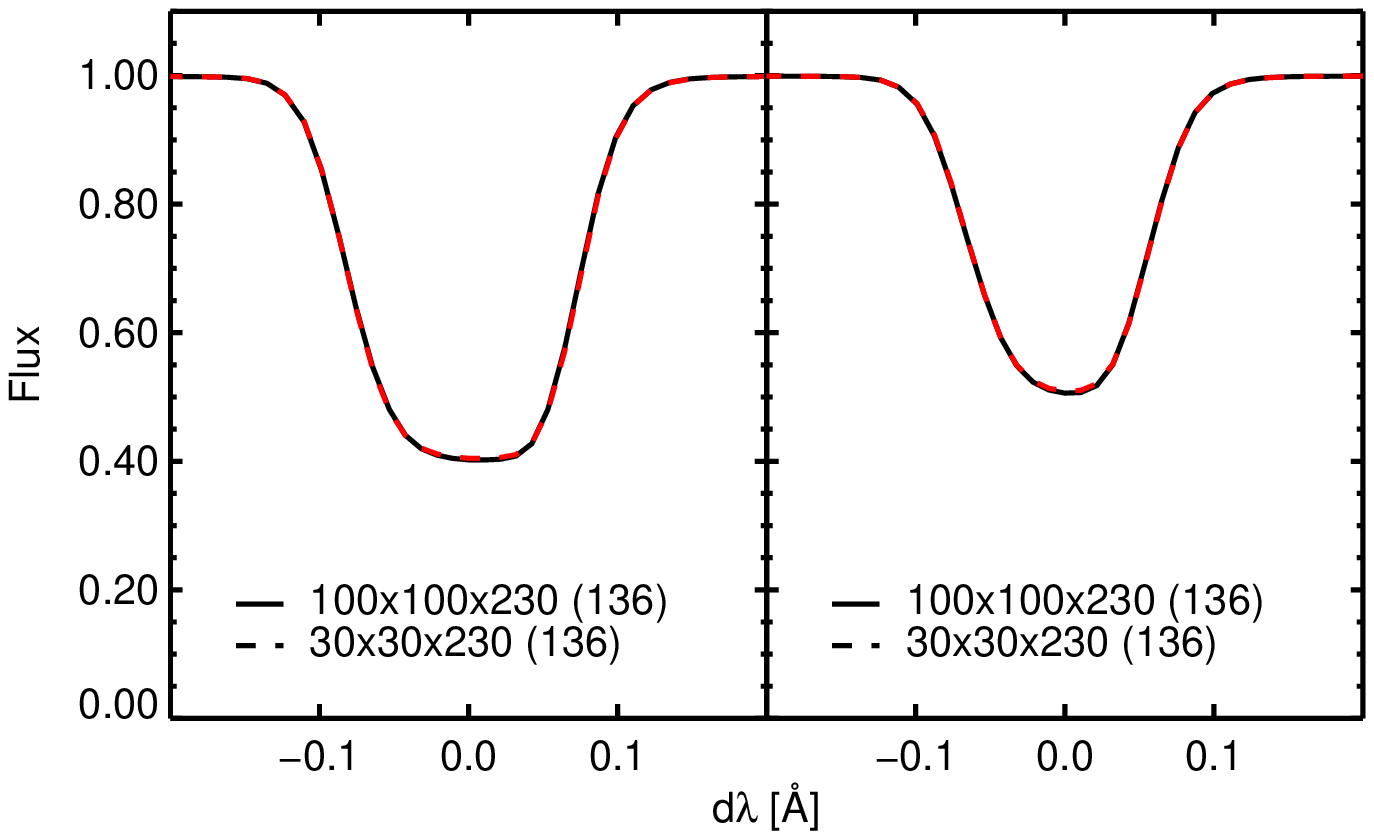}
\includegraphics[width=0.5\textwidth, angle=0]{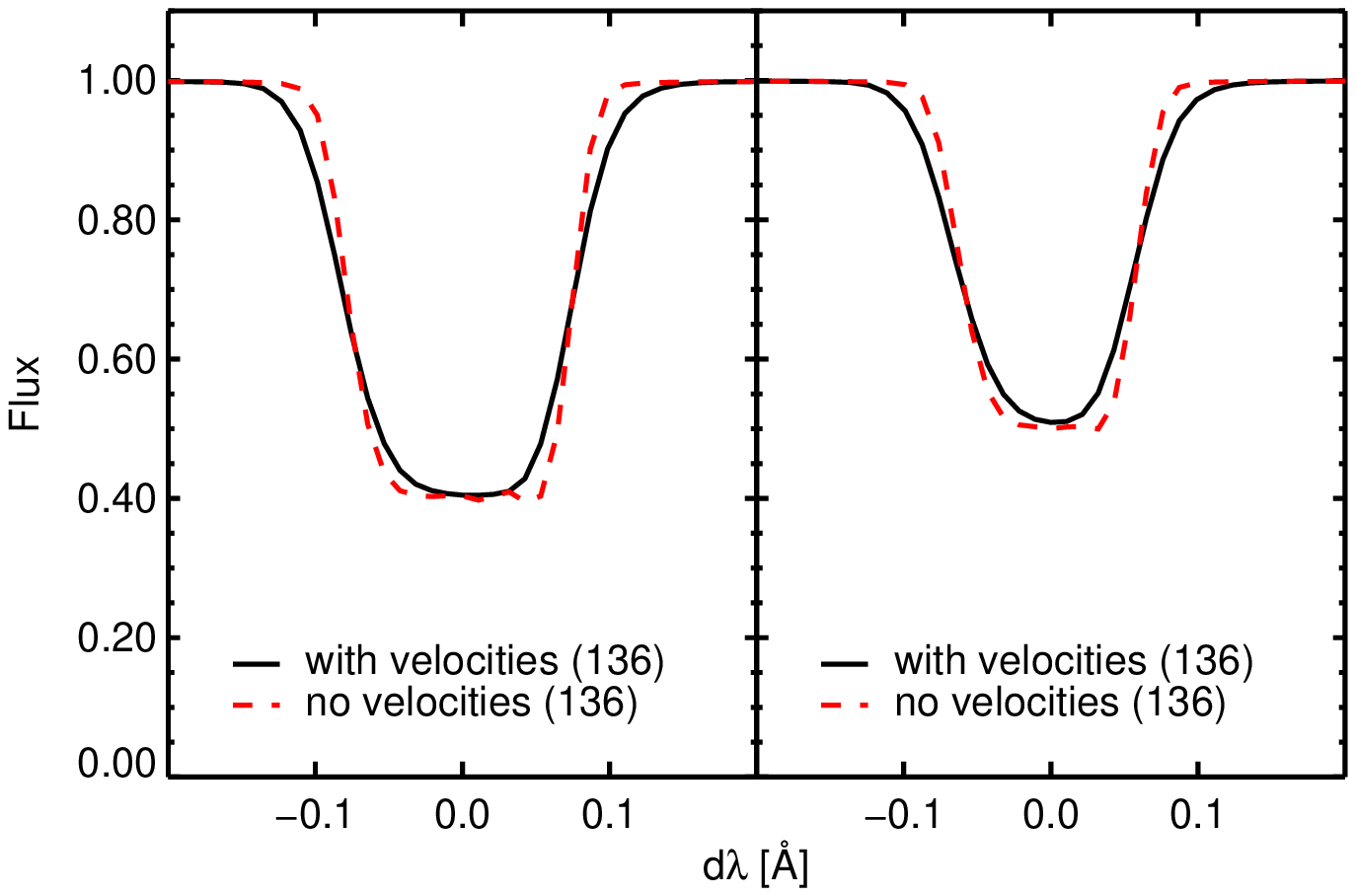}
\includegraphics[width=0.5\textwidth, angle=0]{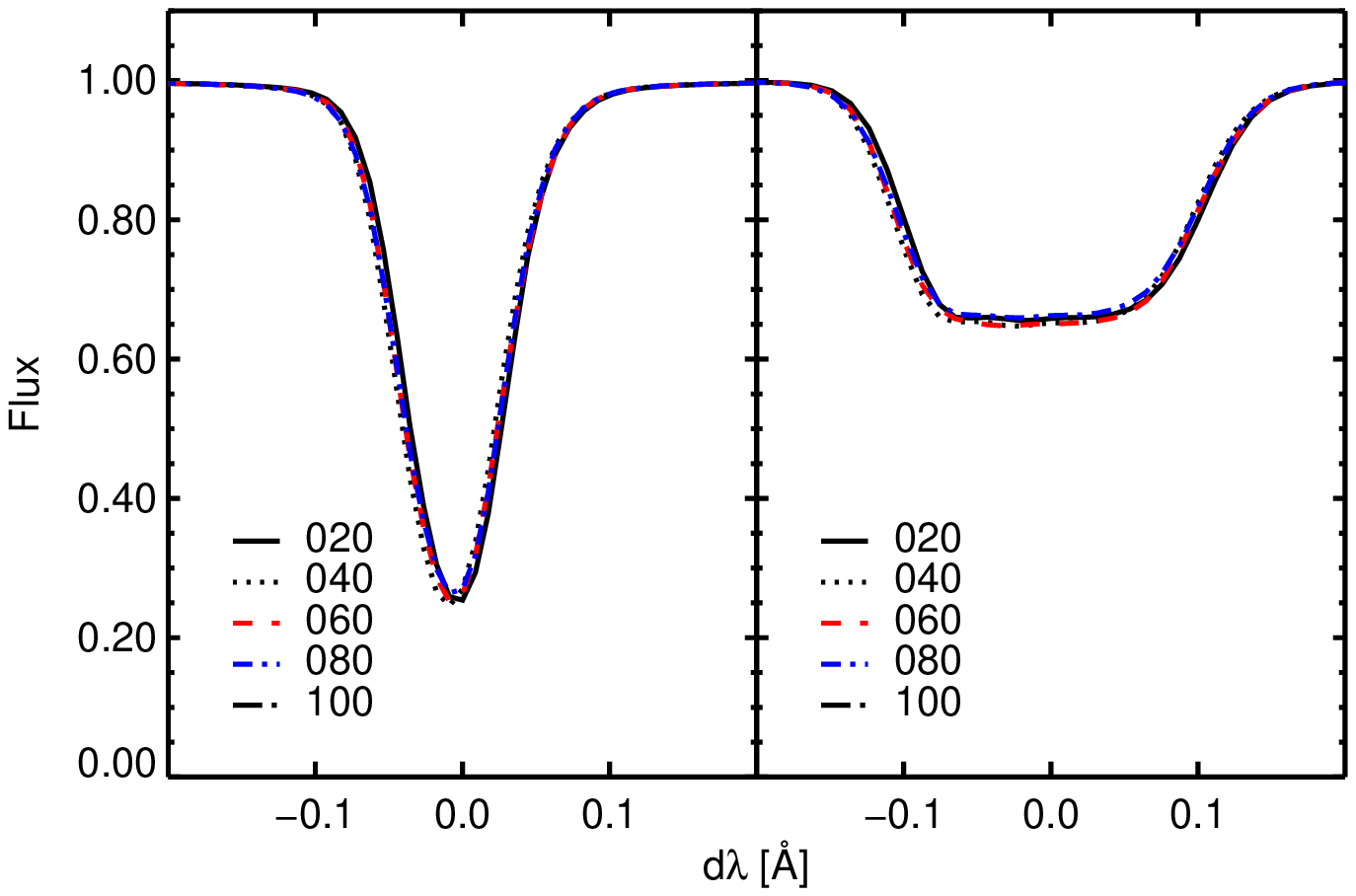}
\caption{A comparison of theoretical profiles for the Mn I lines at 5394 and 5432 \AA\ in the 3D model of the Sun. The impact of the resolution of the 3D atmosphere cube (top panel), radiative transfer calculations with or without velocity field in opacity (middle panel). The bottom panel demonstrates the line profiles computed for $5$ different solar snapshots selected at regular time intervals.}
\label{fig:other3d}
\end{figure}
\newpage
\section{Tables}

\begin{table*}[h]
\caption{LS terms of the target \ion{Mn}{ii} ion included in the close 
coupling expansion.}
\label{fe2target}    
\centering          
\begin{tabular}[scriptsize]{llllllllllll}     
\hline\hline       
Configuration&Term&\multicolumn{2}{c}{Energy (Ryd)} 
& Configuration&Term&\multicolumn{2}{c}{Energy (Ryd)}&  Configuration&Term&\multicolumn{2}{c}{Energy (Ryd)}
\cr
 & & Theory & Exp. & & & Theory & Exp. & & & Theory & Exp. \cr
\\
\hline
3d$^5$($^6$S)4s &$^7$S    &0.0000 &0.0000  & 3d$^5$($^2$G)4s &$^3$G    &0.5430& 0.4508  & 3d$^5$($^4$D)4p &$^3$D$^o$ &0.7630& 0.6688  \cr
3d$^5$($^6$S)4s &$^5$S    &0.1048 &0.0863  & 3d$^5$($^4$F)4s &$^3$F    &0.5491& 0.4544  & 3d$^5$($^4$D)4p &$^3$F$^o$ &0.7678& 0.6720  \cr
3d$^6$  &     $^5$D       &0.1560 &0.1331  & 3d$^5$($^2$F)4s &$^1$F    &0.5495& 0.4290  & 3d$^5$($^2$I)4p &$^3$I$^o$ &0.7847& 0.7137  \cr
3d$^5$($^4$G)4s &$^5$G    &0.2983 &0.2513  & 3d$^5$($^2$H)4s &$^1$H    &0.5607& 0.4698  & 3d$^5$($^4$P)4p &$^3$S$^o$ &0.7881& 0.6735  \cr
3d$^6$       &$^3$H       &0.3230 &0.2793  & 3d$^5$($^2$F)4s &$^3$F    &0.5742& 0.4773  & 3d$^5$($^2$G)4s &$^1$G     &0.7973& 0.6620  \cr
3d$^6$       &$^3$P       &0.3358 &0.2764  & 3d$^5$($^2$G)4s &$^1$G    &0.5760& 0.4798  & 3d$^5$($^4$D)4p &$^3$P$^o$ &0.7976& 0.6907  \cr
3d$^5$($^4$P)4s &$^5$P    &0.3406 &0.2726  & 3d$^6$       &$^3$P       &0.5859& 0.4853  & 3d$^5$($^2$I)4p &$^1$H$^o$ &0.7978& 0.7209  \cr
3d$^6$       &$^3$F       &0.3436 &0.2882  & 3d$^6$       &$^3$F       &0.5929& 0.4900  & 3d$^5$($^2$I)4p &$^3$H$^o$ &0.8043& 0.7263  \cr
3d$^6$       &$^3$G       &0.3488 &0.3178  & 3d$^5$($^2$F)4s &$^1$F    &0.6204& 0.5081  & 3d$^5$(a $^2$D)4p&$^1$D$^o$ &0.8247& 0.7191  \cr
3d$^5$($^6$S)4p&$^7$P$^o$ &0.3564 &0.3518  & 3d$^5$($^2$S)4s &$^3$S    &0.6371& 0.5184  & 3d$^5$(a $^2$D)4p&$^3$F$^o$ &0.8259& 0.7260  \cr
3d$^5$($^4$D)4s &$^5$S    &0.3660 &0.2992  & 3d$^5$($^4$G)4p&$^5$G$^o$ &0.6402& 0.5878  & 3d$^5$($^2$I)4p &$^1$I$^o$ &0.8374& 0.7454  \cr
3d$^5$($^4$G)4s &$^3$G    &0.3828 &0.3027  & 3d$^6$       &$^1$G       &0.6486& 0.5425  & 3d$^5$(a $^2$F)4p&$^1$G$^o$ &0.8449& 0.7407  \cr
3d$^6$        &$^3$D      &0.4022 &0.3448  & 3d$^4$4s$^2$    &$^5$D    &0.6552& 0.5043  & 3d$^5$($^4$F)4p &$^5$G$^o$ &0.8508& 0.7475  \cr
3d$^6$       &$^1$I       &0.4043 &0.3528  & 3d$^5$($^4$G)4p&$^5$H$^o$ &0.6555& 0.5986  & 3d$^5$(a $^2$D)4p&$^3$P$^o$ &0.8524& 0.7409  \cr
3d$^5$($^4$P)4s &$^3$P    &0.4133 &0.3311  & 3d$^5$($^4$G)4p&$^5$F$^o$ &0.6680& 0.6077  & 3d$^5$($^2$F)4p &$^3$G$^o$ &0.8528& 0.7493  \cr
3d$^6$       &$^1$G       &0.4133 &0.3545  & 3d$^5$($^4$P)4p&$^5$D$^o$ &0.6839& 0.6102  & 3d$^5$($^4$F)4p &$^5$F$^o$ &0.8543& 0.7473  \cr
3d$^5$($^6$S)4p&$^5$P$^o$ &0.4331 &0.3959  & 3d$^5$($^4$P)4p&$^5$S$^o$ &0.6858& 0.6099  & 3d5.(a $^2$D)4p  &$^1$F$^o$ &0.8549& 0.7401  \cr
3d$^5$($^2$I)4s &$^3$I    &0.4373 &0.3754  & 3d$^5$($^4$G)4p&$^3$F$^o$ &0.6887& 0.6181  & 3d$^5$($^2$D)4p &$^3$D$^o$ &0.8591& 0.7447  \cr
3d$^5$($^4$D)4s &$^3$D    &0.4449 &0.3628  & 3d$^5$($^4$G)4p&$^3$H$^o$ &0.6925& 0.6181  & 3d$^5$($^4$F)4p &$^5$D$^o$ &0.8604& 0.7536  \cr
3d$^6$       &$^1$S       &0.4478 &0.3714  & 3d$^5$($^2$D)4s &$^3$D    &0.6977& 0.5702  & 3d$^5$(a $^2$F)4p&$^3$F$^o$ &0.8650& 0.7553  \cr
3d$^5$($^2$I)4s &$^1$I    &0.4720 &0.4038  & 3d$^5$($^4$P)4p&$^5$P$^o$ &0.7002& 0.6231  & 3d$^5$(a $^2$F)4p&$^3$D$^o$ &0.8673& 0.7541  \cr
3d$^5$($^4$F)4s &$^5$F    &0.4810 &0.3972  & 3d$^5$($^4$D)4p&$^5$F$^o$ &0.7185& 0.6419  & 3d$^5$(a $^2$G)4p&$^3$H$^o$ &0.8674& 0.7698  \cr
3d$^6$       &$^1$D       &0.4812 &0.3930  & 3d$^5$($^4$P)4p&$^3$P$^o$ &0.7193& 0.6301  & 3d$^5$($^4$F)4p &$^3$G$^o$ &0.8741& 0.7646  \cr
3d$^5$($^2$F)4s &$^3$F    &0.4965 &0.4074  & 3d$^5$($^2$D)4s &$^1$D    &0.7243& 0.5975  & 3d$^5$($^2$H)4p &$^3$I$^o$ &0.8824& 0.7805  \cr
3d$^5$($^2$D)4s &$^3$D    &0.4990 &0.3968  & 3d$^5$($^4$G)4p&$^3$G$^o$ &0.7289& 0.6427  & 3d$^5$(a $^2$D)4p&$^1$P$^o$ &0.8864& 0.7679  \cr
3d$^6$       &$^1$F       &0.5099 &0.4492  & 3d$^5$($^4$D)4p&$^5$P$^o$ &0.7335& 0.6501  & 3d$^5$($^2$H)4p &$^3$G$^o$ &0.8890& 0.7804  \cr
3d$^5$($^2$D)4s &$^1$D    &0.5254 &0.4274  & 3d$^5$($^4$D)4p&$^5$D$^o$ &0.7412& 0.6579  & 3d$^5$(a $^2$F)4p&$^1$D$^o$ &0.8973& 0.7779  \cr
3d$^5$($^2$H)4s &$^3$H    &0.5269 &0.4408  & 3d$^5$($^4$P)4p&$^3$D$^o$ &0.7431& 0.6459  & 3d$^5$($^2$I)4p &$^1$H$^o$ &0.9274& 0.7212  \cr
\hline                  
\end{tabular}
\end{table*}

\begin{table}
\caption{{\sc autostructure} configuration expansions for \ion{Mn}{ii}.\label{table:AUTOconfigs}}
\centering                          
\begin{tabular}{l}        
Configuration expansion\\
\hline\hline
3d$^6$, 3d$^5$4s, 3d$^5$4p, 3d$^5$4d, 3d$^5$5s, 3d$^5$5p,
3d$^4$4s$^2$, \\
3d$^4$4s4p, 3d$^4$4s4d, 3d$^4$4p$^2$\\ 
\hline
$\lambda_{1s}=1.00000,$ 
$\lambda_{2s}=1.25438,$ 
$\lambda_{2p}=1.10633,$
$\lambda_{3s}=1.10274,$\\
$\lambda_{3p}=1.06738,$
$\lambda_{3d}=1.06340,$
$\lambda_{4s}=1.02844,$
$\lambda_{4p}=1.11707,$\\
$\lambda_{4d}=1.46349,$
$\lambda_{5s}= 1.00000,$
$\lambda_{5p}=1.00000$
\\
\hline                                   
\end{tabular}
\end{table}

\begin{table}
\renewcommand{\tabcolsep}{4.5pt}
\caption{Scattering channels correlated to MgH molecular $^5\Sigma^+$ states, asymptotic energies ($J$-average experimental values taken from NIST) with respect to the ground state and electronic bound energies with respect to the ionization limit $\rm{Mn^+(3d^54s\,^5S) + H}$.}
\label{tab:states_MnH}
\begin{center}
\begin{tabular}{llcc}
j & Scattering channels & Asymptotic & Bound\\
  &                     & energy (eV)& energy (eV)\\
\noalign{\smallskip}\hline\noalign{\smallskip} 
 1 & $\rm{Mn(3d^54s^2 \,\,a^6S ) + H}$        & 0.000000 & -8.60854\\
 2 & $\rm{Mn(3d^64s   \,\,a^6D ) + H}$        & 2.145077 & -6.46346\\
 3 & $\rm{Mn(3d^64s   \,\,a^4D ) + H}$        & 2.914772 & -5.69377\\
 4 & $\rm{Mn(3d^54s4p \,\,z^6P^{\circ}) + H}$ & 3.073870 & -5.53467\\
 5 & $\rm{Mn(3d^54s^2 \,\,a^4G ) + H}$        & 3.134155 & -5.47438\\
 6 & $\rm{Mn(3d^54s^2 \,\,b^4D ) + H}$        & 3.768363 & -4.84018\\
 7 & $\rm{Mn(3d^54s4p \,\,z^4P^{\circ}) + H}$ & 3.849324 & -4.75921\\
 8 & $\rm{Mn(3d^54s4p \,\,y^6P^{\circ}) + H}$ & 4.430886 & -4.17765\\
 9 & $\rm{Mn(3d^64s   \,\,b^4G ) + H}$    & 4.663140 & -3.94540\\
10 & $\rm{Mn(3d^54s5s \,\,e^6S ) + H}$    & 5.133436 & -3.47510\\
11 & $\rm{Mn(3d^54s4d \,\,e^6D ) + H}$    & 5.853606 & -2.75493\\
12 & $\rm{Mn(3d^54s5p \,\,w^6P^{\circ}) + H}$ & 5.897434 & -2.71110\\
13 & $\rm{Mn(3d^54s4p \,\,y^6F^{\circ}) + H}$ & 5.972531 & -2.63601\\
14 & $\rm{Mn(3d^54s5s \,\,f^6S ) + H}$     & 6.126726 & -2.48181\\
15 & $\rm{Mn(3d^54s5s \,\,e^4S ) + H}$     & 6.148568 & -2.45997\\
16 & $\rm{Mn(3d^54s4p \,\,v^6P^{\circ}) + H}$ & 6.196287 & -2.41225\\
17 & $\rm{Mn(3d^54s4p \,\,z^4H^{\circ}) + H}$ & 6.209216 & -2.39932\\
18 & $\rm{Mn(3d^54s4p \,\,y^4F^{\circ}) + H}$ & 6.243758 & -2.36478\\
19 & $\rm{Mn(3d^54s4p \,\,x^6F^{\circ}) + H}$ & 6.328611 & -2.27993\\
20 & $\rm{Mn(3d^54s6s \,\,g^6S ) + H}$     & 6.311380 & -2.29716\\
21 & $\rm{Mn(3d^54s4p \,\,x^4P^{\circ}) + H}$ & 6.371961 & -2.23658\\
22 & $\rm{Mn(3d^54s4p \,\,u^6P^{\circ}) + H}$ & 6.466860 & -2.14168\\
23 & $\rm{Mn(3d^54s5d \,\,f^6D ) + H}$     & 6.537766 & -2.07077\\
24 & $\rm{Mn(3d^54s4f \,\,w^6F^{\circ}) + H}$ & 6.568422 & -2.04012\\
25 & $\rm{Mn(3d^54s6p \,\,t^6P^{\circ}) + H}$ & 6.606162 & -2.00238\\
26 & $\rm{Mn(3d^54s7s \,\,h^6S ) + H}$     & 6.752220 & -1.85632\\
27 & $\rm{Mn(3d^54s4d \,\,g^6D ) + H}$     & 6.812434 & -1.79610\\
28 & $\rm{Mn(3d^54s5p \,\,w^4P^{\circ}) + H}$ & 6.868987 & -1.73955\\
29 & $\rm{Mn(3d^54s5f \,\,v^6F^{\circ}) + H}$ & 6.880172 & -1.72837\\
30 & $\rm{Mn(3d^54s7p \,\,r^6P^{\circ}) + H}$ & 6.880244 & -1.72829\\
31 & $\rm{Mn(3d^54s6d \,\,h^6D ) + H}$        & 6.904357 & -1.70418\\
32 & $\rm{Mn(3d^54s5p \,\,t^4P^{\circ}) + H}$ & 6.934452 & -1.67409\\
33 & $\rm{Mn(3d^54s5p \,\,s^6P^{\circ}) + H}$ & 6.944067 & -1.66447\\
34 & $\rm{Mn(3d^65s   \,\,i^6D ) + H}$        & 6.989790 & -1.61875\\
35 & $\rm{Mn(3d^54s6f \,\,u^6F^{\circ}) + H}$ & 7.050626 & -1.55791\\
36 & $\rm{Mn(3d^54s8p \,\,q^6P^{\circ}) + H}$ & 7.058384 & -1.55016\\
37 & $\rm{Mn(3d^54s4p \,\,v^4P^{\circ}) + H}$ & 7.116923 & -1.49162\\
38 & $\rm{Mn(3d^54s7f \,\,t^6F^{\circ}) + H}$ & 7.153545 & -1.45499\\
39 & $\rm{Mn(3d^54s9p \,\,p^6P^{\circ}) + H}$ & 7.156234 & -1.45230\\
40 & $\rm{Mn(3d^54s10p\,\,o^6P^{\circ}) + H}$ & 7.220870 & -1.38767\\
41 & $\rm{Mn(3d^54s11p\,\, ^6P^{\circ}) + H}$ & 7.265446 & -1.34309\\
42 & $\rm{Mn(3d^54s12p\,\, ^6P^{\circ}) + H}$ & 7.297423 & -1.31112\\
43 & $\rm{Mn(3d^54s13p\,\, ^6P^{\circ}) + H}$ & 7.321116 & -1.28742\\
44 & $\rm{Mn(3d^54s14p\,\, ^6P^{\circ}) + H}$ & 7.339133 & -1.26941\\
45 & $\rm{Mn(3d^54s15p\,\, ^6P^{\circ}) + H}$ & 7.351767 & -1.25677\\
46 & $\rm{Mn(3d^54s16p\,\, ^6P^{\circ}) + H}$ & 7.364333 & -1.24421\\
47 & $\rm{Mn(3d^54s17p\,\, ^6P^{\circ}) + H}$ & 7.373379 & -1.23516\\
48 & $\rm{Mn(3d^54s18p\,\, ^6P^{\circ}) + H}$ & 7.380681 & -1.22786\\
49 & $\rm{Mn(3d^54s19p\,\, ^6P^{\circ}) + H}$ & 7.386776 & -1.22176\\
50 & $\rm{Mn(3d^54s20p\,\, ^6P^{\circ}) + H}$ & 7.391888 & -1.21665\\
\noalign{\smallskip}\hline\noalign{\smallskip}            
\end{tabular}                                             
\end{center}                                              
\end{table}                                               
                                                          
\begin{table}                                             
\renewcommand{\tabcolsep}{4.5pt}                          
\begin{center}                                            
\begin{tabular}{llcc}                                      
j & Scattering channel  & Asymptotic  & Bound\\                  
  &                     & energy (eV) & energy (eV)\\                  
\noalign{\smallskip}\hline\noalign{\smallskip}            
51 & $\rm{Mn(3d^54s21p\,\, ^6P^{\circ}) + H}$ & 7.396210 & -1.21233\\
52 & $\rm{Mn(3d^54s22p\,\, ^6P^{\circ}) + H}$ & 7.399904 & -1.20864\\
53 & $\rm{Mn(3d^54s23p\,\, ^6P^{\circ}) + H}$ & 7.403078 & -1.20546\\
54 & $\rm{Mn(3d^54s24p\,\, ^6P^{\circ}) + H}$ & 7.405828 & -1.20271\\
55 & $\rm{Mn(3d^54s25p\,\, ^6P^{\circ}) + H}$ & 7.408231 & -1.20031\\
56 & $\rm{Mn(3d^54s26p\,\, ^6P^{\circ}) + H}$ & 7.410336 & -1.19820\\
57 & $\rm{Mn(3d^54s27p\,\, ^6P^{\circ}) + H}$ & 7.412199 & -1.19634\\
58 & $\rm{Mn(3d^54s28p\,\, ^6P^{\circ}) + H}$ & 7.413848 & -1.19469\\
59 & $\rm{Mn(3d^54s29p\,\, ^6P^{\circ}) + H}$ & 7.415318 & -1.19322\\
60 & $\rm{Mn(3d^54s30p\,\, ^6P^{\circ}) + H}$ & 7.416628 & -1.19191\\
61 & $\rm{Mn(3d^54s31p\,\, ^6P^{\circ}) + H}$ & 7.417802 & -1.19074\\
62 & $\rm{Mn(3d^54s32p\,\, ^6P^{\circ}) + H}$ & 7.418876 & -1.18966\\
63 & $\rm{Mn(3d^54s33p\,\, ^6P^{\circ}) + H}$ & 7.419839 & -1.18870\\
64 & $\rm{Mn(3d^54s34p\,\, ^6P^{\circ}) + H}$ & 7.420701 & -1.18784\\
65 & $\rm{Mn(3d^54s35p\,\, ^6P^{\circ}) + H}$ & 7.421504 & -1.18704\\
66 & $\rm{Mn(3d^54s36p\,\, ^6P^{\circ}) + H}$ & 7.422245 & -1.18629\\
67 & $\rm{Mn(3d^54s37p\,\, ^6P^{\circ}) + H}$ & 7.422891 & -1.18565\\
68 & $\rm{Mn(3d^54s38p\,\, ^6P^{\circ}) + H}$ & 7.423498 & -1.18504\\
69 & $\rm{Mn(3d^54s39p\,\, ^6P^{\circ}) + H}$ & 7.424055 & -1.18448\\
70 & $\rm{Mn(3d^54s40p\,\, ^6P^{\circ}) + H}$ & 7.424578 & -1.18396\\
71 & $\rm{Mn(3d^54s41p\,\, ^6P^{\circ}) + H}$ & 7.425049 & -1.18349\\
72 & $\rm{Mn^+(3d^54s\,^5S) + H^-}$           & 7.854539 & -0.754\\
\noalign{\smallskip}\hline\noalign{\smallskip}
\end{tabular}
\end{center}
\end{table}

\begin{table}
\renewcommand{\tabcolsep}{4.5pt}
\caption{Scattering channels correlated to MgH$^+$ molecular $^6\Sigma^+$ states, asymptotic energies ($J$-average experimental values taken from NIST with respect to the ground state and electronic bound energies with respect to the ionization limit $\rm{Mn^{2+}(3d^5\,^6S) + H}$.}
\label{tab:states_MnH+}
\begin{center}
\begin{tabular}{llcc}
j & Scattering channel  & Asymptotic & Bound\\
  &                     & energy (eV)& energy (eV)\\
\noalign{\smallskip}\hline\noalign{\smallskip} 
 1 & $\rm{Mn^+(3d^54s\,\,a^7S ) + H}$        & 0.000000 & -15.6400\\
 2 & $\rm{Mn^+(3d^54s\,\,a^5S ) + H}$        & 1.174501 & -14.4655\\
 3 & $\rm{Mn^+(3d^6  \,\,a^5D ) + H}$        & 1.808479 & -13.8315\\
 4 & $\rm{Mn^+(3d^54s\,\,a^5G ) + H}$     & 3.418429 & -12.2216\\
 5 & $\rm{Mn^+(3d^54s\,\,b^5D ) + H}$        & 4.070230 & -11.5698\\
 6 & $\rm{Mn^+(3d^54p\,\,z^7P^{\circ}) + H}$ & 4.787521 & -10.8525\\
 7 & $\rm{Mn^+(3d^54p\,\,z^5P^{\circ}) + H}$ & 5.386607 & -10.2534\\
 8 & $\rm{Mn^+(3d^54p\,\,z^5H^{\circ}) + H}$ & 8.145176 & -7.49481\\
 9 & $\rm{Mn^+(3d^54p\,\,z^5F^{\circ}) + H}$ & 8.267108 & -7.37288\\
10 & $\rm{Mn^+(3d^54p\,\,y^5P^{\circ}) + H}$ & 8.476972 & -7.16302\\
11 & $\rm{Mn^+(3d^54p\,\,y^5F^{\circ}) + H}$ & 8.734600 & -6.90539\\
12 & $\rm{Mn^+(3d^54p\,\,x^5P^{\circ}) + H}$ & 8.845407 & -6.79458\\
13 & $\rm{Mn^+(3d^55s\,\,e^7S ) + H}$     & 9.244271 & -6.39572\\
14 & $\rm{Mn^+(3d^55s\,\,e^5S ) + H}$     & 9.469252 & -6.17074\\
15 & $\rm{Mn^+(3d^54d\,\,e^7D ) + H}$     & 9.863794 & -5.77620\\
16 & $\rm{Mn^+(3d^54p\,\,x^5F^{\circ}) + H}$ & 10.16684 & -5.47314\\
17 & $\rm{Mn^+(3d^54d\,\,e^5D ) + H}$     & 10.18466 & -5.45532\\
18 & $\rm{Mn^+(3d^55p\,\,x^7P^{\circ}) + H}$ & 10.66097 & -4.97902\\
19 & $\rm{Mn^+(3d^55p\,\,v^5P^{\circ}) + H}$ & 11.05872 & -4.58127\\
20 & $\rm{Mn^+(3d^56s\,\,f^7S ) + H}$     & 12.11680 & -3.52319\\
21 & $\rm{Mn^+(3d^56s\,\,f^5S ) + H}$      & 12.20133 & -3.43865\\
22 & $\rm{Mn^+(3d^54f\,\,z^7F^{\circ}) + H}$ & 12.20296 & -3.43702\\
23 & $\rm{Mn^+(3d^54f\,\,e^5F^{\circ}) + H}$ & 12.20797 & -3.43202\\
24 & $\rm{Mn^+(3d^55d\,\,f^7D ) + H}$      & 12.38580 & -3.25419\\
25 & $\rm{Mn^+(3d^55d\,\,f^5D ) + H}$      & 12.48374 & -3.15624\\
26 & $\rm{Mn^+(3d^55s\,\,e^5G ) + H}$     & 12.58296 & -3.05702\\
27 & $\rm{Mn^+(3d^54d\,\,e^5I ) + H}$     & 13.20636 & -2.43362\\
28 & $\rm{Mn^+(3d^55s\,\,g^5D ) + H}$     & 13.25794 & -2.38204\\
29 & $\rm{Mn^+(3d^57s\,\,g^7S ) + H}$     & 13.40594 & -2.23404\\
30 & $\rm{Mn^+(3d^55f\,\,y^7F^{\circ}) + H}$ & 13.44114 & -2.19885\\
31 & $\rm{Mn^+(3d^55f\,\,f^5F^{\circ}) + H}$ & 13.44490 & -2.19509\\
32 & $\rm{Mn^+(3d^57s\,\,g^5S ) + H}$     & 13.44576 & -2.19423\\
33 & $\rm{Mn^+(3d^55g\,\,e^7G ) + H}$     & 13.45920 & -2.18079\\
34 & $\rm{Mn^+(3d^55g\,\,f^5G ) + H}$        & 13.45925 & -2.18074\\
35 & $\rm{Mn^+(3d^56d\,\,g^7D ) + H}$     & 13.54475 & -2.09524\\
36 & $\rm{Mn^+(3d^55p\,\,s^5F^{\circ}) + H}$ & 14.09071 & -1.54927\\
37 & $\rm{Mn^+(3d^58s\,\,h^7S ) + H}$     & 14.09672 & -1.54327\\
38 & $\rm{Mn^+(3d^56f\,\,x^7F^{\circ}) + H}$ & 14.11439 & -1.52559\\
39 & $\rm{Mn^+(3d^56g\,\,f^7G ) + H}$     & 14.12574 & -1.51425\\
40 & $\rm{Mn^+(3d^56g\,\,g^5G ) + H}$     & 14.12579 & -1.51420\\
41 & $\rm{Mn^+(3d^56f\,\,g^5F^{\circ}) + H}$ & 14.13750 & -1.50249\\
42 & $\rm{Mn^+(3d^57d\,\,h^5D ) + H}$     & 14.25117 & -1.38882\\
43 & $\rm{Mn^{2+}(3d^5\,^6S) + H^-}$         & 14.88599 & -0.754\\
\noalign{\smallskip}\hline\noalign{\smallskip}
\end{tabular}
\end{center}
\end{table}

\begin{table*}
  \begin{center}
    \caption{NLTE abundance corrections for the MARCS models with $\teff = 4500$ K and $\log g = 1.5$ dex. See Fig. 9 in the main text.}
\setlength{\tabcolsep}{0.02\linewidth}
\begin{tabular}{c c c c c}
\hline
\noalign{\smallskip}
$\lambda \ [\AA]$ & [Fe/H] & CH excl. & CH0 excl. & CH/CH0 incl.  \\ 
\noalign{\smallskip}
\hline
\noalign{\smallskip}
4030                           & 0                              & 0.157                          & 0.176                          & 0.153                           \\ 
4033                           & -1                             & 0.212                          & 0.236                          & 0.222                           \\ 
4034                           & -2                             & 0.378                          & 0.434                          & 0.426                           \\ 
                               & -3                             & 0.473                          & 0.483                          & 0.477                           \\ 
                               & 0                              & 0.198                          & 0.173                          & 0.156                           \\ 
4783                           & -1                             & 0.189                          & 0.214                          & 0.203                           \\ 
4823                           & -2                             & 0.333                          & 0.366                          & 0.358                           \\ 
                               & -3                             & 0.452                          & 0.462                          & 0.456                           \\ 
6013                           & 0                              & 0.203                          & 0.184                          & 0.174                           \\ 
6016                           & -1                             & 0.304                          & 0.288                          & 0.277                           \\ 
6021                           & -2                             & 0.372                          & 0.370                          & 0.364                           \\ 
                               & -3                             & 0.382                          & 0.396                          & 0.392                           \\ 
                               & 0                              & -0.013                         & -0.014                         & -0.013                          \\ 
3488                           & -1                             & -0.035                         & -0.035                         & -0.035                          \\ 
                               & -2                             & -0.078                         & -0.079                         & -0.079                          \\ 
                               & -3                             & 0.076                          & 0.074                          & 0.074                           \\ 
\noalign{\smallskip}
\hline
\end{tabular}
\end{center}
\label{tableN1}
\end{table*}

\begin{table*}
  \begin{center}
        \caption{NLTE abundance corrections for the MARCS models with $\teff = 6000$ K and $\log g = 4.0$ dex. See Fig. 9 in the main text.}

\setlength{\tabcolsep}{0.02\linewidth}
\begin{tabular}{c c c c c}
\hline
\noalign{\smallskip}
$\lambda \ [\AA]$ & [Fe/H] & CH excl. & CH0 excl. & CH/CH0 incl.  \\ 
\noalign{\smallskip}
\hline
\noalign{\smallskip}
4030                           & 0                              & 0.043                          & 0.078                          & 0.042                           \\ 
4033                           & -1                             & -0.043                         & 0.002                          & -0.033                          \\ 
4034                           & -2                             & 0.124                          & 0.183                          & 0.113                           \\ 
                               & -3                             & 0.469                          & 0.529                          & 0.481                           \\ 
                               & 0                              & 0.002                          & 0.038                          & 0.014                           \\ 
4783                           & -1                             & 0.135                          & 0.186                          & 0.154                           \\ 
4823                           & -2                             & 0.272                          & 0.308                          & 0.274                           \\ 
                               & -3                             & 0.315                          & 0.354                          & 0.317                           \\ 
6013                           & 0                              & 0.123                          & 0.109                          & 0.090                           \\ 
6016                           & -1                             & 0.220                          & 0.226                          & 0.205                           \\ 
6021                           & -2                             & 0.280                          & 0.341                          & 0.322                           \\ 
                               & -3                             & 0.288                          & 0.395                          & 0.378                           \\ 
                               & 0                              & -0.017                         & -0.017                         & -0.016                          \\ 
3488                           & -1                             & -0.082                         & -0.080                         & -0.079                          \\ 
                               & -2                             & -0.038                         & -0.041                         & -0.040                          \\ 
                               & -3                             & 0.098                          & 0.093                          & 0.093                           \\ 
\noalign{\smallskip}
\hline
\end{tabular}
\end{center}
\label{tableN2}
\end{table*}

\begin{table*}
  \begin{center}
    \caption{NLTE abundance corrections for the MARCS models with $\teff = 4500$ K and $\log g = 1.5$ dex. See Fig. 10 in the main text.}
\setlength{\tabcolsep}{0.02\linewidth}
\begin{tabular}{c c c c}
\hline
\noalign{\smallskip}
$\lambda \ [\AA]$ & [Fe/H] & reference atom & no Kaulakys  \\ 
\noalign{\smallskip}
\hline
\noalign{\smallskip}
4030                           & 0                              & 0.138                          & 0.153                           \\ 
4033                           & -1                             & 0.223                          & 0.222                           \\ 
4034                           & -2                             & 0.401                          & 0.426                           \\ 
                               & -3                             & 0.396                          & 0.477                           \\ 
                               & 0                              & 0.132                          & 0.156                           \\ 
4783                           & -1                             & 0.225                          & 0.203                           \\ 
4823                           & -2                             & 0.383                          & 0.358                           \\ 
                               & -3                             & 0.438                          & 0.456                           \\ 
6013                           & 0                              & 0.168                          & 0.174                           \\ 
6016                           & -1                             & 0.292                          & 0.277                           \\ 
6021                           & -2                             & 0.416                          & 0.364                           \\ 
                               & -3                             & 0.442                          & 0.392                           \\ 
                               & 0                              & -0.012                         & -0.013                          \\ 
3488                           & -1                             & -0.035                         & -0.035                          \\ 
                               & -2                             & -0.078                         & -0.079                          \\ 
                               & -3                             & 0.075                          & 0.074                           \\ 
\noalign{\smallskip}
\hline
\end{tabular}
\end{center}
\label{tableN3}
\end{table*}

\begin{table*}
  \begin{center}
    \caption{NLTE abundance corrections for the MARCS models with $\teff = 6000$ K and $\log g = 4.0$ dex. See Fig. 10 in the main text.}
\setlength{\tabcolsep}{0.02\linewidth}
\begin{tabular}{c c c c}
\hline
\noalign{\smallskip}
$\lambda \ [\AA]$ & [Fe/H] & reference atom & no Kaulakys  \\ 
\noalign{\smallskip}
\hline
\noalign{\smallskip}
4030                           & 0                              & 0.029                          & 0.042                           \\ 
4033                           & -1                             & -0.015                         & -0.033                          \\ 
4034                           & -2                             & 0.078                          & 0.113                           \\ 
                               & -3                             & 0.423                          & 0.481                           \\ 
                               & 0                              & 0.013                          & 0.014                           \\ 
4783                           & -1                             & 0.118                          & 0.154                           \\ 
4823                           & -2                             & 0.236                          & 0.274                           \\ 
                               & -3                             & 0.292                          & 0.317                           \\ 
6013                           & 0                              & 0.048                          & 0.090                           \\ 
6016                           & -1                             & 0.134                          & 0.205                           \\ 
6021                           & -2                             & 0.269                          & 0.322                           \\ 
                               & -3                             & 0.344                          & 0.378                           \\ 
                               & 0                              & -0.016                         & -0.016                          \\ 
3488                           & -1                             & -0.079                         & -0.079                          \\ 
                               & -2                             & -0.040                         & -0.040                          \\ 
                               & -3                             & 0.093                          & 0.093                           \\ 
\noalign{\smallskip}
\hline
\end{tabular}
\end{center}
\label{table}
\end{table*}

\end{appendix}
%
%
%
%
\end{document}